\documentclass[fleqn,usenatbib]{mnras}

\usepackage{newtxtext}

\usepackage[T1]{fontenc}

\newcommand{\pD}[2]{\frac{\partial #2}{\partial #1}}

\newcommand{\D}[2]{\frac{{\rm d} #2}{{\rm d} #1}}

\newcommand\bb[1]{\mbox{\boldmath{$#1$}}}
\newcommand\grad{\bb{\nabla}}
\newcommand\bcdot{\,\bb{\cdot}\,}
\newcommand\btimes{\,\bb{\times}\,}
\newcommand{\mc}[1]{\mathcal{#1}}
\newcommand{\mcb}[1]{\bb{\mathcal{#1}}}
\newcommand{\msf}[1]{\mathsf{#1}}
\newcommand{\msb}[1]{\bb{\mathsf{#1}}}

\newcommand{\rmd}{{\rm d}}
\newcommand{\rme}{{\rm e}}
\newcommand{\const}{{\rm const}}

\newcommand{\rvmp}{Rev.~Mod.~Phys.}
\newcommand{\jcomp}{J.~Comput.~Phys.}

\DeclareRobustCommand{\VAN}[3]{#2}
\let\VANthebibliography\thebibliography
\def\thebibliography{\DeclareRobustCommand{\VAN}[3]{##3}\VANthebibliography}

\usepackage{graphicx}	
\usepackage{amsmath}	
\usepackage{amssymb}	

\usepackage[dvipsnames]{xcolor}
\defcitealias{XK21}{Paper~I}

\newcommand{\revision}[1]{{#1}}

\title[Protostellar disc formation and evolution]{Formation and evolution of protostellar accretion discs.\\
II. From 3D simulation to a simple semi-analytic model of Class 0/I discs}

\author[W.~Xu and M.~W.~Kunz]{Wenrui Xu$^1$\thanks{Contact e-mail: \href{mailto:wenruix@princeton.edu}{wenruix@princeton.edu}}
 and Matthew W.~Kunz$^{1,2}$\\
$^1$Department of Astrophysical Sciences, Princeton University, Peyton Hall, Princeton, NJ 08544, USA\\
$^2$Princeton Plasma Physics Laboratory, PO Box 451, Princeton, NJ 08543, USA
}

\date{Accepted XXX. Received YYY; in original form ZZZ}

\pubyear{2021}

\begin{document}
\label{firstpage}
\pagerange{\pageref{firstpage}--\pageref{lastpage}}
\maketitle

\begin{abstract}
We use a 3D radiative non-ideal magnetohydrodynamic (MHD) simulation to investigate the formation and evolution of a young protostellar disc from a magnetized pre-stellar core. The simulation covers the first ${\sim}10~{\rm kyr}$ after protostar formation, and shows a massive, weakly magnetized disc with radius that initially grows and then saturates at ${\sim}30~{\rm au}$. 
The disc is gravitationally unstable with prominent large-amplitude spiral arms.
We use our simulation results and a series of physical arguments to construct a predictive and quantitative physical picture of Class 0/I protostellar disc evolution from several aspects, including (i) the angular-momentum redistribution in the disc, self-regulated by gravitational instability to make most of the disc marginally unstable; (ii) the thermal profile of the disc, well-approximated by a balance between radiative cooling and accretion heating; and (iii) the magnetic-field strength and magnetic-braking rate inside the disc, regulated by non-ideal magnetic diffusion.
Using these physical insights, we build a simple 1D semi-analytic model of disc evolution.
We show that this 1D model, when coupled to a computationally inexpensive simulation for the evolution of the surrounding pseudodisc, can be used reliably to predict disc evolution in the Class 0/I phase. The predicted long-term evolution of disc size, which saturates at ${\sim}30~{\rm au}$ and eventually shrinks, is consistent with a recent observational survey of Class 0/I discs.
Such hierarchical modelling of disc evolution circumvents the computational difficulty of tracing disc evolution through Class 0/I phase with direct, numerically converged simulations.
\end{abstract}




\section{Introduction}

The formation and subsequent evolution of accretion discs around young (Class 0/I) protostars are very important events in the formation of stars and planets.
Such discs constitute an evolutionary bridge between the earliest stages of star formation, during which a pre-stellar core fragments out of its natal molecular cloud and commences dynamical contraction, and the later (Class II) stage, when an optically visible pre-main-sequence star is surrounded by a protoplanetary disc, with planet formation likely already underway.

In \citet[hereafter \citetalias{XK21}]{XK21} we reviewed the various physical factors relevant to protostellar disc formation, such as the initial conditions (core magnetization, molecular cloud turbulence, magnetic-field--rotation misalignment), chemistry (especially the dust-size distribution and its evolution), non-ideal magnetohydrodynamic (MHD) effects (ambipolar, Hall, Ohmic), and radiation.
We also briefly reviewed existing numerical simulations that focus on the consequences of one or more of these physical factors, and pointed out two key issues. First, in terms of physical understanding, there lacks a clear, quantitative, and physically complete picture of protostellar disc formation and evolution. Second, in terms of numerical modeling, the trade-off between resolution and computational cost (mainly real-world time) makes it difficult to achieve numerical convergence in long-term (${\gtrsim}10$ kyr after protostar formation, or covering a significant fraction of the Class 0/I phase) simulations. The second problem, at least in part, contributes to the first problem by making it difficult to extract a complete physical picture from simulations.

We further analyzed in \citetalias{XK21} a series of 2D (axisymmetric) and 3D non-ideal MHD simulations (with the same physical setup, but with different numerical treatments and resolutions), and showed that convergence can be achieved with an affordable resolution if carefully chosen, physically motivated inner boundary conditions are employed. Armed with a numerically reliable and relatively well-resolved 3D simulation, we obtained some basic physical understanding regarding the angular-momentum budget of the protostar-disc system and how the mass and size of the disc is self-regulated by gravitational instability (GI). In this paper, we improve and extend the 3D simulation in \citetalias{XK21} by including a radiation model that produces realistic thermal evolution (\citetalias{XK21} adopted a barotropic equation of state for simplicity) and integrating the system for long enough to cover a significant fraction of the Class 0/I phase. From this simulation, we construct a predictive physical picture and use it to develop a 1D semi-analytic disc model. This model circumvents the high computational cost of direct 3D simulations and enables a large parameter survey (to be presented in the next paper of this series), thereby affording a better understanding of how the various physical factors mentioned earlier together determine the outcome of disc formation.

This paper is organized as follows. In Section \ref{sec:method}, we present our simulation setup, with a focus on the new treatment of thermodynamics. In Section \ref{sec:overview}, we provide an overview of the simulation results. We then construct a physical picture of disc evolution by focusing on several aspects, one at a time. In Section \ref{sec:AMbudget}, we discuss the angular-momentum budget of the disc and how disc growth and protostar accretion are self-regulated by GI. In Section \ref{sec:Tprofile}, we study the thermal evolution of the disc, and show how the disc temperature profile can be estimated using simple analytic arguments. In Section \ref{sec:Bevolution}, we focus on the evolution of the magnetic field, which is modulated by non-ideal MHD effects, and provide estimates for the field strength and magnetic braking efficiency inside the disc. Section \ref{sec:model} assembles these ingredients to construct a simple, semi-analytic 1D model of disc evolution. We show that coupling such a model to a computationally inexpensive, large-inner-boundary pseudodisc simulation produces reliable predictions for disc evolution. Section \ref{sec:compare} compares our results with previous works, and explains the difference between them. Section \ref{sec:summary} summarizes our key results and foreshadows future work in this series.

\section{Method of solution}\label{sec:method}

We perform a 3D non-ideal MHD simulation using the code \verb'Athena++' (\citealt{Stone2020}; with a few new modules and algorithmic adjustments developed for this series of studies) that follows the evolution of a self-gravitating, magnetic, poorly ionized pre-stellar core until ${\sim}10$ kyr after protostar (i.e., point-mass) formation. This evolution includes the formation and early evolution of a massive, rotationally supported, protostellar accretion disc. The simulation setup is largely similar to the fiducial 3D simulation in \citetalias{XK21}, with the main difference being that this study includes a radiation model that provides realistic estimates for radiative cooling and thermal-energy diffusion (whereas in \citetalias{XK21} we adopted a barotropic equation of state). This radiation model is described (and tested) in Section \ref{sec:radmodel} and Appendix \ref{A:radiation}. We also improve the speed of the code by a factor of a few by adopting super-time-stepping for the magnetic diffusion \citep{Meyer2014}. Other minor modifications to our numerical approach are documented in Appendix \ref{A:numerical}.

\subsection{Initial condition}

We start our simulation with an initially spherical pre-stellar core with uniform temperature $T_0 = 10~{\rm K}$ and radially ($r$) dependent number density of neutrals given by
\begin{equation}
    n_{\rm n}(r) = \max \left\{\frac{n_0}{1+(r/r_0)^2}, n_\infty\right\},
\end{equation}
with initial central density $n_0 = 10^4~{\rm cm}^{-3}$ and characteristic scale $r_0=0.1~{\rm pc}$. The background density is $n_\infty = 500~{\rm cm}^{-3}$; it represents the ambient density in the parent molecular cloud and is excluded from the calculation of the self-gravitational potential. We also exclude $n_\infty$ from all integrated diagnostics (e.g., a `density-weighted' integration in the $z$ direction is weighted only by $\max\{0, n_{\rm n}-n_\infty\}$), so that the diagnostics are unaffected by material outside of the pre-stellar core. The mean mass per neutral particle $m_{\rm n}=2.33m_{\rm p}$ (accounting for molecular hydrogen with 20\% He by number), so that the neutral mass density $\rho_{\rm n} = m_{\rm n}n_{\rm n}$ and the initial isothermal sound speed $c_{\rm s0}\equiv (k_{\rm B}T_0/m_{\rm n})^{1/2}\simeq 0.188~{\rm km~s}^{-1}$. The total self-gravitating mass within $r_0$ is then $1.43~{\rm M}_\odot$. We set the core into uniform rotation with an initial angular frequency $\Omega_0 = 0.2 c_{\rm s0}/r_0\simeq 1.22\times 10^{-14}~{\rm rad}~{\rm s}^{-1}$ for $R<R_0$ (where $R=r\sin\theta$ is the cylindrical radius) and $\Omega = \Omega_0(R/r_0)^{-2}$ for $R>r_0$. The core is threaded by a uniform magnetic field with strength $B_0=25~\mu{\rm G}$ that is aligned with the rotation axis, giving an initial mass-to-flux ratio in the central flux tubes of the core of ${\simeq}1.5$ times the critical value for collapse, $(3/2)(63{\rm G})^{-1/2}$ \citep{ms76}. This initial condition is representative of a typical NH$_3$ core \citep{BarrancoGoodman1998,Jijina1999,Crutcher1999}, and is identical to that used in \citetalias{XK21} (section 2.1) except for how $\Omega(R)$ is modulated outside of $r_0$.

\subsection{Non-ideal MHD diffusivities}\label{sec:adod}

We include ambipolar diffusion and Ohmic dissipation in our simulation; the Hall effect is neglected. As in \citetalias{XK21}, the associated diffusivities are calculated using an equilibrium chemical network that includes electrons, atomic and molecular ions, and a distribution of (neutral, singly negatively charged, and singly positively charged) spherical dust grains. The CR ionization rate is $\zeta_{\rm CR}=10^{-17}~{\rm s}^{-1}$. The grain size distribution is that of a truncated MRN \citep{mrn77} distribution with minimum grain size $0.1~\mu{\rm m}$, divided into 5 size bins. These choices are identical to those made in \citetalias{XK21} (section 2.3), except that temperature and density are now treated as independent variables in the calculation of the species abundances.

\subsection{Thermal evolution and radiation model}\label{sec:radmodel}

The thermal evolution of the system is captured by an ideal equation of state with a temperature-dependent ratio of specific heats $\gamma\equiv P/(\rho u)+1$,  where $P$ and $\rho u$ are the thermal pressure and internal energy density, respectively. The latter evolves according to
\begin{equation}
    \pD{t}{(\rho u)} + \grad\bcdot(\rho u\bb{v}) = -P\grad\bcdot\bb{v} + \Gamma_{\rm diff} + Q_{\rm rad} ,    
\end{equation}
where $\bb{v}$ is the velocity of the bulk-neutral fluid, $\Gamma_{\rm diff}$ represents diffusive heating due to ambipolar diffusion and Ohmic dissipation, and $Q_{\rm rad}$ (given below) captures the effect of radiative cooling and transport. The temperature dependence of $\gamma$ captures the vibrational and rotational degrees of freedom of H$_2$, and is the same as described in \citet[see their fig.~1]{KM09}.

We model the radiation term $Q_{\rm rad}$ as follows. In the optically thin limit (optical depth $\tau\rightarrow 0$), the gas should cool down towards the ambient temperature at $r\to \infty$, $T_0=10~{\rm K}$. The corresponding radiative cooling rate is
\begin{equation}
    Q_{\tau\to 0} = -4\rho\kappa_{\rm P}\sigma (T^4-T_0^4), 
\end{equation}
where $\kappa_{\rm P}$ is the Planck mean opacity and $\sigma$ is the Stefan--Boltzmann constant. In the opposite limit ($\tau\rightarrow\infty$), in which the gas is optically thick in all directions and the length scale of the temperature variation is ${\gg}(\rho\kappa_{\rm R})^{-1}$ (where $\kappa_{\rm R}$ is the Rosseland mean opacity), the radiative energy transport is diffusive and may be modeled by
\begin{equation}
    Q_{\tau\to\infty} = -\grad\bcdot\mcb{F}_{\tau\to\infty}
\end{equation}
with the radiative energy flux
\begin{equation}
    \mcb{F}_{\tau\to\infty} = -\frac{4}{3} (\rho\kappa_{\rm R})^{-1}\sigma \grad T^4.
\end{equation}
Because most of the system (per volume) is in one of these two limits, we may obtain a good estimate of $Q_{\rm rad}$ by interpolating between them using the optical depth:
\begin{equation}
    Q_{\rm rad} \approx \rme^{-\tau}Q_{\tau\to 0} - \grad\bcdot ( \rme^{-1/\tau}\mcb{F}_{\tau\to\infty} ).
\end{equation}
This provides a simple estimate of the rate of radiative cooling and diffusion without performing computationally expensive radiative transfer calculations. We expect this estimate to perform reasonably well, mainly because most of the system has either $\tau\ll 1$ (envelope and pseudodisc) or $\tau\gg 1$ (disc), and the region with $\tau\sim 1$ (only a thin layer on the disc surface) constitutes a very small volume (cf.~Fig.~\ref{fig:rt_comparison}).

In Appendix \ref{A:radiation} we provide the opacity table we use, describe how we estimate $\tau$ in each cell of our simulation data, and perform a few tests of our radiation model.
In summary, we compare our estimate of $\tau$ in our simulation with the corresponding $\tau$ obtained via ray tracing, finding that the errors in our estimates of $\rme^{-\tau}$ and $\rme^{-1/\tau}$ are ${\ll}1$ except in a few cells. We also compare our model's prediction with full radiative transfer for a test problem, where we estimate the temperature profile in thermal equilibrium for a small patch of a geometrically thin disc (so the problem is effectively 1D in the vertical direction) subject to constant heating. We varied the heating rate and disc optical depth to survey a large parameter space, and 
found that our model accurately predicts the mean (density-weighted) temperature of the disc, with an error converging to zero when $\tau\ll 1$ or ${\gg}1$ and a maximum error ${\sim}25\%$ at intermediate optical depths. Additionally, where our simulation overlaps with previous simulations of protostellar core collapse using flux-limited diffusion to solve for the radiative transfer \citep[e.g.,][]{KM10,Tomida2015}, we find very similar temperatures. 

Our radiation model does not include the effect of protostellar irradiation. This should not significantly affect the protostellar disc evolution for the evolutionary phases we simulate, because the innermost part of the disc generally has a larger aspect ratio $H/R$ than the rest of the disc and of the pseudodisc farther out (hence shielding them from any protostellar irradiation). That being said, it is possible for protostellar irradiation to affect the evolution of the outflow cone and the protostellar envelope high above the midplane (mainly through photoionization), and this may affect outflow propagation (but not launching, since disc surface is well shielded) and envelope dispersal (which should become important only towards the end of the Class I phase). Addressing these topics is beyond the scope of the current paper, but will be enabled by future improvements of and additions to our radiation and chemistry models.

\subsection{Computational domain and spatial resolution}

We use a spherical-polar grid ($r,\theta,\phi$) with an inner radial boundary at $1~{\rm au}$ and an outer radial boundary at $10^5~{\rm au}$. Material flowing through the inner boundary is irreversibly accreted by a point-mass `protostar'. The radial ($r$) grid is log-uniform with 240 cells, with $\rmd r/r \approx 0.05$. To reduce computational cost, we simulate only $\theta\in [0,\pi/2]$, with a reflecting boundary condition at $\theta=\pi/2$, and $\phi\in [0,\pi]$, with periodic boundaries at $\phi=0$ and $\pi$. The polar $(\theta)$ grid is non-uniform with 24 cells, with the grid spacing decreasing towards the midplane to be $1/3$ of that at the pole. This gives a midplane angular resolution of ${\simeq}0.036~{\rm rad}$. The azimuthal ($\phi$) grid is uniform with 16 cells for $\phi\in [0,\pi]$.

This setup is similar to that used for our fiducial simulation in \citetalias{XK21}. We showed in \citetalias{XK21} that this setup has sufficient resolution in $(r,\theta)$ and,
provided certain boundary conditions are used, a sufficiently small inner boundary size to achieve numerical convergence (see \citetalias{XK21}, sections 6.2 and 6.3). In Appendix \ref{A:convergence} we show that our simulation results are converged with respect to $\phi$ resolution as well.

\subsection{Boundary conditions}\label{sec:bc}

The boundary conditions we use are the same as those used in \citetalias{XK21} (section 2.5). One slightly unconventional feature of these boundary conditions is that we do not allow angular momentum to be advected through the inner boundary. This is because the amount of angular momentum that can be accreted onto the protostar (or lost at sufficiently small radii where thermal ionization allows efficient angular-momentum removal from the disc via magnetic braking or outflow launching) is very small and, in reality, the net angular-momentum flux going through the disc at $r\sim 1~{\rm au}$ (our typical inner boundary size) should be small compared to the angular-momentum flux allowed by an open boundary condition. In section 6.2 of \citetalias{XK21}, we demonstrated that this choice is very helpful for numerical convergence (with respect to decreasing inner boundary size), especially when the inner boundary is comparable to the circularization radius of accreted material (otherwise a significant amount of angular momentum would be  lost unphysically through the inner boundary).

\subsection{Definitions}

\textbf{The protostar-disc system}: We use this term to define the region that is supported against gravity (either centrifugally or by pressure), plus the point mass enclosed within the inner radial boundary of the computational domain. When the protostar first forms, this region is the pressure-supported `first hydrostatic core'; later on, it consists of a protostar (inside the inner boundary) and a disc (which is mainly rotationally supported). In practice, we define the radial boundary of the protostar-disc system as the radius where the kinetic energy (averaged over a spherical shell) becomes dominated by the contribution from the azimuthal motion.

\textbf{Protostar, accretor, and disc}: Although the boundary of the protostar-disc system is easy to define, defining the boundary between the protostar and the inner disc can be tricky. For simplicity, we identify the protostar mass as the mass of the point-mass accretor within the inner boundary and count the remainder of the mass in the protostar-disc system as that of the disc. Note that this definition will count most of the first hydrostatic core as being part of the disc.

\textbf{Pseudodisc}: We use the term `pseudodisc' to refer to the pre-stellar material that is pressure-supported (and thus flattened) along magnetic-field lines but is not rotationally supported in the cylindrical-radial direction.

\textbf{$\bb{\Omega_{\rm K}}$ and $\bb{Q_{\rm K}}$}: At a given radius, we define the rotation rate required for full (Keplerian) rotational support as $\Omega_{\rm K}(R)\equiv \sqrt{-g_R R}$, where $g_R$ is evaluated at the disc midplane. Inside of the disc, we generally find $\Omega\approx \Omega_{\rm K}$. Because $\Omega_{\rm K}$ exhibits significantly less fluctuations than does $\Omega$, it is useful to replace $\Omega$ with $\Omega_{\rm K}$ when evaluating some variables. We use a subscript `K' to denote variables evaluated using $\Omega_{\rm K}$. For example, in this paper we often characterize the strength of GI with the Toomre $Q$ parameter,
\begin{equation}
    Q = \frac{c_{\rm s}\kappa}{\pi{\rm G}\Sigma},
\end{equation}
where $c_{\rm s}$ is the (density-weighted average) sound speed, $\kappa$ is the epicyclic frequency, and $\Sigma$ is the column density. Because $\kappa$ is very sensitive to spatial fluctuations in $\Omega$, we often approximate $Q$ with $Q_{\rm K}$, in which case we replace $\Omega$ by $\Omega_{\rm K}$ when evaluating the epicyclic frequency.

\section{Overview of evolution}\label{sec:overview}

\begin{figure*}
    \centering
    \includegraphics[scale=0.66]{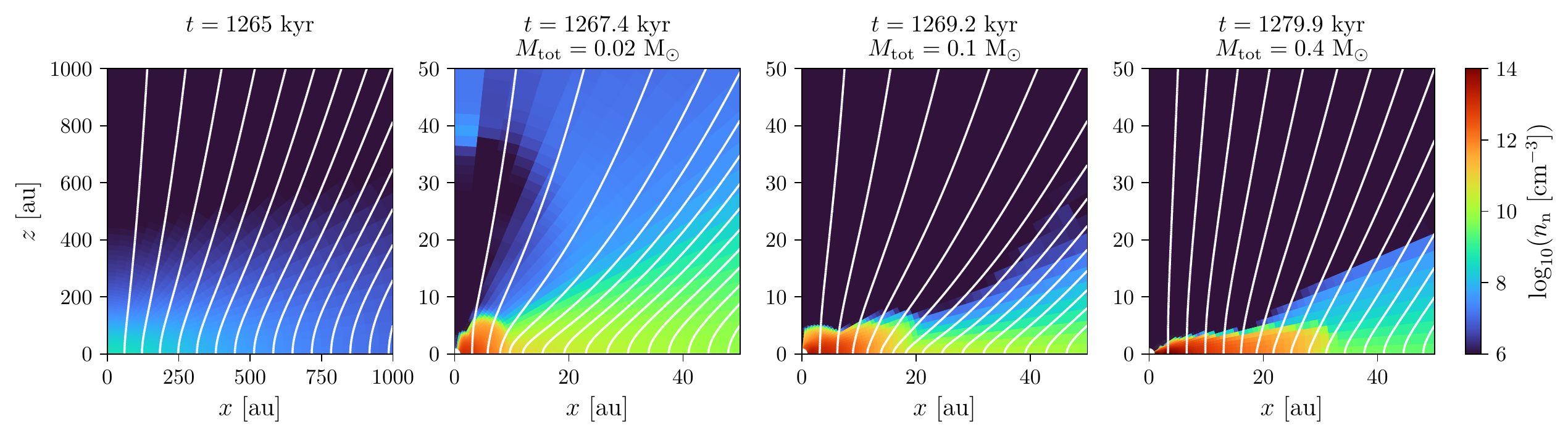}
    \caption{Zoom-in snapshots of the azimuthally averaged density profile (colour) and magnetic-field lines (white) at a few different epochs. For snapshots after protostar formation, the total mass of the protostar and rotationally supported disc $M_{\rm tot}$ (which is frequently used as a proxy of time in this paper)  is given.}
    \label{fig:snapshot}
\end{figure*}

\begin{figure*}
    \centering
    \includegraphics[scale=0.66]{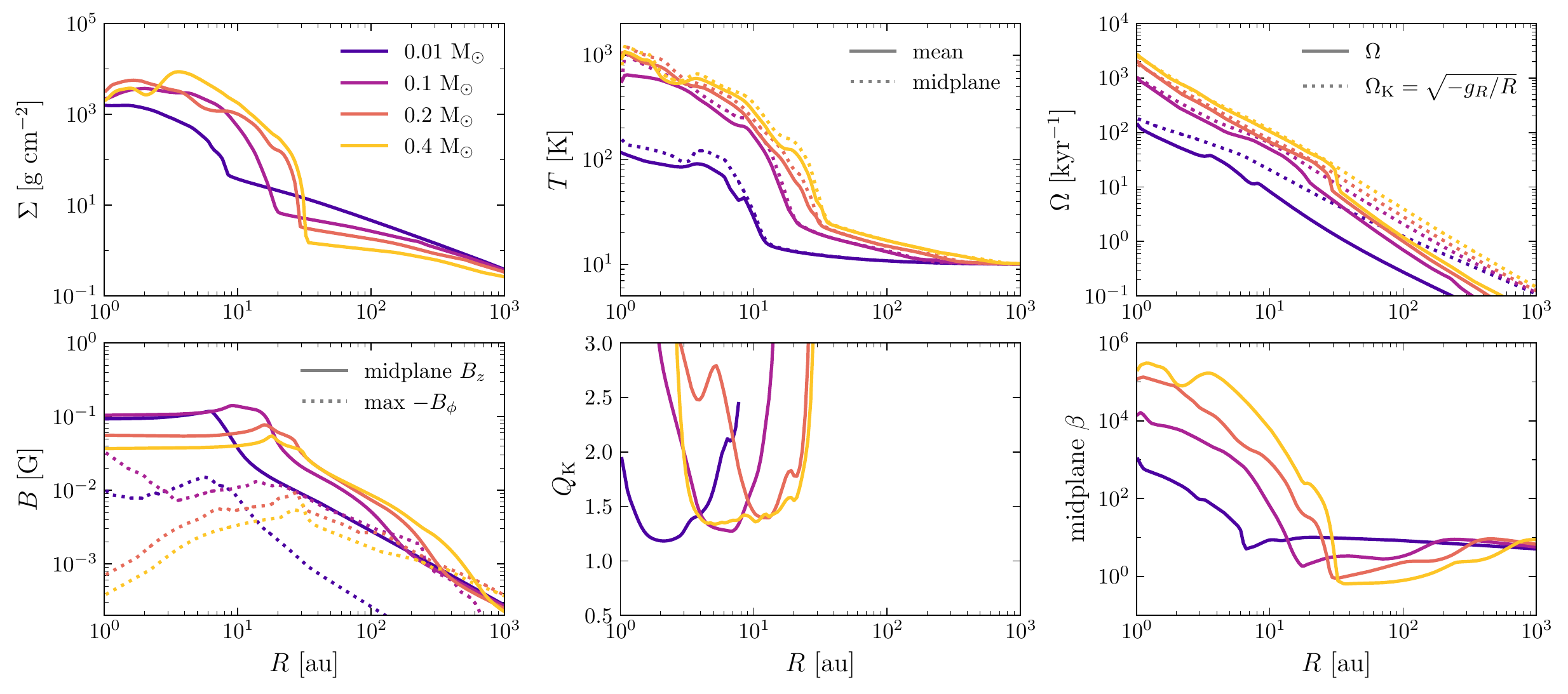}
    \caption{Azimuthally averaged column density (top left), mean (density-weighted) and midplane temperatures (top center), mean (density-weighted) rotation rate (top right), magnetic field (bottom left), $Q_{\rm K}$ (a proxy of Toomre $Q$, bottom center), and midplane plasma $\beta$ (bottom right) in the innermost $10^3~{\rm au}$ of the core at a few different epochs (labeled by $M_{\rm tot}$). The disc that forms within $R\lesssim 30~{\rm au}$ is hot, near-Keplerian, and gravitationally unstable, with an approximately uniform, straight, and dynamically unimportant magnetic field.}
    \label{fig:disc_1d}
\end{figure*}

\begin{figure*}
    \centering
    \includegraphics[scale=0.66]{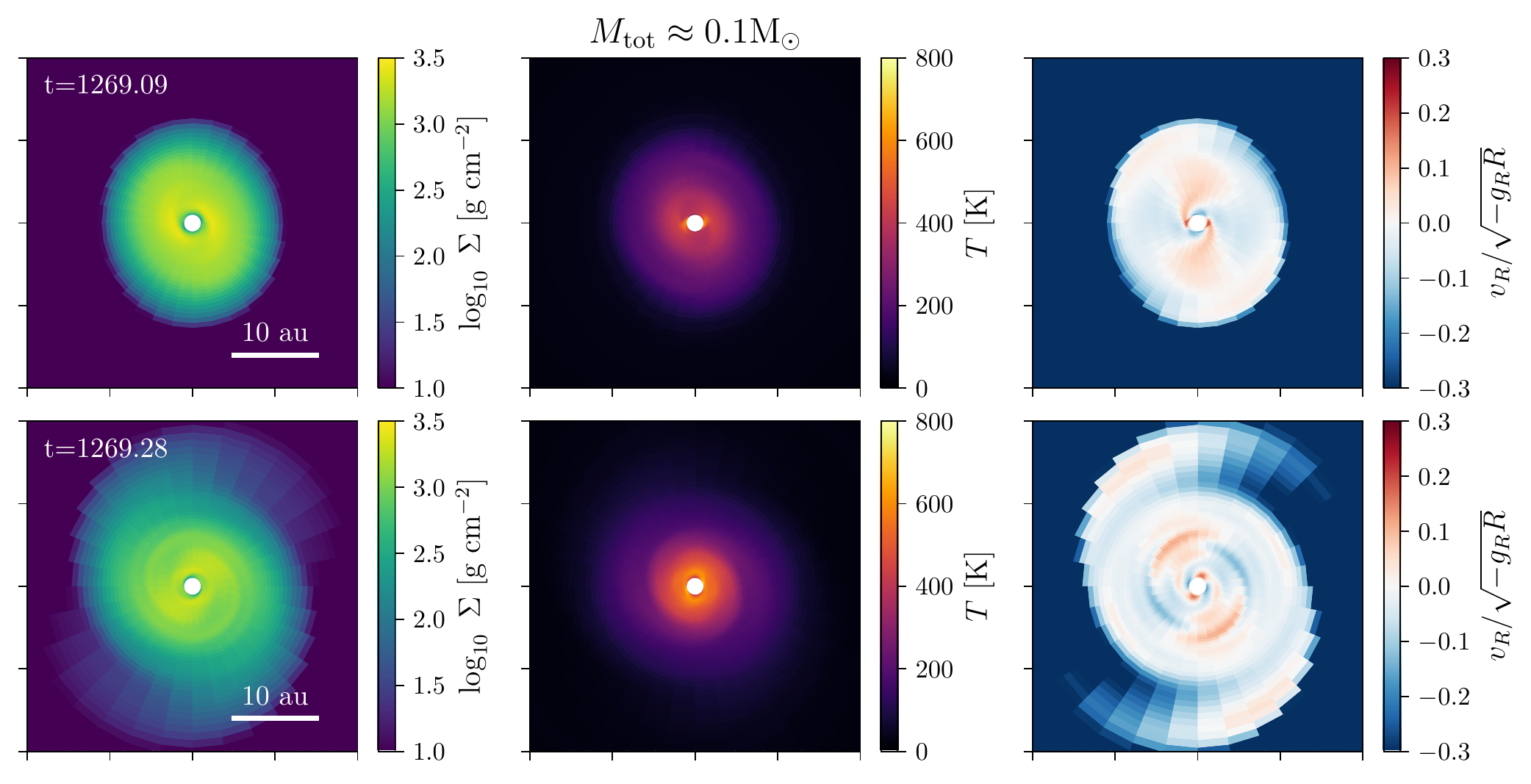}\\
    \includegraphics[scale=0.66]{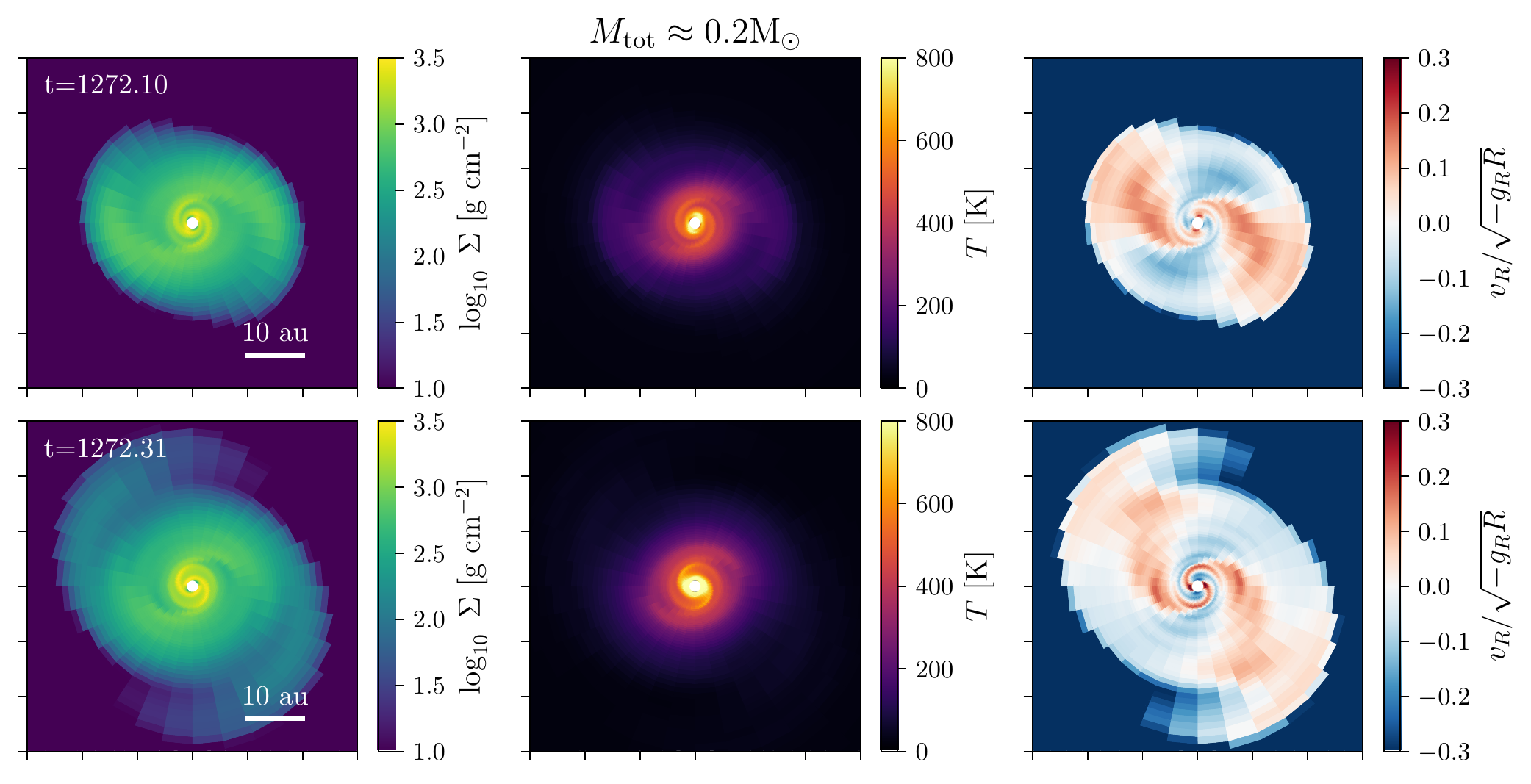}\\
    \includegraphics[scale=0.66]{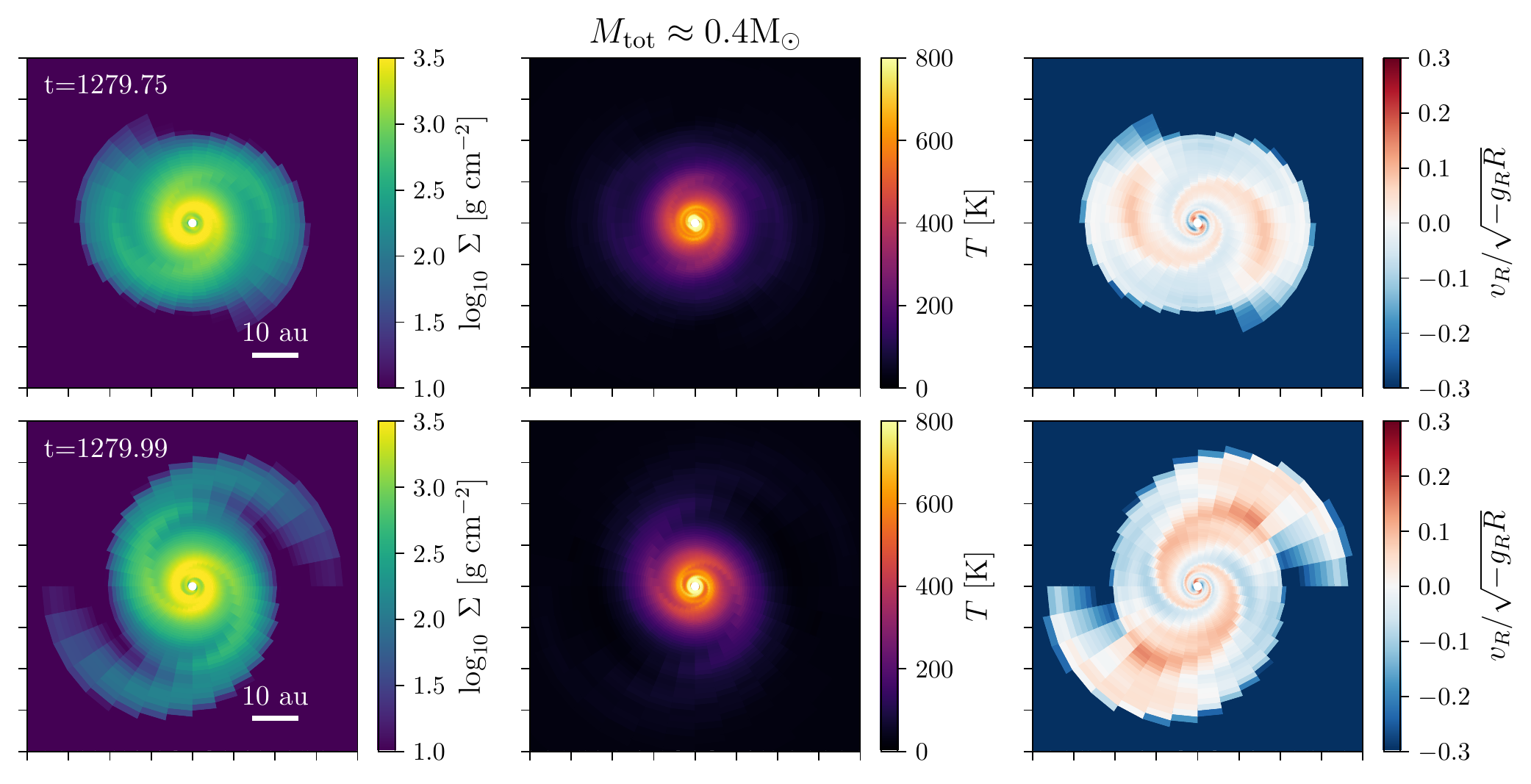}\\
    \caption{Disc column density (left), mean temperature (centre), and radial velocity (right) as viewed from above. Each row corresponds to a different epoch; to provide an unbiased selection, for each $M_{\rm tot}$ we plot the snapshots that exhibit the minimum and maximum disc size measured around that $M_{\rm tot}$. The disc size undergoes significant oscillations as spirals emerge and disperse. A rich display of substructures, mostly spirals, is visible.}
    \label{fig:midplane}
\end{figure*}

\begin{figure}
    \centering
    \includegraphics[scale=0.66]{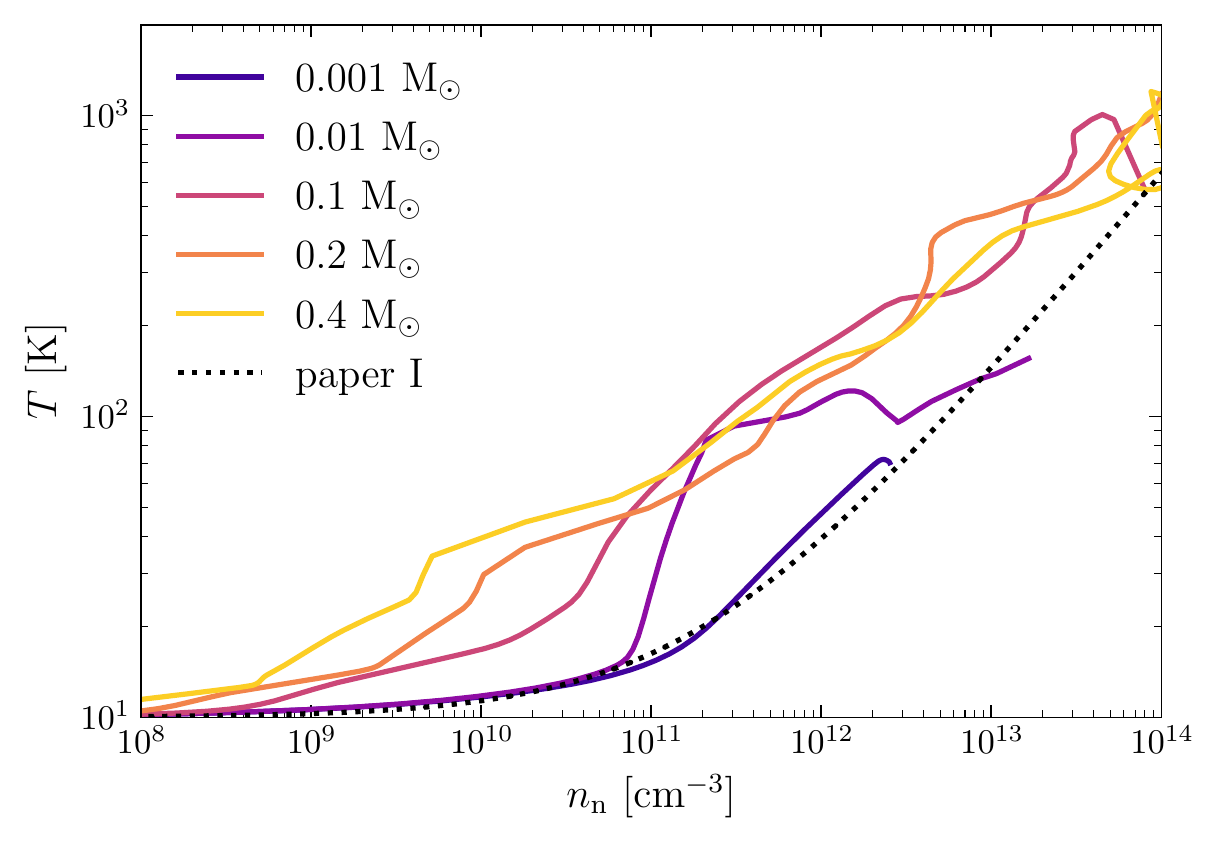}
    \caption{Relation between azimuthally averaged midplane temperature and density at a few different epochs. The barotropic equation of state used in \citetalias{XK21} is shown for reference, and fits well the temperature evolution at early times. With our treatment of radiative cooling, the temperature in the disc is far from barotropic.}
    \label{fig:nT}
\end{figure}

\begin{figure}
    \centering
    \includegraphics[scale=0.66]{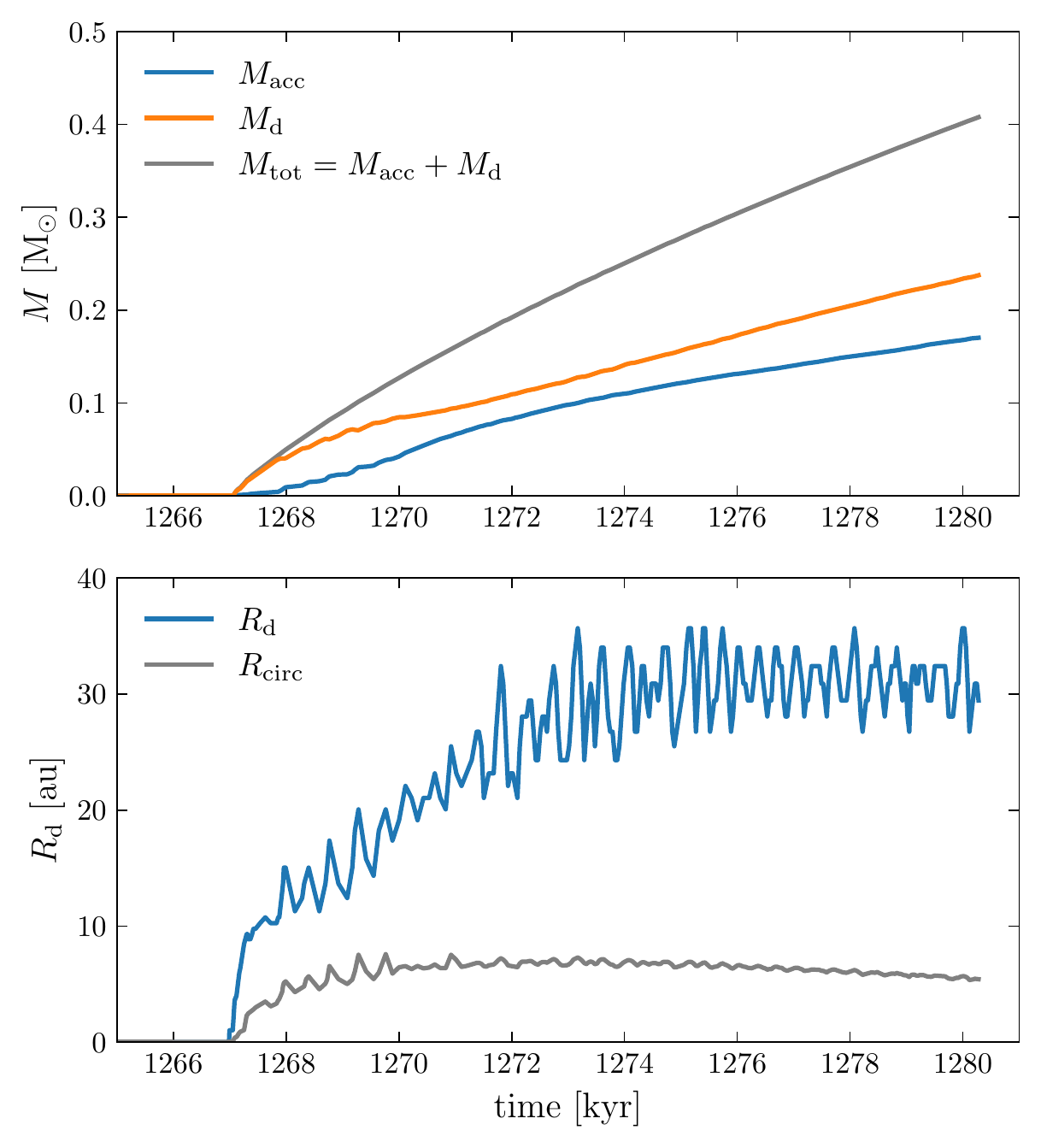}
    \caption{Time evolution of disc and accretor (protostar) mass (top panel) and disc radius (bottom panel). The circularization radius of accreted material, $R_{\rm circ} \equiv l^2/({\rm G}M_{\rm tot})$, where $l$ is the specific angular momentum of material accreted onto the disc, is shown for reference. The disc remains slightly more massive than the protostar. The disc radius is significantly larger than the circularization radius of incoming material, as a result of angular-momentum transport (redistribution) within the disc.}
    \label{fig:M_and_Rd}
\end{figure}

In this section we provide an overview of the evolution of the protostellar core (and the disc formed within) using Figs \ref{fig:snapshot}--\ref{fig:M_and_Rd}, which show azimuthally averaged (Figs~\ref{fig:snapshot} and \ref{fig:disc_1d}) and midplane (Fig.~\ref{fig:midplane}) profiles at different epochs, the evolution of midplane temperature with density (Fig.~\ref{fig:nT}), and the time evolution of the protostellar disc mass and size (Fig.~\ref{fig:M_and_Rd}). We focus on describing the simulation results qualitatively; the detailed (quantitative) physical picture of disc formation and evolution is discussed in Sections \ref{sec:AMbudget}--\ref{sec:Bevolution}.

\subsection{Pre-stellar collapse and disc formation}

The evolution up to the formation of a rotationally supported disc is nearly identical to that found in the fiducial simulation of \citetalias{XK21} (see section 3 and figs 1 and 2 there). During the first ${\sim}1~{\rm Myr}$ of evolution, the pre-stellar core flattens along magnetic-field lines into a (not rotationally-supported) pseudodisc and radially contracts dynamically (though slower than free fall). The magnetic field is dragged along with the collapsing core and becomes hourglass-shaped (first panel of Fig.~\ref{fig:snapshot}). The pseudodisc acquires a flat column density $\Sigma$ in its central region where thermal pressure smooths out any structure and a near-self-similar profile close to $\Sigma \propto R^{-1}$ outside this region. Likewise, the magnetic field is approximately uniform in the central region, outside of which its vertical component falls off nearly as $B_z \propto R^{-1}$ (with a slight deviation due to residual ambipolar diffusion during dynamical contraction). During this phase, the angular momentum of the infalling gas is reduced only slightly by magnetic braking, because the braking timescale is longer than the dynamical time. The temperature remains close to $10~{\rm K}$ (until the central density reaches $n_{\rm n} \sim 10^{11}~{\rm cm}^{-3}$), as the core remains optically thin. These processes affect the mass and angular-momentum budget of the protostar-disc system, which we discuss further in Section \ref{sec:AMbudget} (see also sections 3.1, 4.2, and 4.3 of \citetalias{XK21}).

When the central column density becomes sufficiently large, the gas in the innermost few au becomes optically thick and the temperature increases (see Figs~\ref{fig:disc_1d} and \ref{fig:nT} at ${\leq}0.01~{\rm M}_{\odot}$). The additional pressure support leads to the formation of a first hydrostatic core with $n_{\rm n}\gtrsim 10^{12}~{\rm cm}^{-3}$. At the same time, ambipolar diffusion `reawakens' to become increasingly important as the charged species are adsorbed onto the dust grains, thereby decoupling all species but the electrons from the magnetic field and causing the magnetic flux to pile up just outside of the hydrostatic core (see, e.g., \citealt{TM07b} and \citealt{KM10}). The first hydrostatic core also marks the epoch of point-mass formation, which creates an expanding region (reaching ${\gtrsim}10^3~{\rm au}$ at the end of our simulation) in which gravity is dominated by the point-mass instead of the pseudodisc. Within this region, the pseudodisc profile flattens from the pre-stellar profile with $\partial\ln\Sigma/\partial\ln R \approx -1$ to one having $\partial\ln\Sigma/\partial\ln R \approx -1/2$ (Fig.~\ref{fig:disc_1d}; see also discussion in \citealt{DBK12}, section 7.1). The accelerated infall of the gas and the consequent pinching of the magnetic-field lines  push the location of magnetic decoupling (and the resulting pile-up of magnetic flux) to larger radii (Section \ref{sec:decoupling}; see also \citealt{ck98} and \citealt{cck98}).

The first hydrostatic core continues to accrete mass and angular momentum from the pseudodisc, and the accumulation of angular momentum soon shapes it into a rotationally supported torus (second panel of Fig.~\ref{fig:snapshot}). This torus quickly becomes gravitationally unstable (with $Q_{\rm K}\sim 1.5$; see Fig.~\ref{fig:disc_1d}), and the non-axisymmetric perturbations associated with GI (see Fig.~\ref{fig:midplane}) transport angular momentum outwards to allow the formation of a central protostar and a growing, rotationally supported disc with $\Omega\approx\Omega_{\rm K}$ (see Fig.~\ref{fig:disc_1d}). The above evolution takes place within the first $\sim$kyr after protostar formation.

\subsection{Disc evolution}\label{sec:discevolution}

After the rotationally supported disc appears, the disc remains gravitationally unstable and the disc mass remains comparable to the protostar mass. Meanwhile, the disc size undergoes substantial growth and saturates at ${\sim}30~{\rm au}$ (Figs~\ref{fig:midplane} and \ref{fig:M_and_Rd}; see also the last two panels of Fig.~\ref{fig:snapshot}). The saturation of the disc size occurs mainly because increased magnetic braking in the pseudodisc at later times prevents the disc from accumulating enough angular momentum to grow further (see Section \ref{sec:discbudget}).

The disc size is significantly larger than the circularization radius of incoming material (Fig.~\ref{fig:M_and_Rd}; this also leads to the sudden change in $\Omega$ at the disc edge seen in Fig.~\ref{fig:disc_1d}). This is because the accreted material leaves most of its angular momentum in the disc before being accreted by the protostar, which can hold approximately zero angular momentum. This angular momentum is then transported outwards by GI to cause disc spreading (see Section \ref{sec:GItransport}). GI also gives rise to various substructures, such as prominent gravitationally excited spiral arms (see Fig.~\ref{fig:midplane}).

The temperature inside the disc is typically a ${\rm few}\times 10^2~{\rm K}$ (Fig.~\ref{fig:disc_1d}). While the barotropic equation of state we employed in \citetalias{XK21} fits well the temperature evolution at early times (before disc formation), the relation between temperature and density is far from barotropic once the protostellar disc forms (Fig.~\ref{fig:nT}), with the temperature being generally larger than predicted by the barotropic relation (which is based on central density-temperature relation in radiative non-rotating core collapse simulations). This suggests that using a barotropic equation of state to model young protostellar disc evolution cannot produce the correct thermal evolution. The increased disc temperature compared to \citetalias{XK21} also implies that the disc has to reach a higher column density to be gravitationally unstable, which explains the larger disc-to-star mass ratio. We discuss what determines the disc temperature profile in Section \ref{sec:Tprofile}.

The magnetic field inside the disc is approximately uniform, and the ratio between thermal and magnetic pressure $\beta\equiv 8\pi P/B^2 \gg 1$ (Fig.~\ref{fig:disc_1d}).
This weak and nearly uniform field is mainly a result of non-ideal magnetic diffusion (Ohmic and ambipolar) decoupling the field from the gas in the radial direction, which we discuss in Section \ref{sec:Bevolution}.
The field is also approximately straight inside the disc, with a very small azimuthal component (Fig.~\ref{fig:disc_1d}); we discuss the generation of an  azimuthal field and the strength of magnetic braking inside the disc in Section \ref{sec:discbraking}.

\revision{The magnetic field threading the disc also launches a relatively weak outflow, which removes little mass and angular momentum from the disc and barely affects disc evolution (Section \ref{sec:discbudget}).
We refrain from a more detailed discussion of outflow properties in this paper, as the propagation of the outflow far above the disc might be significantly impacted by numerical factors such as the low resolution in the polar region, the use of a density floor (cf. \citetalias{XK21} appendix A2), and the omission of protostellar irradiation (Section \ref{sec:radmodel}).}

\section{Angular-momentum budget and transport}\label{sec:AMbudget}

In this section we study the angular-momentum budget of the disc. Considering the protostar and the disc as a whole, we separate the problem into a study of how much mass and angular momentum enters and leaves the protostar-disc system (Section \ref{sec:discbudget}), and a study of how angular momentum is redistributed within the protostar-disc system to allow accretion and disc growth (Section \ref{sec:GItransport}). We also include a discussion of how the redistribution of angular momentum by GI leads to formation of radial substructures and the spread of the gravitationally unstable region (Section \ref{sec:GIsize}). We note that results in Sections \ref{sec:discbudget} and \ref{sec:GItransport} are very similar to those in sections 4 and 5 of \citetalias{XK21}, which offers a more detailed discussion.

\subsection{Mass and angular-momentum budget of the protostar-disc system}\label{sec:discbudget}

\begin{figure}
    \centering
    \includegraphics[scale=0.66]{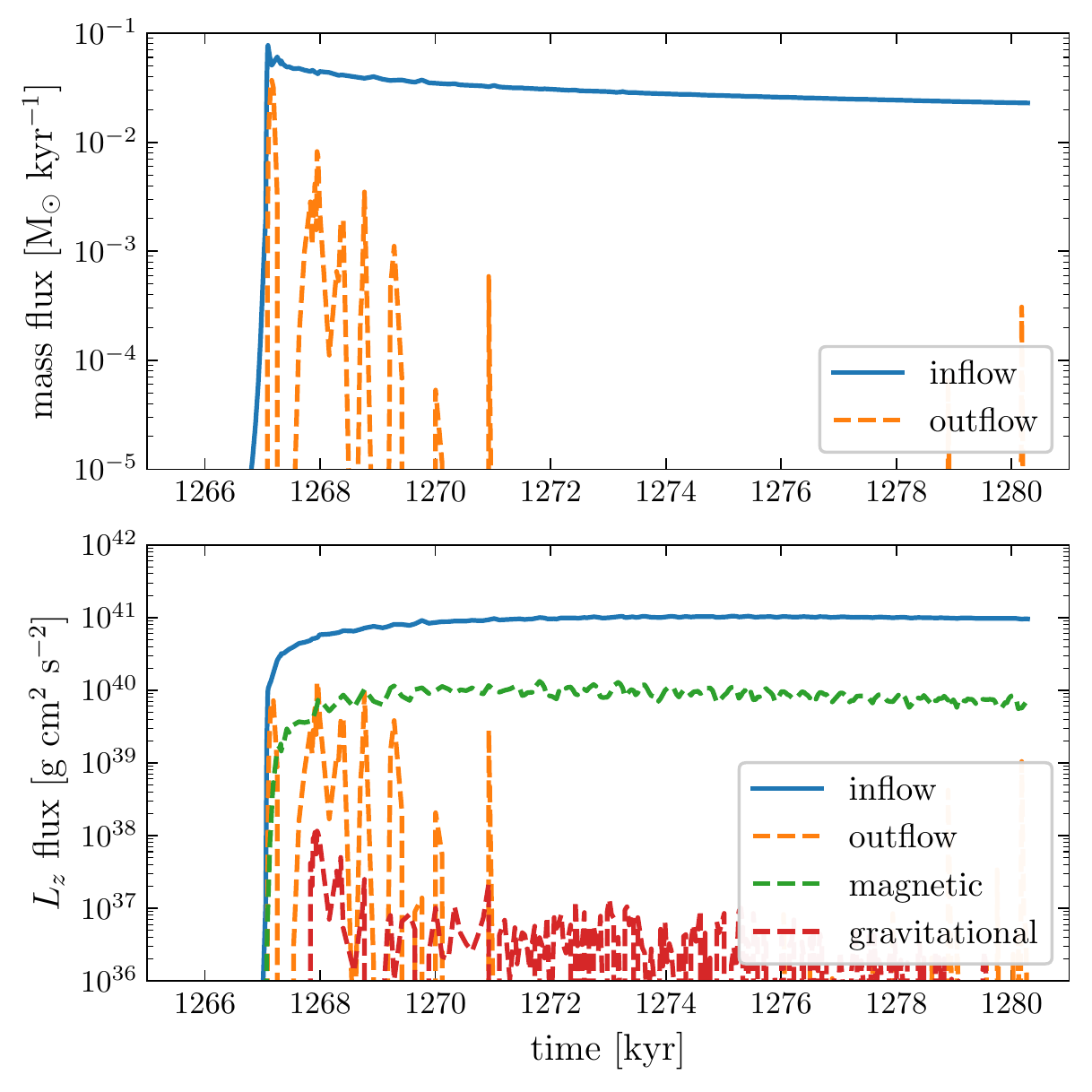}
    \caption{Mass (top panel) and angular-momentum (bottom panel) fluxes going into (solid) and out of (dashed) the protostar-disc system via different mechanisms. These fluxes are evaluated on a sphere with radius slightly larger than $R_{\rm d}$. Both the mass and angular-momentum fluxes are dominated by inflow (accretion). Note that $c^3_{{\rm s}0}/{\rm G} \simeq 1.6~(T/10~{\rm K})^{3/2}~{\rm M}_\odot~{\rm Myr}^{-1}$.}
    \label{fig:ML_flux}
\end{figure}

Mass and angular momentum enters and leaves the protostar-disc system through several channels. Fig.~\ref{fig:ML_flux} plots the mass and angular-momentum flux through a sphere of radius slightly larger than $R_{\rm d}$ from different channels; these are the mass flux from inflow/accretion\footnote{\revision{The accretion rate here refers to accretion from the pseudodisc onto the protostar-disc system, rather than from the disc onto the protostar; the rate corresponding to the latter is similar to the former but exhibits much higher variability \citep[cf.][]{Machida2019}. In this paper we forego any discussion of the variability of the accretion rate onto the central protostar, as it can be sensitive to the treatment of the inner boundary.}} (region with $v_r<0$) and outflow (region with $v_r>0$),
and the angular-momentum flux from radial advection ($R\rho v_r v_\phi$) by inflow and outflow and from transport by magnetic and gravitational stresses ($-R B_r B_\phi/4\pi$, $ R g_r g_\phi/4\pi{\rm G}$). Both the mass and angular-momentum fluxes are dominated by accretion (inflow), which happens mostly through the pseudodisc instead of the envelope (because the latter has negligible density). The angular-momentum flux from the magnetic stress, which is approximately the rate of angular-momentum removal by disc magnetic braking, remains about one order of magnitude lower than the angular momentum injection by accretion; other mechanisms of mass and angular-momentum removal are even weaker compared to accretion.
Therefore, nearly all mass and angular momentum accreted by the protostar-disc system stays within the protostar-disc system.

As a result, the angular-momentum budget of the protostar-disc system can be understood mostly in terms of the accretion of mass and angular momentum from the pseudodisc. One informative diagnostic is the circularization radius of the material accreted from the pseudodisc, which is shown in the bottom panel of Fig.~\ref{fig:M_and_Rd}. In the absence of angular-momentum transport within the disc, this circularization radius would define the size of the disc.
We see a roughly constant circularization radius, which implies that the specific angular momentum of material accreted at a given $M_{\rm tot}$ scales roughly as $l\propto M_{\rm tot}^{1.5}$. If angular momentum is conserved during the infall, this scaling should have a slope between $5/3$ (when magnetic field is relatively weak and infall is approximately in the spherical-radial direction) and 2 (when magnetic field is very strong and all material first fall along magnetic field lines to the midplane and then radially collapse). The observed slope deviates from this range suggests the presence of increasing magnetic braking in the pseudodisc.

\begin{figure}
    \centering
    \includegraphics[scale=0.66]{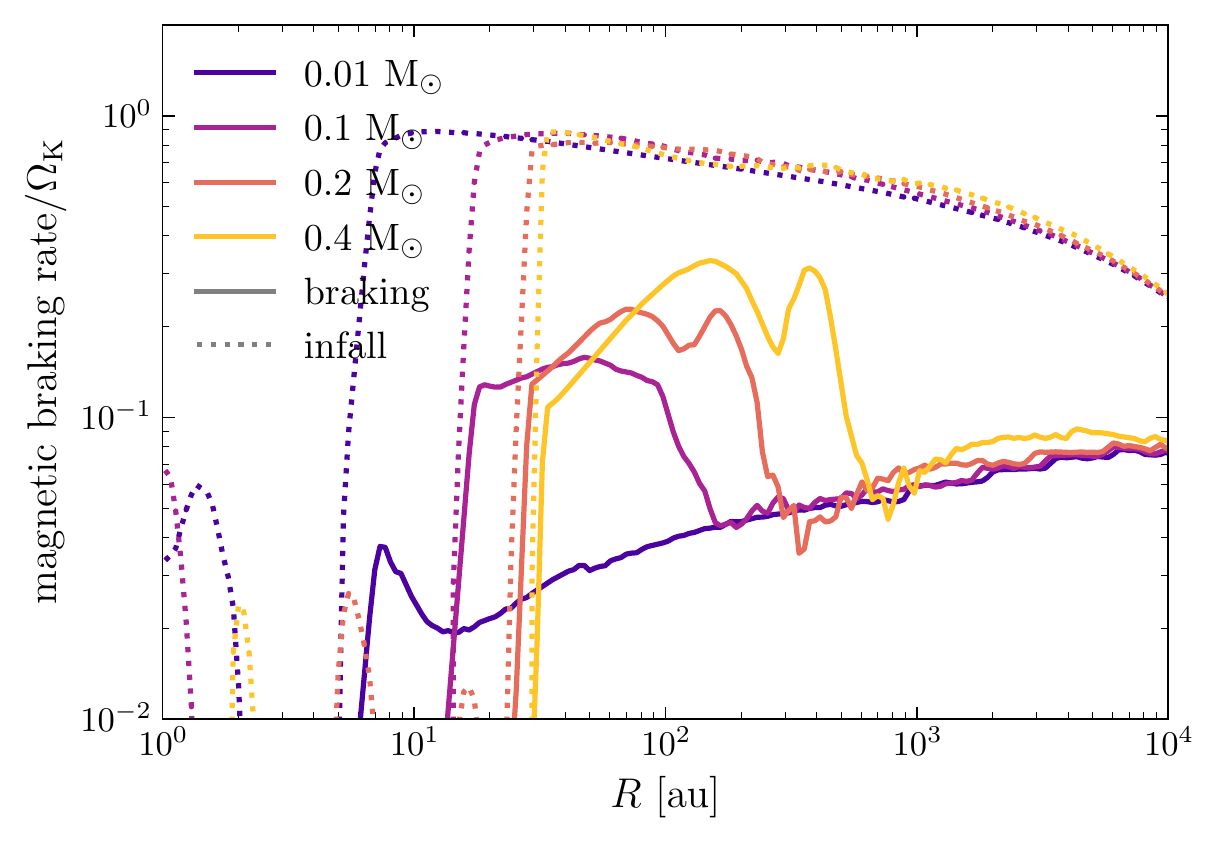}
    \caption{Comparison between pseudodisc magnetic-braking rate (magnetic torque divided by angular momentum) and infall rate ($-v_r/r$), both normalized by $\Omega_{\rm K}(R)$ (which is also approximately the local free-fall rate), at a few different epochs. (See Section \ref{sec:discbudget} for how these quantities are obtained.)  Magnetic braking is relatively weak in the outer pseudodisc. In the inner pseudodisc, however, the magnetic-braking rate can be comparable to the infall rate, and angular-momentum removal by magnetic braking becomes increasingly important at late times.}
    \label{fig:braking_rate}
\end{figure}

In Fig.~\ref{fig:braking_rate} we compare the rate of magnetic braking with the rate of infall. To compute the braking rate, we first label each $(R,\phi,z)$ location with the amount of magnetic flux threading a disc of radius $R$ at height $z$, denoted $\Phi_B(R,z)$. [This procedure gives the same $\Phi_B$ for all $\phi$ at a given $(R,z)$.] The braking rate at a given radius $R_0$ is then found by integrating the magnetic torque on an infinitesimal flux tube with $\Phi_B(R,z)$ between $\Phi_B(R_0,0)\pm \rmd\Phi_B$; because the field is close to axisymmetry, this flux corresponds approximately to the region along the field lines starting from $R_0\pm\rmd R$. However, we do not integrate through the entire flux tube -- magnetic braking involves the transport of angular momentum along field lines from the collapsing pre-stellar core (including the disc that forms within the core) to the low density, magnetically dominated, non-collapsing background, and integrating through the entire flux tube would have the torques in these two regions cancel out. Instead, we separate the core from the background, and integrate only the magnetic torque in the core. To do so, we define the core as the region where the magnetic energy is less than the sum of the kinetic and internal energies (i.e., not magnetically dominated). The braking rate at each cylindrical radius is then defined as the ratio of the magnetic torque and the angular momentum, both integrated in the `core' part of each corresponding flux tube. The main advantage of this definition is that it focuses on angular-momentum removal from the core and is unaffected by angular-momentum transport along magnetic-field lines within the core (which should barely affect the angular-momentum budget of the system). The infall rate is calculated as a density-weighted average of $-v_r/r$ in the same (`core' flux-tube) region.

The magnetic braking rate shown in Fig.~\ref{fig:braking_rate} is consistent with the evolution of $R_{\rm circ}$ we observe.
The braking rate generally increases in time, and at $0.4~{\rm M}_\odot$, braking is only a factor of ${\approx}3$ slower than infall in the inner part of the pseudodisc, which allows magnetic braking to affect the angular-momentum budget significantly.
In Section \ref{sec:decoupling} we show that the increased braking at late times can be explained by the radial decoupling of the magnetic field from the gas, which leads to a pile-up of magnetic flux in the inner part of the pseudodisc. We also show in Section \ref{sec:modelcomparison} that we expect this trend of increased braking to continue, and that the pseudodisc magnetic braking should eventually become strong enough to affect significantly the angular-momentum budget and disc size evolution.

In summary, the mass and angular-momentum budgets of the disc are determined mainly by accretion from the pseudodisc. The angular-momentum budget of the disc is affected by magnetic braking in the pseudodisc, which is initially weak but becomes increasingly important at later times; the increased magnetic braking at later times is partly responsible for the saturated disc size observed in Figure \ref{fig:M_and_Rd}.

%
%
%
\subsection{Angular-momentum transport within the protostar-disc system: GI}\label{sec:GItransport}

Material in the disc has to lose angular momentum in order to be accreted by the protostar. Since mechanisms that remove angular momentum from the disc all appear to be inefficient, accretion requires some mechanism that transports angular momentum radially outwards (which also leads to disc spreading).
In our simulation, such angular-momentum transport is facilitated mainly by GI, as in \citetalias{XK21} (section 5.1).\footnote{The magnetorotational instability, another popular mechanism for radial angular-momentum transport in differentially rotating discs \citep{bh98}, is fully suppressed by the strong magnetic diffusion ($\eta \gg \Omega H^2$) in our disc \citep[cf.][]{Kawasaki2021}.}

\begin{figure*}
    \centering
    \includegraphics[scale=0.66]{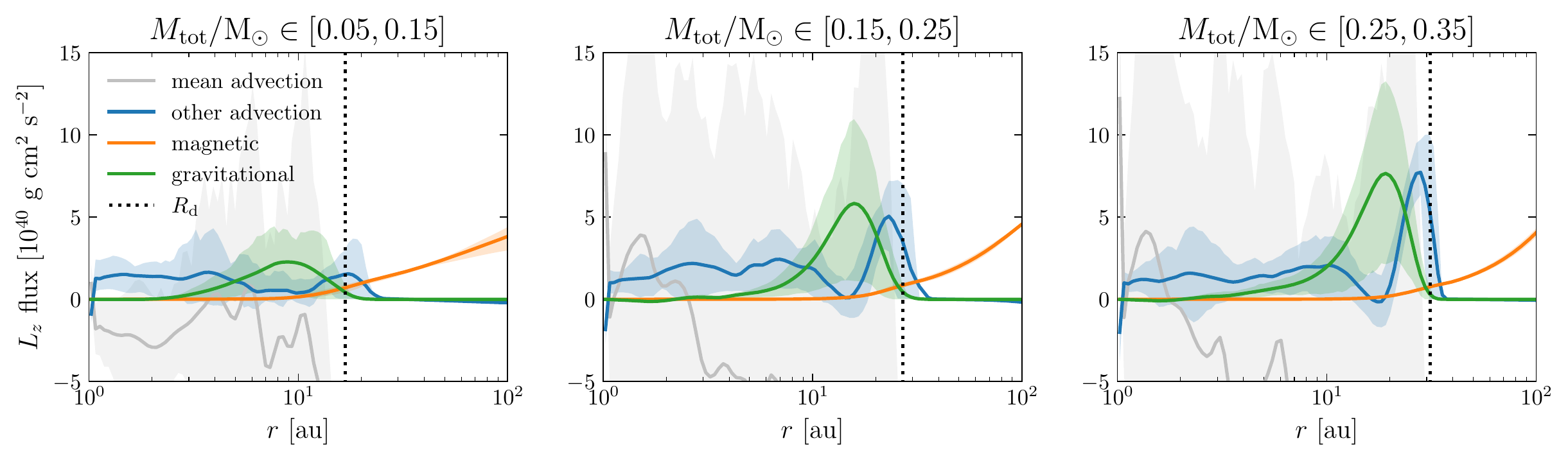}\\
    \includegraphics[scale=0.66]{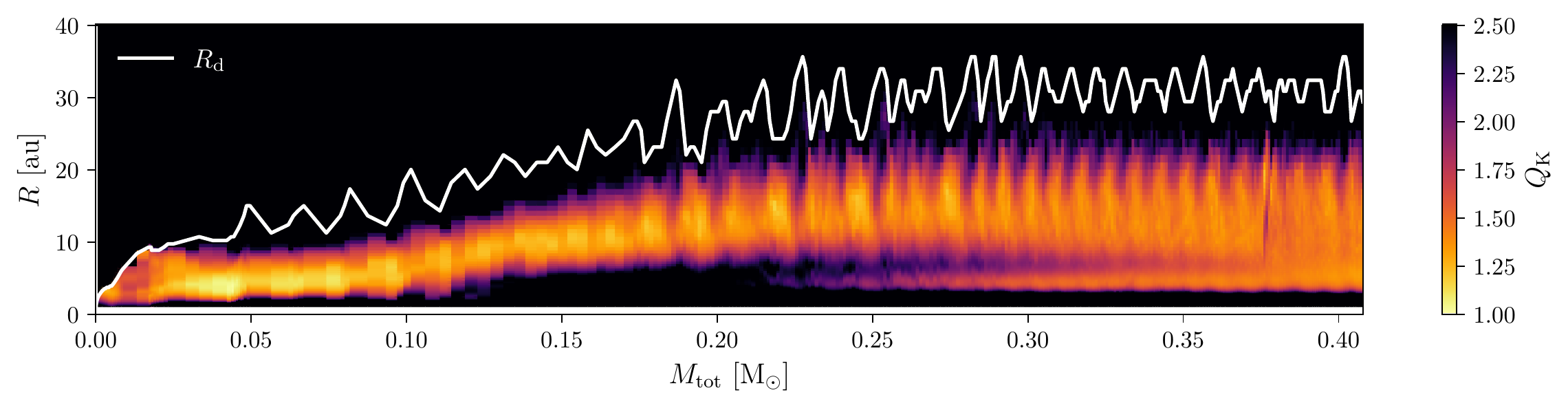}
    \caption{Top: Radial profiles of angular-momentum flux due to different mechanisms. Each line traces the mean value across all snapshots in the stated $M_{\rm tot}$ range; the shaded region indicates $1\sigma$ deviations about this mean. We decompose the `advection' contribution into that associated with the mean flow (viz., mass flux times density-weighted mean specific angular momentum; grey) and the residual `other advection' (blue), which is due mainly to transport by the spiral waves. Inside the disc, the angular-momentum transport not associated with mean advection is driven primarily by GI-induced gravitational stresses (green) and spiral waves (blue). Magnetic braking (orange) remains weak inside the disc. Bottom: Space-time evolution of $Q_{\rm K}$ (a proxy of Toomre $Q$) inside the disc; the disc edge $R_{\rm d}$ is plotted for reference. When the total protostar-disc mass $M_{\rm tot}\in [0.12,0.24]~{\rm M}_\odot$, the inner part of the disc is gravitationally stable and angular-momentum transport there is due to spiral waves excited radially farther out. Later on, a second gravitationally unstable region appears at smaller radii, and most of the disc eventually being marginally gravitationally unstable (see Section \ref{sec:GIsize}).}
    \label{fig:Q_and_L_transport}
\end{figure*}

In the top panel of Fig.~\ref{fig:Q_and_L_transport} we plot the angular-momentum flux through spherical ($r$) shells due to different mechanisms. Angular-momentum transport (coloured lines) inside the disc is dominated by turbulent advection (Reynolds stress, blue) and gravitational stress (green), which can both be interpreted as the outcome of spiral waves excited by GI. Note that the gravitational stress is confined mostly to regions that are locally gravitationally unstable (cf.~bottom panel of Fig.~\ref{fig:Q_and_L_transport}), but the turbulent advection is not. The non-negligible amount of turbulent advection seen in the gravitationally stable regions arises from propagation of spiral waves into these regions (see Section \ref{sec:GIsize}).

The strength of GI, which determines the rate of angular-momentum transport, can be quantified using the Toomre $Q$ parameter, with smaller $Q$ giving stronger instability and faster angular-momentum transport, and hence leading to larger accretion rates. On the other hand, stronger instability and faster accretion tend to increase disc temperature (due to additional turbulent heating as well as additional release of gravitational energy via accretion) and reduce column density (by reducing disc mass and increasing disc size); this causes $Q$ to increase. As a result, the disc should settle into a self-regulated state in which the disc is marginally unstable and GI produces just the right amount of angular-momentum transport (and heating) to maintain approximately constant $Q$ \citep{VB07,KratterLodato2016}. Additionally, the strength of GI scales steeply with $Q$: at $Q\sim 2$, GI is turned fully off, but at $Q\sim 1$, GI becomes so strong that the resulting angular-momentum transport is generally much faster than needed for the self-regulated equilibrium. Therefore, without knowing any details of the evolution, the value of $Q$ at this self-regulated state can be estimated to within a factor of 2.

An important implication of this idea of gravitational self-regulation is that one can easily estimate the disc profile (including disc size and mass) by assuming that the whole disc has some constant $Q$, if one knows the total mass and angular momentum of the disc (discussed in the previous subsection) and the thermal profile of the disc (Section \ref{sec:Tprofile}). We discussed this idea in section 5 of \citetalias{XK21} using the example of an isothermal disc; we also apply this idea to build a simple 1D disc model in Section \ref{sec:model}.

One caveat, however, is that the disc being gravitationally self-regulated does not directly imply that most of the disc needs to be marginally unstable. This is mainly because GI excites spiral waves that can propagate into gravitationally stable regions and transport angular momentum there. In the next subsection we discuss this process in more detail, and argue that most of the disc should eventually be marginally gravitationally unstable.

\subsection{Size of the gravitationally unstable region}\label{sec:GIsize}

While we expect GI to account for most of the angular-momentum transport everywhere in the disc, this does not require the entire disc to be gravitationally unstable. Instead, as we show in Fig.~\ref{fig:Q_and_L_transport} and discuss further below, it is possible for a wave excited in one unstable region to propagate to a stable region and deposit some of its angular momentum there.
Furthermore, because the angular-momentum flux needed to facilitate accretion generally decreases towards smaller radii, the spiral waves excited in the outer part of the disc in principle have sufficient angular-momentum flux to provide all angular-momentum transport needed at smaller radii. (Note that the angular-momentum flux of a spiral wave remains constant as the wave propagates if the wave is not damped through dissipative processes such as shocks.) Therefore, while the need for GI to transport angular momentum requires a gravitationally unstable outer disc, it is not certain whether the inner part of the disc needs to be unstable as well.

In our simulation, we do see the inner disc to be gravitationally stable for a period of time around $M_{\rm tot} \sim 0.2~{\rm M}_\odot$. However, a second gravitationally unstable region appears later on at ${\approx}4~{\rm au}$ and eventually merges with the first, outer unstable region to make most of the disc unstable (Fig.~\ref{fig:Q_and_L_transport}).
In the discussion below, we explain this behaviour and argue that the appearance of additional gravitationally unstable regions at yet smaller radii should be a generic behaviour and that, eventually, most of the disc should be unstable.

First, let us consider the evolution of column density at a locally stable patch of the disc in which angular-momentum transport is facilitated by spiral waves excited at some unstable region at larger radii. Using mass and angular-momentum conservation, we have
\begin{equation}\label{eqn:fMfL}
    \partial_t (2\pi R\Sigma) = -\partial_R F_M,\quad
    \partial_t (2\pi R \Sigma l) = -\partial_R (F_L + F_M l).
\end{equation}
Here $F_L$ is the (positive) angular momentum flux by spirals though radius $r$, and $F_M=-\dot M(R)$ is the mass flux. Let us assume for simplicity that the specific-angular-momentum profile $l(R)$ stays constant in time, which is equivalent to assuming that the timescale of column-density evolution is ${\gg}\Omega^{-1}$ and ${\ll}M_{\rm tot}/\dot M_{\rm tot}$. This produces a simple relation between the accretion rate (mass flux) and the derivative of $F_L$; using equation \eqref{eqn:fMfL},
\begin{equation}\label{eqn:fMfL2}
    -F_M = (\partial_R l)^{-1}\partial_R F_L.
\end{equation}
In a gravitationally stable region, $F_L$ stays constant when the amplitude of the spiral wave is small and damping (mainly through shock formation) is negligible, and $\partial_R F_L$ is mainly due to the damping of the wave.
Therefore, in a gravitationally stable region, accretion (inward mass flux) through a radius is directly proportional to the amount of spiral wave damping at this radius.

Now we consider the radial variation of spiral wave damping and how that affects the disc mass distribution.
A spiral wave excited in the gravitationally unstable region generally has a pattern speed $\Omega_{\rm p}$ comparable to $\Omega$ at the location of its excitation.
Let us assume either that the unstable region is narrow or that the location where waves propagating outside the GI region are excited is close to the inner boundary of the GI region (which is a relatively good assumption when the GI region is extended, because spiral waves cannot propagate far inside a GI region; cf. \citealt{Bethune2021}).
In this case, the inner Lindblad resonance (ILR), which occurs at $m(\Omega-\Omega_{\rm p})=\kappa$ (with $m=2$ for the dominant mode), lies outside the GI region.
The location of the ILR ($R_{\rm ILR}$) is important because we expect an increase in wave-disc interaction (which makes the wave more nonlinear and, for instance, more prone to shock formation) and wave damping around the resonance. Since mass flux is proportional to spiral-wave damping [equation \eqref{eqn:fMfL2}], the mass flux should also increase here, and so mass should tend to pile up at radii slightly smaller than $R_{\rm ILR}$.

\begin{figure}
    \centering
    \includegraphics[scale=0.66]{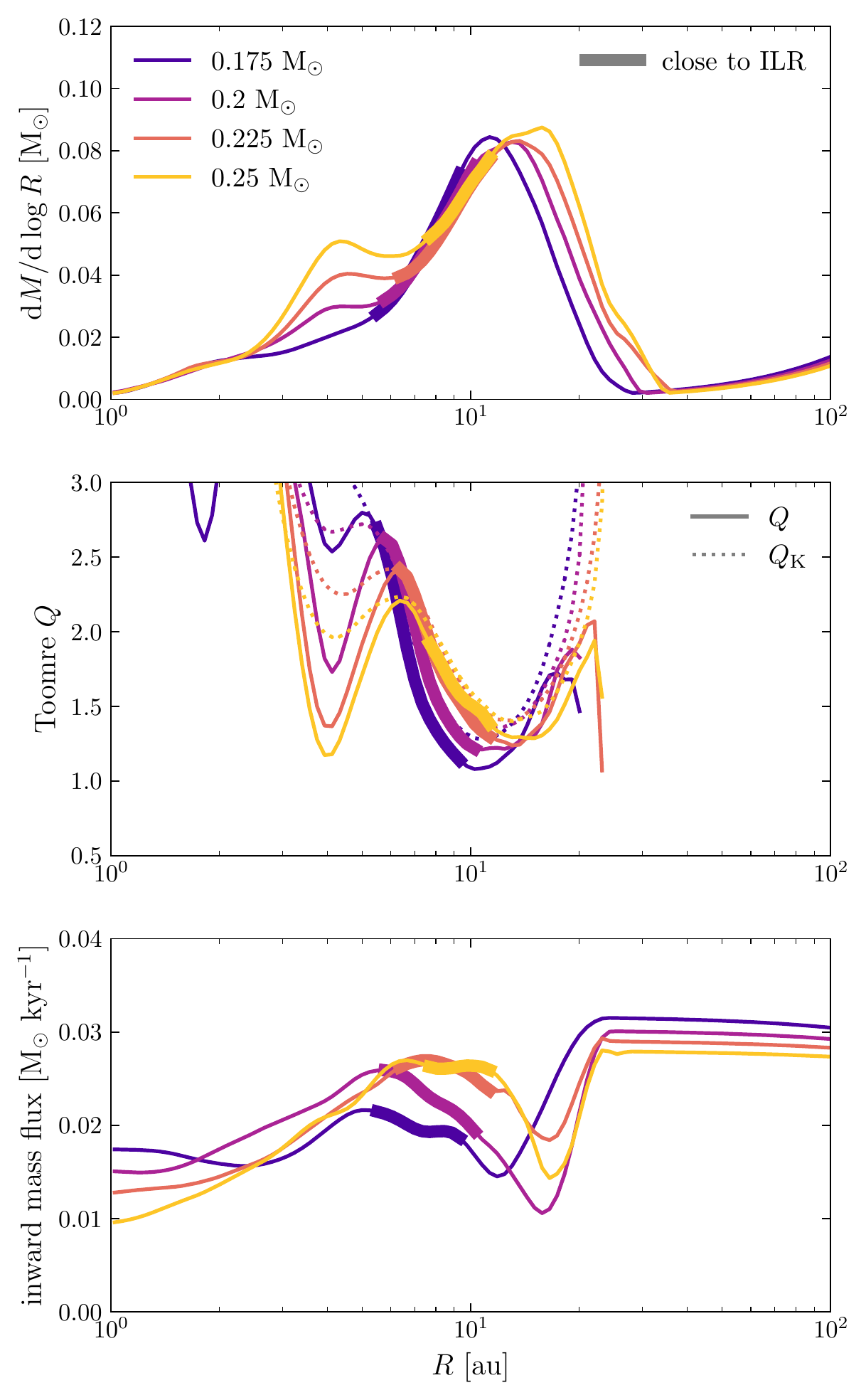}
    \caption{Mass distribution, Toomre $Q$ profile (with $Q_{\rm K}$ plotted for reference), and inward mass flux during the development of the second gravitationally unstable region. Each curve is averaged using 10 snapshots to suppress random fluctuations. The pattern speed of the spirals, $\Omega_{\rm p}$, is measured directly from multiple snapshots. The region within 20\% of the inner Lindblad resonance [ILR, where $m(\Omega-\Omega_{\rm p})=\kappa$ with $m=2$ for the dominant spirals] is marked with a thick line. (The region around the ILR is marked to account for fluctuations in $\kappa$.)
    Enhanced spiral-wave damping around the ILR increases the accretion mass flux, causing mass pileup at slightly smaller radii; this eventually leads to the formation of the second GI region.}
    \label{fig:lindblad}
\end{figure}

Fig.~\ref{fig:lindblad} shows that the profile of the mass flux and the evolution of the mass distribution is consistent with this theory: the region around the ILR remains close to the local maximum of the mass flux, and there is a pile-up of mass at smaller radiii, which eventually becomes the second gravitationally unstable region seen in Fig.~\ref{fig:Q_and_L_transport}.
Note that the ILR is not the only factor that affects the damping of spiral waves (and thus the mass flux). For example, we expect the waves to be more nonlinear and to undergo more damping (via steepening and shocking) when the column density is low (e.g., between $5$ and $7~{\rm au}$ in Fig.~\ref{fig:lindblad}); this explains why the peak in the mass-flux profile is slightly to the left of the ILR marked in the bottom panel of Fig.~\ref{fig:lindblad}.

Eventually, the mass pile-up at $R\lesssim R_{\rm ILR}$ causes that region to become gravitationally unstable as well. 
This changes how the column-density evolution is regulated, because now the excitation (or amplification) of waves by GI also affects $\partial_R F_L$. Additionally, waves are now excited at multiple locations, with different pattern speeds.
We therefore expect the feature around $R_{\rm ILR}$ to be less significant, if not to eventually disappear, thus providing a reason for the eventual merging of the two GI regions we see near the end of the simulation.

In reality this process would likely repeat itself: after the two GI regions merge, a significant portion of the angular-momentum transport at yet smaller radii should come from spiral waves excited around the inner region of that extended GI region. Mass can then pile up beyond the ILR of this wave and create another unstable region, which again should merge with the outer GI region. Given enough time, we expect this process to repeat indefinitely and eventually the whole disc (except for the outer edge, where transport is dominated by the dissipation of outwardly propagating waves) should become unstable. It is difficult to verify this idea directly with our simulation, however, because the new location of the ILR is too close to the inner radial boundary of the computational domain.

The mass pile-up process we describe above happens at a characteristic timescale ${\sim}\pi R^2\Sigma / \dot M(R)$, which can be comparable to the timescale of evolution of the entire system, ${\sim}M_{\rm tot}/\dot M_{\rm tot}$, when the disc-to-star mass ratio is order unity. If the evolution shown in our simulation is indeed generic, then it should be possible to find systems showing a gravitationally stable inner disc or exhibiting radial substructure in their (azimuthally averaged) column-density profile.

\section{Energy budget and disc temperature profile}\label{sec:Tprofile}

The disc temperature profile is determined by a balance between heating and radiative cooling. We assess this balance by first estimating the heating rate (which is also approximately the rate of radiative cooling; Section \ref{sec:heatcool}) and then finding the relationship between the cooling rate and the disc temperature (Sections \ref{sec:coolT} and \ref{sec:spiralconv}).

\subsection{Rate of heating and cooling}\label{sec:heatcool}

We begin our estimate of the heating and cooling rates in the disc with a simple argument. As material in the disc moves radially inwards to be accreted, gravitational energy is released at a rate of $-g_R\dot M(R)$ per unit radius. A fraction of this energy (half of it, in the case of a Keplerian disc) is turned into increased rotational kinetic energy, and the rest should be mostly turned into heat.
Therefore, the heating rate per unit radius is of order $-g_R\dot M(R)\approx -g_R\dot M_{\rm tot}$. The cooling rate should be similar, because the timescale of heating and cooling is generally short compared to the timescale of disc evolution and the disc should be in approximate thermal equilibrium.

In Appendix \ref{A:thermal_budget}, we perform a more rigorous analysis of the thermal budget of the disc. It turns out that the intuition above is only partially correct: the energy flux from turbulent perturbations (including spiral waves) can cause a significant amount of radial energy transport and affect the rate of heating (or, more accurately, energy injection).
However, in Appendix \ref{A:thermal_budget} we also show that the energy injection due to such a turbulent flux is at most comparable to the energy injection by accretion, and $-g_R\dot M_{\rm tot}$ remains a good order-of-magnitude estimate of the disc heating and cooling rates.

\begin{figure}
    \centering
    \includegraphics[scale=0.66]{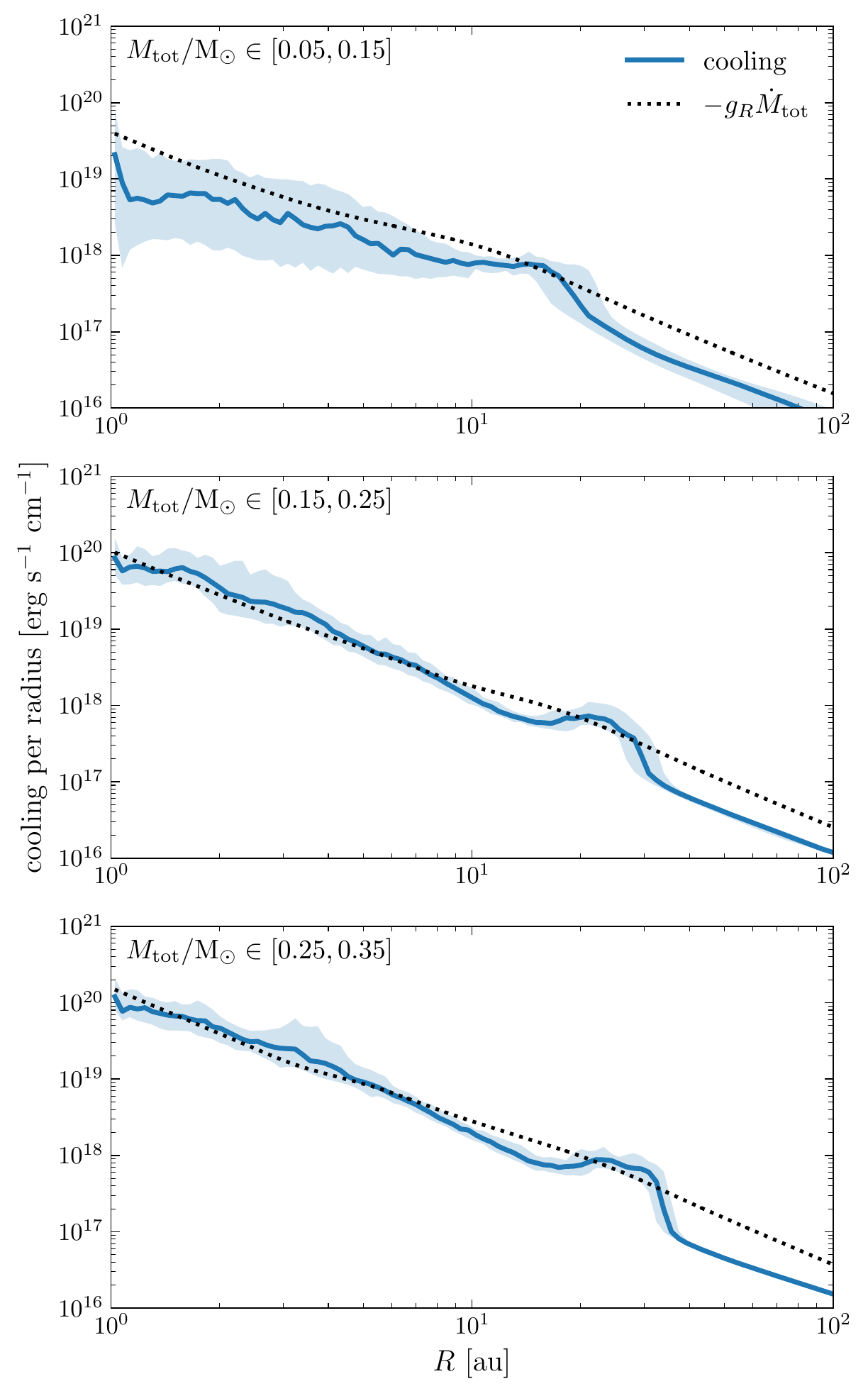}
    \caption{Cooling rate per radius at different epochs. Shaded area marks $1\sigma$ deviations. Inside the disc, the cooling rate required by thermal equilibrium is consistent with the (order-of-magnitude) analytic estimate $-g_R\dot M_{\rm tot}$.}
    \label{fig:energy_budget}
\end{figure}

We check the validity of this estimate by comparing it with the cooling rate measured from our simulation in Fig.~\ref{fig:energy_budget}. The actual cooling rate agrees relatively well with $-g_R\dot M_{\rm tot}$ inside the disc.

\subsection{Relation between cooling rate and disc temperature}\label{sec:coolT}

Now that we have an estimate of the disc cooling rate, the disc temperature profile can be estimated as well if the relation between cooling and temperature is known. Here we discuss this relation.

\begin{figure}
    \centering
    \includegraphics[scale=0.66]{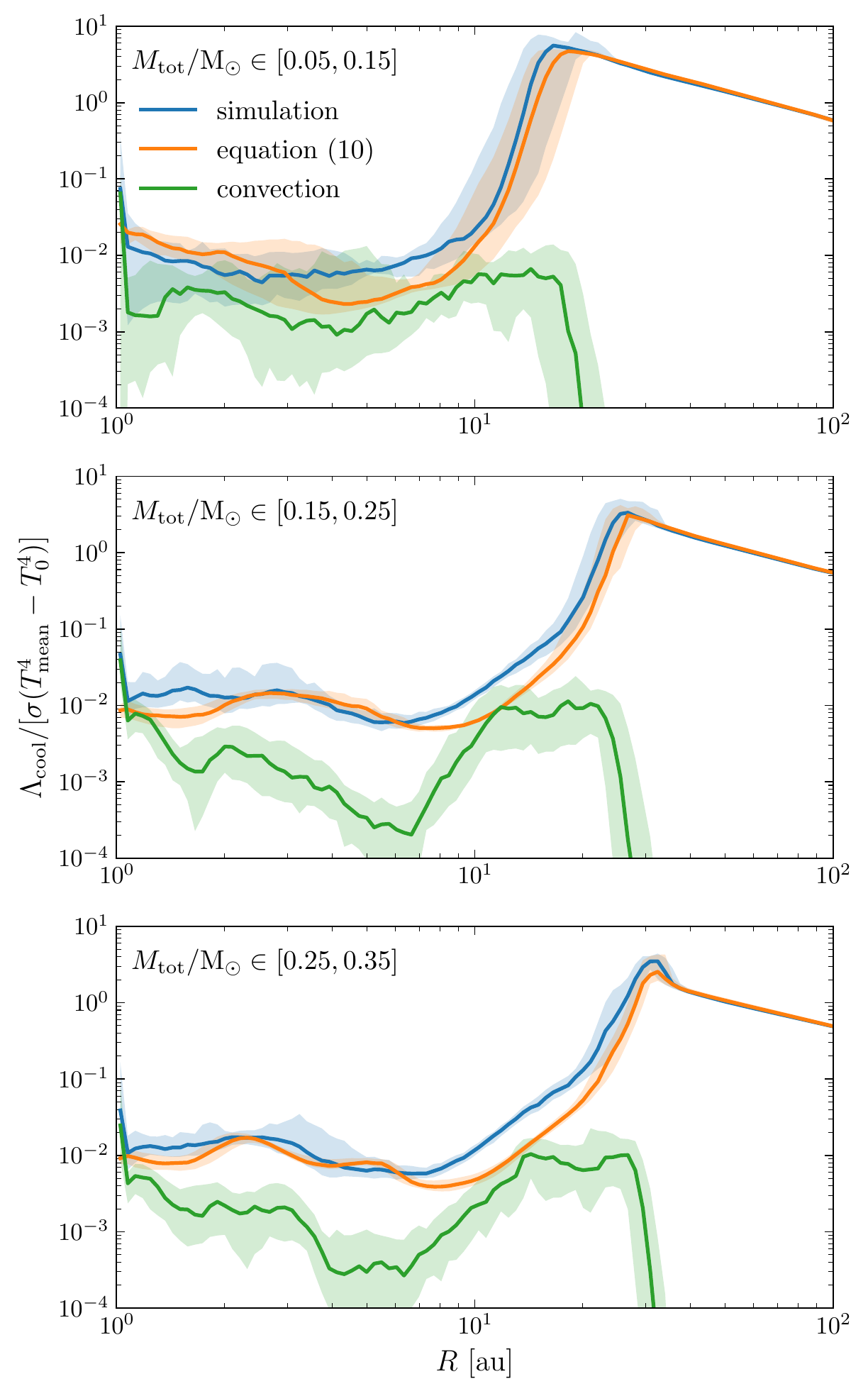}
    \caption{Normalized cooling rate per area at a few different epochs. The orange curve corresponds to the estimate in Eq. \eqref{eq:cooling_est}, which ignores the effect of turbulence (including spirals). This estimate fits reasonably well to simulation.
    The green curve measures vertical convection: We measure the vertical convective internal energy flux $\langle\rho u  v_z'\rangle$ (where $u$ is the specific internal energy and $v_z'$ is the difference between $v_z$ and its density-weighted azimuthal average) and take twice its maximum in $z$ (at a given $R$; the factor of 2 accounts for both sides of the disc) as an estimate for the cooling due to vertical convection. Convection is always weak compared to the total cooling.}
    \label{fig:cooling_rate}
\end{figure}

Again, we begin with a rough estimate, which ignores the effect of turbulent fluctuations altogether. In this case, the cooling rate (per area) of a disc in thermal equilibrium can be estimated as follows:\footnote{
In the literature, $(16/3\tau)\sigma T^4$ is frequently quoted as the cooling rate for an optically thick disc. However, the $T$ in this expression should be treated as the mid-plane temperature; the mean (density-weighted average) temperature is smaller than the mid-plane temperature by a factor of 0.874 (in the limit  $\tau\to \infty$), if we assume the disc has constant opacity \citep[eq. 3.11]{Hubeny1990}. The $8/(16/3 \times 0.874^4)=0.875$ factor in equation \eqref{eq:cooling_est} takes this into account.
}
\begin{equation}\label{eq:cooling_est}
    \Lambda_{\textrm{no-conv}} \sim \frac{8\tau_{\rm P}}{1+0.875 \tau_{\rm P}\tau_{\rm R}} \,\sigma (T_{\rm mean}^4-T_0^4) .
\end{equation}
Here $\tau_{\rm P}$ and $\tau_{\rm R}$ are the Planck and Rosseland optical depths at the midplane. Equation \eqref{eq:cooling_est} is exact when $\tau_{\rm P}\ll 1$ or $\tau_{\rm P},\tau_{\rm R}\gg 1$ and provides a smooth transition between these two limits. We compare this estimate with the actual cooling rate in Fig.~\ref{fig:cooling_rate}. While there are some small differences (at most by a factor of ${\sim}2$), this estimate stays very close to the actual cooling rate.

\begin{figure}
    \centering
    \includegraphics[scale=0.66]{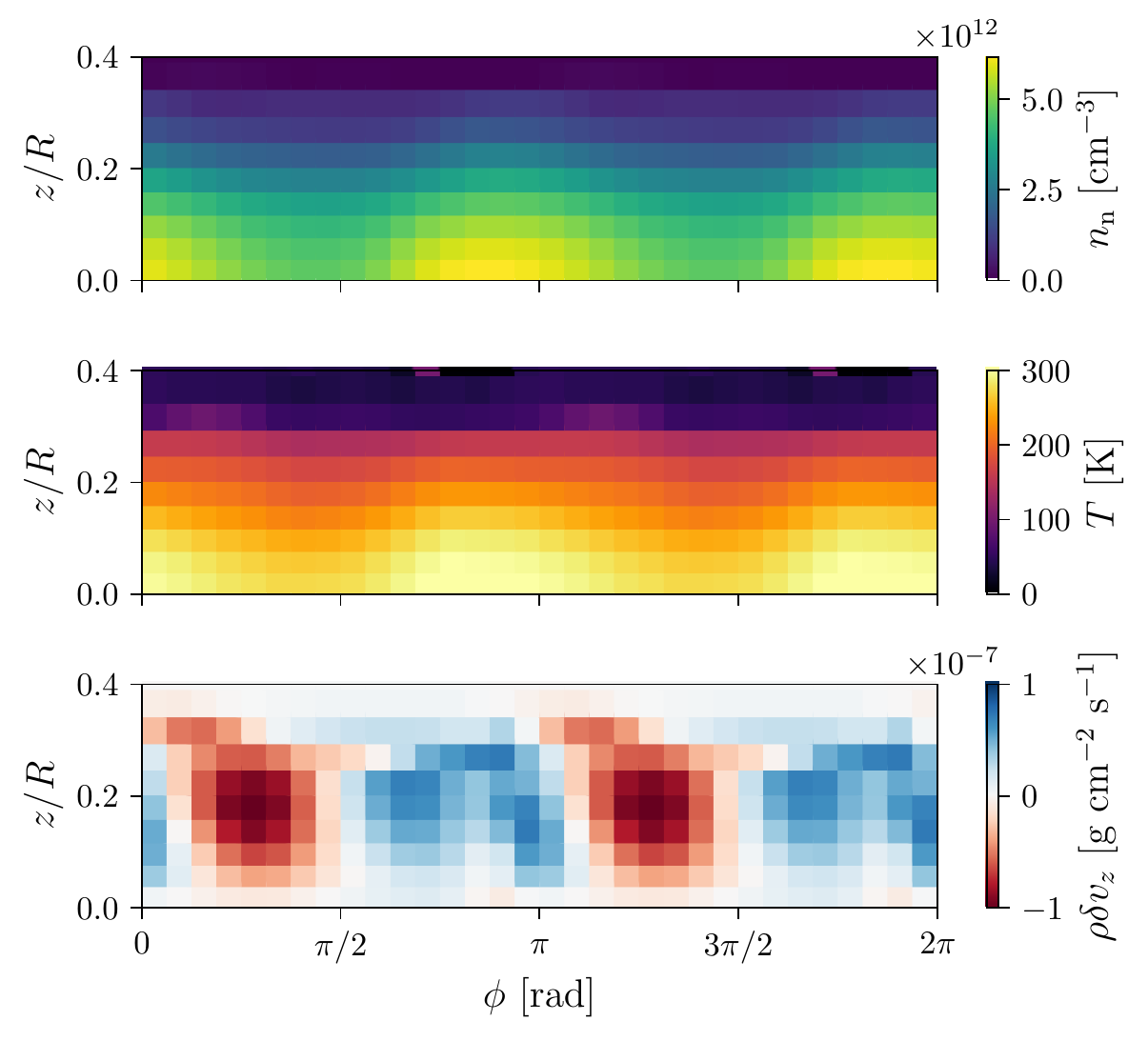}
    \caption{Density, temperature, and vertical velocity in the disc at $10~{\rm au}$, when $M_{\rm tot}=0.3~{\rm M}_\odot$. Both the temperature fluctuations and vertical motions correlates with the density fluctuations; this leads to convective heat transport that enhances cooling. However, as we show in Fig.~\ref{fig:cooling_rate} and Section \ref{sec:spiralconv}, such an effect is never significant.}
    \label{fig:spiral_convection}
\end{figure}

Next we consider how the inclusion of turbulent fluctuations -- mainly the spirals -- might affect the cooling rate. First, the temperature and optical depth are no longer uniform at a given radius. But we anticipate the effect of this variation to be relatively unimportant, since these fluctuations, and the variation in cooling rate they induce, should at most be of order unity. A second, and potentially more important, effect of the spirals is the vertical convection associated with them \citep[e.g.,][]{RiolsLatter2018, Bethune2021}.\footnote{This should not be confused with convection due to a vertical entropy gradient; our disc is generally stable against the latter, since for typical disc temperatures (a few 100~K) we have $\gamma\approx 1.4$ and $\kappa_{\rm P}\sim\const$ \citep[eqn.~4]{Rafikov2007}.} An example of such convective motions is shown in Fig.~\ref{fig:spiral_convection}.
In our 3D simulation, we find that this kind of convection is generally weaker than the total radiative cooling by a factor of a few or more (Fig.~\ref{fig:cooling_rate}, green curve), and thus does not significantly affect the cooling rate.
In the next subsection we argue that the same should be true for spiral convection in any Class 0/I disc.
Therefore, even in the presence of turbulence and spirals, equation \eqref{eq:cooling_est} should remain a reasonably good estimate for the relation between disc temperature and cooling rate.

\subsection{Convection driven by spirals}\label{sec:spiralconv}

In this subsection we derive an analytic estimate of the convection by spirals and use that to argue that such convection should not dominate cooling in a Class 0/I disc.

First, because there is a strong correlation between vertical velocity and temperature (as both correlate with the spirals), the amount of convective cooling should be proportional to the characteristic amplitude of the velocity perturbations times the amplitude of fluctuations in the internal energy density:
\begin{equation}
    \Lambda_{\rm conv} \sim \frac{1}{2} U(\delta v_z/H)(\delta u/u).
\end{equation}
Here $U$ is the vertically-integrated internal energy, $\delta v_z$ is the (density-weighted) rms velocity perturbation, $H$ is the scale height (measured by dividing the density-weighted rms sound speed by $\Omega$), and $\delta u$ is the (density-weighted) rms specific internal energy perturbation. The factor of $1/2$ is chosen to fit the simulation result (see Appendix \ref{A:spiral_convection}).

Next we estimate $\delta v_z$ and $\delta u/u$.
The vertical motion is driven mainly by the pressure gradient associated with temperature perturbations (note that spiral waves in an isothermal disc do not generate vertical motion; see \citealt{RiolsLatter2018}); increased temperature leads to vertical expansion of the disc by increasing the scale height needed for hydrostatic equilibrium $H_{\rm eq}$.
The frequency at which the temperature is perturbed by the spirals is (in the frame corotating with the gas)
\begin{equation}
\Omega_{\rm s} \equiv |\Omega_{\rm p}-\Omega|.
\end{equation}
Meanwhile, the characteristic rate at which the disc returns to vertical hydrostatic equilibrium can be estimated using the vertical sound crossing rate, $c_{\rm s}/H \sim \Omega$.
Comparing $\Omega$ with the `forcing' frequency $\Omega_{\rm s}$ gives two different regimes. When $\Omega\gtrsim\Omega_{\rm s}$, the disc can adjust itself into vertical hydrostatic equilibrium quickly, and the amplitude of vertical oscillation is just $\delta H_{\rm eq}\sim H(\delta u/u)$. This gives
\begin{equation}
    \delta v_z \sim \Omega_{\rm s} \delta H_{\rm eq} \sim c_s  (\Omega_{\rm s}/\Omega) (\delta u/u)\quad({\rm for}~\Omega\gtrsim\Omega_{\rm s}).
    \label{eq:dvz_fast}
\end{equation}
When $\Omega\lesssim\Omega_{\rm s}$, the disc cannot adjust to vertical hydrostatic equilibrium, and the amplitude of the vertical acceleration remains $a_z\sim g_z(\delta u/u) \sim \Omega^2 H (\delta u/u)$. The typical vertical velocity is then
\begin{equation}
    \delta v_z \sim \Omega_{\rm s}^{-1} a_z \sim c_s  (\Omega/\Omega_{\rm s}) (\delta u/u) \quad({\rm for}~\Omega\lesssim\Omega_{\rm s}).
    \label{eq:dvz_slow}
\end{equation}
Combining equations \eqref{eq:dvz_fast} and \eqref{eq:dvz_slow}, we get
\begin{equation}
    \delta v_z \sim c_{\rm s} \min(\Omega_{\rm s}/\Omega, \Omega/\Omega_{\rm s}) (\delta u/u) .
    \label{eq:dvz_est}
\end{equation}

Finally, we estimate $\delta u/u$ by assuming that all of the heating happens near the spiral shock, while cooling is more uniformly distributed in $\phi$. Gas in the disc encounters spiral shocks at a frequency of $\Omega_{\rm s}/\pi$ (for $m=2$ spirals), so the energy injection (per disc area) at each encounter is $\Lambda_{\rm heat}(\pi/\Omega_{\rm s})$, and the jump in internal energy across the shock is $\delta u \sim \Lambda_{\rm heat}(\pi/\Omega_{\rm s}) \Sigma^{-1}$.
We therefore expect
\begin{equation}
    \delta u/u \sim \min\left(\frac 12, \frac{\Lambda_{\rm heat}(\pi/\Omega_{\rm s})}{U}\right) .
    \label{eq:du_est}
\end{equation}
Here we cap $\delta u/u$ to $1/2$, corresponding to the case where under-dense material contains nearly zero specific internal energy.

Altogether then, we have
\begin{equation}
    \Lambda_{\rm conv} \sim \frac 12 U \Omega_{\rm s} \min(1,\Omega^2/\Omega_{\rm s}^2) \min\left(\frac 12, \frac{\pi\Lambda_{\rm heat}}{U\Omega_{\rm s}}\right)^2.
    \label{eq:convection_est}
\end{equation}
This gives $\Lambda_{\rm conv}\lesssim (\pi/4)\Lambda_{\rm {heat}}$, with the equality being achieved only when $\Omega_{\rm s}\sim\Omega$ (i.e., at radii $\lesssim$ the radius of spiral excitation) and $U\Omega \sim \Lambda_{\rm heat}$ (i.e., the heating/cooling timescale is comparable to orbital timescale).
In Appendix \ref{A:spiral_convection} we compare the above analytic estimates with the convective energy transport measured from the simulation, and show that these estimates model the simulation measurement accurately. 
The relation between vertical velocity and heating rate proposed above is also consistent with the result observed in the higher resolution simulation (of an initially gravitationally unstable disc) from \citet{Bethune2021}, which shows $\delta v_z\sim 0.1 c_{\rm s}$ for $\Lambda_{\rm heat}\sim 0.1U$ and $\Omega\sim\Omega_{\rm s}$.

In summary, in Class 0/I discs, spiral convection should not produce enough vertical-energy transport to account for most of the cooling.

\section{Magnetic field evolution}\label{sec:Bevolution}

\subsection{Magnetic decoupling and field strength in the disc}\label{sec:decoupling}

In the ideal-MHD limit, the magnetic field is frozen into and advected with the gas, and the field strength is simply proportional to the column density. In reality, however, the presence of non-ideal magnetic diffusion (Ohmic dissipation and ambipolar diffusion for our simulation) allows the field to become decoupled from the gas and left behind as the bulk-neutral fluid flows inwards. Neglecting the Hall effect, this decoupling is described by the non-ideal induction equation
\begin{equation}\label{eqn:induction}
\partial_t \bb{B} = \grad\btimes\left(\bb{v}\btimes\bb{B} - \tfrac{4\pi}{c}\eta_{\rm O}\bb{j} - \tfrac{4\pi}{c}\eta_{\rm A}\bb{j}_\perp\right) ,
\end{equation}
where $\bb{j} = (c/4\pi)\grad\btimes\bb{B}$ is the current density and $\eta_{\rm O}$ and $\eta_{\rm A}$ are the Ohmic and ambipolar diffusivities. For simplicity, we ignore the contribution from the field-parallel current $\bb{j}_\parallel$ to the Ohmic term, which is a good approximation because $\bb{j}$ is nearly perpendicular (`$\perp$') to $\bb{B}$ everywhere and $\eta_{\rm O}\ll\eta_{\rm A}$ in most of the protostellar core. Under this approximation, equation \eqref{eqn:induction} may be rewritten as
\begin{equation}
\partial_t \bb{B} = \grad\btimes[(\bb{v} + \bb{v}_{\rm d})\btimes\bb{B}],
\end{equation}
where the non-ideal drift velocity $\boldsymbol v_{\rm d}$ is defined by
\begin{equation}
\bb{v}_{\rm d} \equiv - (\eta_{\rm O} + \eta_{\rm A}) \frac{\bb{B}\btimes(\grad\btimes\bb{B})}{ B^{2}} .
\label{eq:v_drift}
\end{equation}
The combination $\bb{v}+\bb{v}_{\rm d}$ may be interpreted as the velocity of the field lines. Comparing the drift velocity $\bb{v}_{\rm d}$ with the gas velocity $\bb{v}$ therefore allows us to quantify the decoupling between the field and the gas. Note that the second part of equation \eqref{eq:v_drift} describes how non-uniform the field is; having the same form as the Lorentz force, it goes to zero when the field is straight and uniform, and increases as the field strength becomes more non-uniform or the field lines become more pinched.

\subsubsection{Different regimes of field decoupling}

\begin{figure}
    \centering
    \includegraphics[scale=0.66]{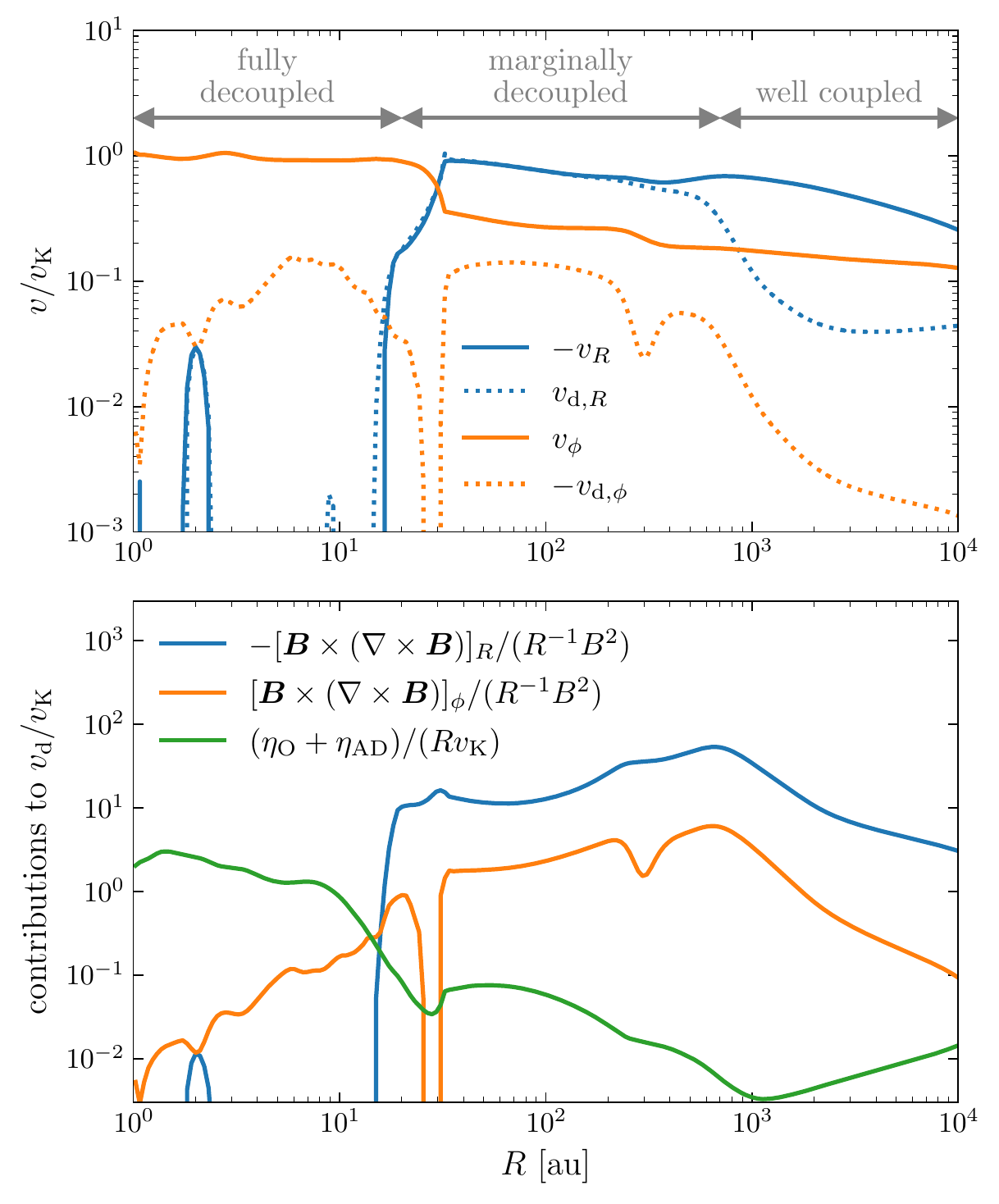}
    \caption{Top panel: Comparison between gas velocity $v$ (solid) and non-ideal drift velocity $v_{\rm d}$ (dotted) measured in the midplane in the radial (blue) and azimuthal (orange) directions, all azimuthally averaged (weighted by midplane $B_z$) and normalized by $v_{\rm K}\equiv \sqrt{-g_R R}$ (top panel). In the radial direction, the magnetic field is well coupled in the outer pseudodisc, marginally decoupled in the inner pseudodisc, and fully decoupled in the disc (as marked in the figure); in the  azimuthal direction, the magnetic field remains relatively well coupled in the whole domain. Bottom panel: Contributions to the midplane $v_{\rm d}/v_{\rm K}$ from magnetic-field non-uniformity (blue, orange) and from non-ideal magnetic diffusivity (green), all azimuthally averaged. Both panels are evaluated when $M_{\rm tot} \approx 0.4~{\rm M}_\odot$, with the behaviour being qualitatively similar at other epochs.}
    \label{fig:drift}
\end{figure}

In Fig.~\ref{fig:drift} we show the radial drift velocity at the midplane and compare it with the radial velocity of the (predominantly neutral) gas. We divide the domain into three different regimes:

\textbf{Outer pseudodisc: well-coupled field.} At large radii, $|v_{{\rm d},R}|\ll |v_R|$ and the field is well-coupled to the gas. The flux evolution is determined by (near) flux freezing, with a residual amount of ambipolar diffusion.

\textbf{Inner pseudodisc: marginally decoupled field.}
In the inner part of the pseudodisc, as the field becomes more pinched and the ambipolar diffusivity increases, $v_{{\rm d},R}$ becomes comparable to $-v_R$ and the field radially decouples from the gas.
The decoupling of the magnetic field tends to straighten out the field lines (reducing field non-uniformity), which in turn reduces $v_{{\rm d},R}$ and makes the field better coupled to the gas.
Therefore, the system tends towards a self-regulated, marginally decoupled state in which $v_{{\rm d},R}$ is kept comparable to $-v_R$ by regulating field non-uniformity. This idea is consistent with the profiles of drift velocity and field non-uniformity shown in Fig.~\ref{fig:drift}.

The decoupling also slows the inward advection of the magnetic field, leading to a local enhancement in magnetic-field strength, which is visible in the $B_z$ panel of Fig.~\ref{fig:disc_1d}.
Since the magnetic field remains relatively well coupled in the azimuthal direction, this increased field strength can help explain the increased braking in the inner pseudodisc seen in Fig.~\ref{fig:braking_rate}.

We note in passing that this local increase in magnetic-field strength due to decoupling is physically similar to the `magnetic wall' \revision{suggested} in \citet{LiMcKee1996}, although in our case the wall is caused by ambipolar diffusion rather than Ohmic dissipation (as in \citealt{cck98}, \citealt{ck98}, and \citealt{TM05b}) \revision{and there is no clear evidence of magnetic interchange instability occurring behind the wall. In our simulation, the maximum growth rate of the interchange \citep{ls95}, which is proportional to the gradient of the local mass-to-flux ratio, is always smaller than the local infall rate (which is close to free-fall). Meanwhile, significant asymmetries due to interchange have been observed in some simulations of protostar formation \citep[e.g.,][]{Krasnopolsky2012,Zhao2018, Machida2019}, and so its occurrence (i.e., whether it can occur on a timescale shorter than both infall and magnetic diffusion) likely depends upon the details of the simulation setup and various physical conditions.}

\textbf{Protostellar disc: fully decoupled field.} In the protostellar disc, the even higher density further increases the non-ideal diffusion; more importantly, the typical radial velocity is significantly lower compared to the pseudodisc due to increased column density. As a result, the field is fully decoupled radially and becomes nearly uniform and straight.

We comment that the marginally decoupled regime and the fully decoupled regime are essentially controlled by the same physics, in which the magnetic field self-regulates its non-uniformity (for a given magnetic diffusivity) to give a drift velocity that approximately cancels the gas velocity. The main difference between these two regimes is just whether the field non-uniformity required in this self-regulated state is negligibly small.

\subsubsection{Where does decoupling begin?}

One interesting trend we observe is that the location where magnetic decoupling begins (i.e., the boundary between the well-coupled regime and the marginally decoupled regime) shifts outwards in time.
This trend causes the expansion of the region in the inner pseudodisc with enhanced $B_z$ (Fig.~\ref{fig:disc_1d}) and increased magnetic braking (Fig.~\ref{fig:braking_rate}), thereby contributing to the increased importance of pseudodisc magnetic braking at later times.
Here we discuss the origin of this trend.

We can obtain a rough estimate for the location of decoupling by considering the curvature of the pseudodisc magnetic field in the ideal-MHD limit. Decoupling happens approximately when the drift velocity corresponding to this ideal-MHD field non-uniformity becomes comparable to the infall velocity. From the perspective of the pseudodisc, the central protostar-disc system is effectively a point mass. We therefore divide the pseudodisc into an outer, `pre-stellar' region, where gravity is dominated by the self-gravity of the pseudodisc [$M({<}R)\gg M_{\rm tot}$], and an inner, `post-stellar' region, where gravity is dominated by the point mass [$M({<}R)\approx M_{\rm tot}$].\footnote{The post-stellar region is often called the `expansion wave' region in the literature, following \citet{Shu1977}. Here we avoid this terminology because in our case the outer (pre-stellar) envelope is not a near-hydrostatic singular isothermal sphere \citep[as in][]{Shu1977}, but rather a dynamically contracting pseudodisc. Moreover, the boundary between the pre-stellar region and the post-stellar region (`expansion wave') is not a physical, outward propagating wave/shock but a transition between two self-similar profiles.
}
The pre-stellar region acquires a near-self-similar profile during the pre-stellar collapse, one which holds as well during the Class 0/I phase, with (approximately) $\Sigma \propto R^{-1}$ and $M({<}R)\propto R$. The post-stellar region also converges to a near-self-similar profile, but with $\Sigma \propto R^{-1/2}$ \citep[e.g.,][]{cck98,DBK12}.
The boundary between these two regions, $R_{\rm post}$, occurs roughly at $M({<}R)\sim 2M_{\rm tot}$ and expands outwards as $M_{\rm tot}$ increases.

We use these profiles to obtain the scaling of the field non-uniformity $-[\bb{B}\btimes(\grad\btimes\bb{B})]_R/B^2$ in the pseudodisc midplane (assuming ideal-MHD).
Because the pseudodisc is geometrically thin, this non-uniformity is dominated by the field-line curvature, and so
\begin{equation}
-\frac{[\bb{B}\btimes(\grad\btimes\bb{B})]_R}{B^2} \approx \frac{1}{B_z} \pD{z}{B_R} \sim \frac{B_R^{\rm d}}{B_z H}.
\end{equation}
Here $B_R^{\rm d}$ is the strength of the radial field at the pseudodisc surface; the pseudodisc scale height $H$ may be estimated as
\begin{equation}
H \sim \frac{c_{\rm s}^2}{g_z}.
\end{equation}
The sound speed $c_{\rm s}\approx c_{\rm s0}$, because disc is approximately isothermal in the regions of interest here. The vertical component of the gravitational acceleration $g_z$ is ${\sim}2\pi {\rm G}\Sigma$ in the pre-stellar region, and ${\sim}({\rm G}M_{\rm tot}/R^2) (H/R)$ in the post-stellar region. Meanwhile, previous semi-analytic calculations \citep[e.g.,][]{cck98,galli06} predict, in the flux-freezing limit, $B_z, B_R^{\rm d} \propto R^{-1}$ in the pre-stellar region and $B_z \propto R^{-1/2}$, $B_R^{\rm d} \propto R^{-2}$ in the post-stellar region, the latter corresponding to a split monopole. Therefore, at a given time, the field curvature should scale as $R^{-1}$ and $R^{-5/2}$ for the pre-stellar and post-stellar regions, respectively.
Since the field curvature must be continuous around $R_{\rm post}$ and the pre-stellar profile remains approximately constant during the Class 0/I stage (because the local dynamical timescale is $\gtrsim$ the timescale after point-mass formation), we find
\begin{equation}
\frac{B_R^{\rm d}}{B_z H}\text{~(ideal-MHD)} \propto
\left\{
\begin{array}{ll}
      R^{-1} & (R\gtrsim R_{\rm post}) \\
      R^{-5/2}R_{\rm post}^{3/2} &(R\lesssim R_{\rm post})
\end{array} 
\right. .
\label{eq:curvature_est}
\end{equation}
This profile suggests that, at a given location $R$, the field curvature remains constant until $R_{\rm post}$ reaches this radius, and quickly increases (${\propto} R_{\rm post}^{3/2}$) afterwards (until decoupling sets in). The infall velocity also increases after $R_{\rm post}$ exceeds $R$, but it increases more slowly, with a scaling ${\propto} R_{\rm post}^{1/2}$ at a given radius.
Therefore, the growth of $R_{\rm post}$ (driven by accretion onto the point-mass $M_{\rm tot}$) promotes decoupling by increasing field curvature and thereby lowering the magnetic diffusivity required for decoupling. This argument is similar to that made in appendix C of \citet{ck98} (see also \citealt{cck98} and \citealt{kk02}). While the increase in field curvature is the main cause of decoupling in our pseudodisc, we note that in general magnetic decoupling may also be driven mainly by increased diffusivity \citep[e.g.,][]{DM01} and/or decreased radial velocity (e.g., our protostellar disc).

An example of decoupling driven by field curvature increase in the post-stellar region is shown in the bottom panel of Fig. \ref{fig:drift}, with the transition between the pre- and post-stellar regions occurring at $R_{\rm post} \sim 2000~{\rm au}$ and decoupling starting around ${\sim}800~{\rm au}$, after field curvature has increased by a factor of a few in the post-stellar region.
A similar correlation between the location of decoupling and $R_{\rm post}$ is observed throughout our simulation, suggesting that the expansion of the marginally decoupled region (and the increased importance of braking at later times -- see Section \ref{sec:discbraking}) is largely due to the expansion of the post-stellar region.

\subsubsection{What determines the field strength in the protostellar disc?}

Now we discuss how to obtain a rough estimate of the protostellar disc magnetic-field strength using the field strength in the pseudodisc, which will be used for our semi-analytic disc model in Section \ref{sec:model}.

One complicating factor is that the boundary between the marginally decoupled regime and the fully decoupled regime does not lie exactly at the disc boundary (defined as the location where the kinetic energy becomes dominated by its azimuthal component), but is rather slightly inside the disc. For example, in Fig.~\ref{fig:drift} the disc boundary is at $R_{\rm d}\sim 30~{\rm au}$ while the boundary of the fully decoupled regime is at $R_{\rm dec}\sim 20~{\rm au}$.
This is because the increase in magnetic diffusivity and column density (which produces the decrease in typical radial velocity) happens gradually across the transition region between the dense, gravitationally self-regulated region and the pseudodisc.

For $R < R_{\rm dec}$, the magnetic-field strength should be nearly constant. Between $R_{\rm dec}$ and $R_{\rm d}$, the magnetic field is non-uniform and slightly pinched to provide the finite drift velocity necessary for canceling the gas velocity. This leads to the bump in field strength at the disc edge shown in Fig.~\ref{fig:disc_1d}.
The transition region between $R_{\rm dec}$ and $R_{\rm d}$ is generally narrow, with $R_{\rm dec}/R_{\rm d}\sim 1$; the difference in magnetic-field strength between these two locations should be of order unity.
Therefore, we argue that the field strength in the pseudodisc around $R_{\rm d}$ can serve as a reasonable estimate of the field strength inside the disc, with an order-unity uncertainty due to the marginally decoupled region that occurs between $R_{\rm dec}$ and $R_{\rm d}$.

\subsection{Magnetic braking in the disc}\label{sec:discbraking}

Another important aspect of magnetic-field evolution is the generation of an azimuthal field component and the resulting magnetic braking. We showed earlier that angular-momentum removal from the protostar-disc system by magnetic braking is weak compared to angular-momentum injection by accretion. In this subsection we aim to understand what determines the strength of magnetic braking in the disc. This could help us determine in the future whether disc magnetic braking should be similarly unimportant in other systems.

Inside the disc, while the magnetic field is decoupled from the gas in the radial direction, it is relatively well-coupled to the gas in the azimuthal direction, with a non-ideal drift velocity significantly smaller than the azimuthal velocity (Fig.~\ref{fig:drift}).
Consider a field line threading the disc; the rotation rate of the field line near the disc should then be much greater than its rotation rate at $z\to\infty$.
\revision{Traditionally, the winding of field lines (and the associated angular momentum transport) due to this differential rotation between the disc and infinity is often considered as the main source of magnetic braking. In this case, the strength of the braking may be estimated using the steady-state model presented in \citet{bm94}, in which one effectively assumes that the region of a flux tube casually connected to the disc by torsional Alfv\'en waves is brought into corotation with the disc.
This kind of model may not apply to our disc, however, given that the gas dynamics occurring along the field lines -- a combination of outflow launching near the disc and pre-stellar infall further out -- is quite different than is assumed in this model, and the magnetic diffusivity in the outflow cavity and envelope could also have a nontrivial impact on how the field lines wind up. 
In fact, the method from \citet{bm94} provides an estimate for the magnetic-braking rate that is 1--2 orders of magnitude greater than what we measure in the simulation.}

We propose a different model to explain the strength of magnetic braking we observe, which assumes that the main origin of disc magnetic braking is not the differential rotation between the disc and infinity, but rather the (vertical) differential rotation {\em within} the disc. Under this assumption, the disc magnetic braking rate can be estimated as follows.

We observe the magnetic field to be dynamically unimportant inside the disc, as magnetic energy remains well below the kinetic and thermal energy. Therefore, the rotation of the gas is approximately the same as that in a hydrodynamic disc, and so along a field line at $R$ (which is approximately vertical inside the disc) the amount of differential rotation inside the disc (due to the vertical variation of radial gravity and pressure gradient) is approximately
\begin{equation}
\Delta\Omega \sim (z_{\rm d}/R)^2 \Omega.
\end{equation}
Here $\Omega$ is the rotation rate of disc at $R$, and $z_{\rm d}$ is the half-thickness of the disc.\footnote{$z_{\rm d}$ is defined as the height where magnetic energy exceeds the sum of kinetic and thermal energy, and should be comparable to the scale height $H=c_{\rm s}/\Omega$. However, since the braking rate estimate has a steep dependence on $z_{\rm d}$, we do not use $H$ to approximate $z_{\rm d}$ here. For our simulation, $z_{\rm d}$ is typically ${\sim}1.5 H$.}

In steady state, the field must rotate uniformly along each field line, and so the vertical differential rotation of the gas in the disc needs to be canceled by \revision{the vertical variation of} the non-ideal drift velocity. 
\revision{
The amplitude of the drift velocity then has to be at least
\begin{equation}
    v_{{\rm d},\phi} \sim R\Delta\Omega.\label{eq:drift_est_1}
\end{equation}
Meanwhile, using equation \eqref{eq:v_drift}, the azimuthal drift velocity is approximately
\begin{equation}
    v_{{\rm d},\phi} \sim -\frac{B_{\phi}^{\rm d}}{B_{\rm z} z_{\rm d}}\eta,\label{eq:drift_est_2}
\end{equation}
where $B_\phi^{\rm d}$ is $B_\phi$ evaluated at disc surface and $\eta$ is the typical $\eta_{\rm O}+\eta_{\rm AD}$ inside the disc, which can be estimated using $\eta_{\rm O}+\eta_{\rm AD}$ at the disc midplane.
Equations \eqref{eq:drift_est_1} and \eqref{eq:drift_est_2} provide an estimate of $B_\phi^{\rm d}$.}
The resulting braking rate per unit area is
\begin{equation}
F_{L,\rm brake} = -2RB_\phi^{\rm d} B_z \sim 2 B_z^2 \eta^{-1} \Omega z_{\rm d}^3.
\label{eq:braking_est}
\end{equation}

\begin{figure}
    \centering
    \includegraphics[scale=0.66]{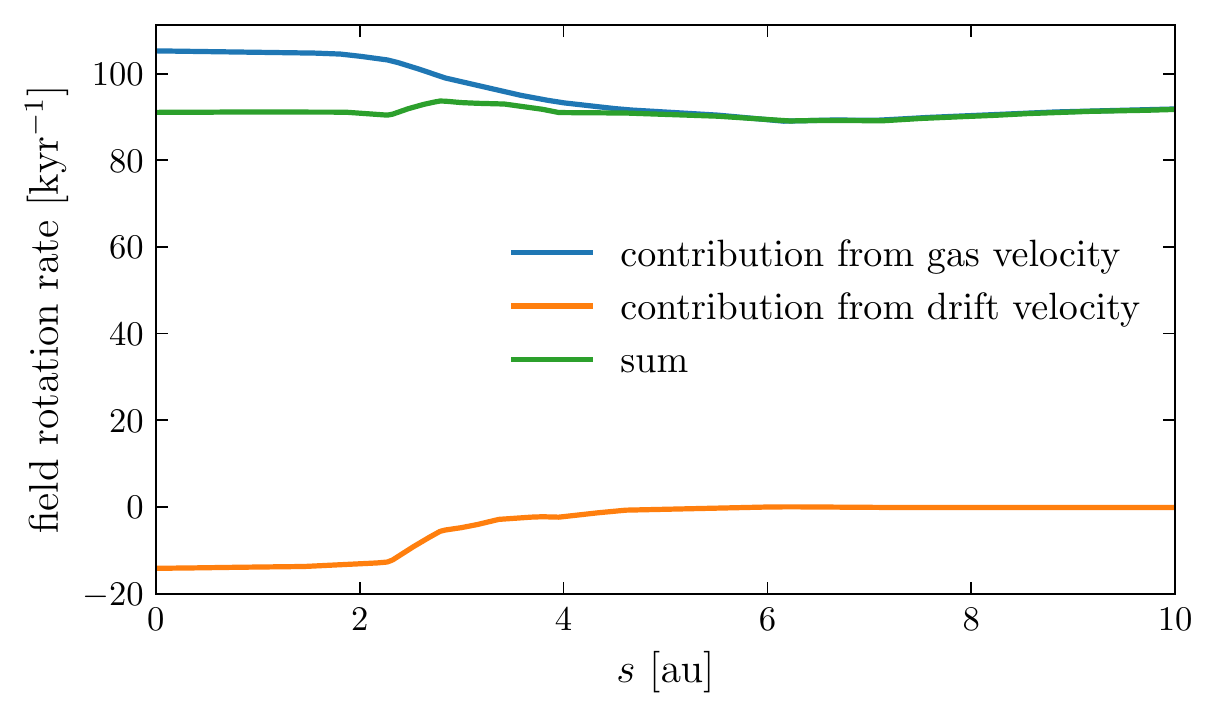}
    \caption{Azimuthally averaged field-rotation rate (green; see definition in Section \ref{sec:discbraking}) and its contributions from gas velocity (blue) and drift velocity (orange), evaluated when $M_{\rm tot} = 0.4~{\rm M}_\odot$ along a field line going through the midplane at $10~{\rm au}$. The abscissa is distance along the field line ($s=0$ at midplane). The field is in nearly uniform rotation, and the vertical differential rotation of the gas within the disc ($s\lesssim 3~{\rm au}$) is countered by the non-ideal drift velocity.}
    \label{fig:field_rot}
\end{figure}

We verify this physical picture in Fig.~\ref{fig:field_rot}, which shows the rotation rate of the field along a field line (green), and its contribution from the gas velocity (blue) and the drift velocity (orange). 
To compute the rotation rate of the field, we first find the horizontal component of the field velocity, $\bb{v}_{\rm f}^{\rm h}$, which is defined such that $v_{{\rm f},z}^{\rm h}=0$ and $\bb{v}_{\rm f}-\bb{v}_{\rm f}^{\rm h}$ (where $\bb{v}_{\rm f}\equiv \bb{v}+\bb{v}_{\rm d}$ is the field velocity) is parallel to the magnetic field (so $\bb{v}_{\rm f}\btimes\bb{B} = \bb{v}_{\rm f}^{\rm h}\btimes\bb{B}$).
The field rotation rate can then be defined as $\Omega_{\rm f}\equiv v_{{\rm f},\phi}^{\rm h}/R$.
The contribution to $\Omega_{\rm f}$ from the gas velocity and the drift velocity can be computed similarly by replacing $\bb{v}_{\rm f}$ with $\bb{v}$ or $\bb{v}_{\rm d}$. We comment that inside the disc, both $\bb{v}$ and $\bb{v}_{\rm d}$ are nearly horizontal, therefore we do not have to distinguish the velocities from their horizontal components in the earlier discussion.
Fig. \ref{fig:field_rot} shows that the field is indeed in nearly uniform rotation, and the differential gas rotation inside the disc is offset by non-ideal drift velocity.

\begin{figure}
    \centering
    \includegraphics[scale=0.66]{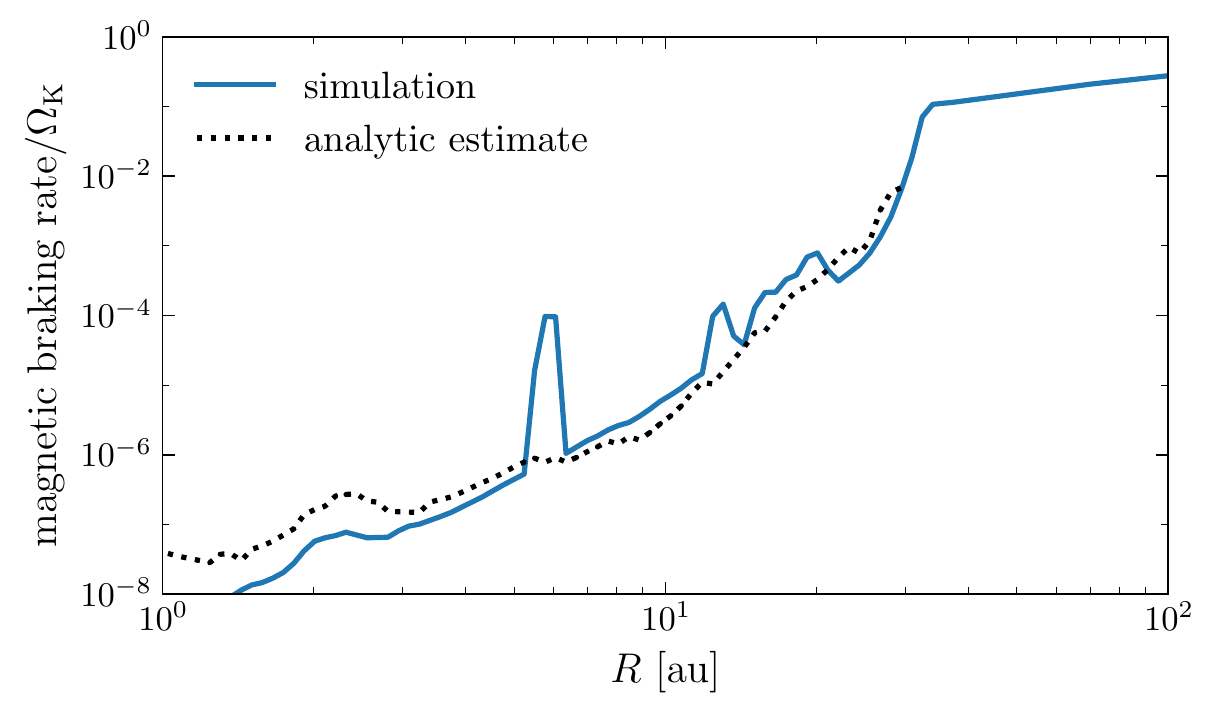}
    \caption{Comparison between the magnetic-braking rate when $M_{\rm tot} = 0.4~{\rm M}_\odot$ (same definition as in Fig.~\ref{fig:braking_rate}) and the analytic estimate \eqref{eq:braking_est}. The two show good agreement inside the disc. \revision{The occasional peaks in the measured braking rate (such as the one at ${\sim}6$ au) are random fluctuations likely associated with turbulent motions and spiral waves in the disc.}}
    \label{fig:braking_est}
\end{figure}

We also compare the estimated braking rate using equation \eqref{eq:braking_est} with the actual braking rate inside the disc in Fig.~\ref{fig:braking_est}. Our model predicts the braking rate relatively well, suggesting that it is likely the main mechanism determining the strength of disc magnetic braking in our simulation.

\section{Modeling disc evolution with a semi-analytic disc model}\label{sec:model}

With the physical insights acquired from the previous sections, we are now in a position to construct a simple analytic model of disc evolution (Section \ref{sec:1dmodel}), which, when used together with a computationally inexpensive 2D pseudodisc simulation, allows a relatively accurate prediction for the long-term disc evolution (Sections \ref{sec:1dpseudo} and \ref{sec:modelcomparison}).

\subsection{A semi-analytic 1D disc model}\label{sec:1dmodel}

\subsubsection{Basic assumptions}

Consider a gravitationally self-regulated disc similar to the one found in our simulation, with most of the disc marginally unstable. Let $R_{\rm GI}$ be the outer boundary of the gravitationally unstable region. While we expect the disc size $R_{\rm d}$ to be ${\sim}R_{\rm GI}$ from gravitational self-regulation (Sections \ref{sec:GItransport} and \ref{sec:GIsize}), there is still some order-unity difference between $R_{\rm GI}$ and $R_{\rm d}$, as the region between $R_{\rm GI}$ and $R_{\rm d}$ serves as a transition region for spiral waves excited within $R_{\rm GI}$ to disperse (Fig.~\ref{fig:Q_and_L_transport}). For simplicity, we assume
\begin{equation}
R_{\rm d} = f_{\rm d} R_{\rm GI}
\end{equation}
with $f_{\rm d}$ being a constant of order unity. To match the typical value of $R_{\rm d} / R_{\rm GI}$ realised in our 3D simulation, we set $f_{\rm d}=1.5$.
We also assume that most of disc mass and angular momentum are contained in $R\leq R_{\rm GI}$ (because the column density falls off rapidly outside $R_{\rm GI}$) and that the Toomre $Q$ parameter is approximately constant for $R\leq R_{\rm GI}$, i.e.,
\begin{equation}
Q(R\leq R_{\rm GI}) \approx Q_0 .
\end{equation}
We set $Q_0=1.5$ by default. $f_{\rm d}$ and $Q_0$ will be the only free parameters of our model.

\subsubsection{Local model for temperature and column density}

Next we model the mean temperature $T_{\rm mean}$ and column density $\Sigma$ at a given location inside the gravitationally self-regulated region ($R\leq R_{\rm GI}$).
We assume the disc to be near Keplerian, so that
\begin{equation}
Q \approx Q_{\rm K} = Q_0 .\label{eq:1D_Q}
\end{equation}
(Recall that $Q_{\rm K}$ is Toomre's $Q$ evaluated with $\Omega = \Omega_{\rm K} \equiv \sqrt{-g_R/R}$.) We estimate $g_R$ as $g_R\approx -{\rm G}M({<}R)/R^2$ where $M({<}R)$ is the total mass enclosed within $R$, including the protostar. Meanwhile, we showed earlier that thermal equilibrium in the disc requires a cooling rate of order $-g_R\dot{M}_{\rm tot}$ with $\dot{M}_{\rm tot}$ being the accretion rate of the protostar-disc system (Section \ref{sec:heatcool}), and that the relation between the cooling rate and mean temperature is approximated well by equation \eqref{eq:cooling_est} (Section \ref{sec:coolT}). Therefore, we can estimate $T_{\rm mean}$ by requiring
\begin{equation}
2\pi R \,\frac{8\tau_{\rm P}}{1+0.875\tau_{\rm P}\tau_{\rm R}} \,\sigma(T_{\rm mean}^4-T_0^4) = -g_R\dot M_{\rm tot}.\label{eq:1D_thermal}
\end{equation}
Here, the midplane optical depths $\tau_{\rm P,R}$ can be estimated using $\tau_{\rm P,R}\approx \Sigma\, \kappa_{\rm P,R}(T_{\rm mean})/2$, where $\kappa_{\rm P,R}(T)$ is the Planck/Rosseland mean opacity evaluated at temperature $T$ and $T_0=10~{\rm K}$ is the ambient temperature. Note that the steep dependence of the cooling rate on $T_{\rm mean}$ ensures that we get a reasonable temperature estimate even though $-g_R\dot{M}_{\rm tot}$ is only an order-of-magnitude estimate of the cooling rate (cf.~Fig.~\ref{fig:cooling_rate}). Using the two constraints above, we can then solve for $\Sigma$ and $T_{\rm mean}$ if $M({<}R)$ and $\dot{M}_{\rm tot}$ are known.

\subsubsection{Estimating disc size and disc profile}

The local model above cannot be directly used to predict the disc profile because $M({<}R)$ depends on $\Sigma(R)$ when the disc-to-star mass ratio is finite.
Here we describe how $\Sigma(R)$, $M({<}R)$ and $R_{\rm GI}$ can be calculated using the total mass ($M_{\rm tot}$), angular momentum ($L_{\rm tot}$), and accretion rate ($\dot M_{\rm tot}$) of the protostar-disc system.

For a given $\Sigma(R)$ and $M({<}R)$, we can solve for $R_{\rm GI}$ by requiring that
\begin{equation}
\int_0^{R_{\rm GI}} 2\pi R\rmd R~ \Sigma(R) R^2\Omega_{\rm K} = L_{\rm tot}. \label{eq:1D_RGI}
\end{equation}
Meanwhile, for a given $\Sigma(R)$ and $R_{\rm GI}$, we can also obtain $M({<}R)$ by solving
\begin{equation}
\D{R}{} M({<}R) = 2\pi R\Sigma(R) \label{eq:1D_MR}
\end{equation}
with the boundary condition $M({<}R_{\rm GI}) = M_{\rm tot}$.

To solve for $\Sigma(R)$, $M({<}R)$ and $R_{\rm GI}$, we perform the following iteration:
\begin{enumerate}
\item Initialize $M({<}R)$ to $M_{\rm tot}$;
\item Calculate $\Sigma(R)$ (and also $T_{\rm mean}$) using $M({<}R)$ and equations \eqref{eq:1D_Q} and \eqref{eq:1D_thermal}, and solve $R_{\rm GI}$ from equation \eqref{eq:1D_RGI};
\item Update $M({<}R)$ using equation \eqref{eq:1D_MR};
\item Repeat (ii) and (iii) until the disc profile converges.
\end{enumerate}
The disc size is simply $R_{\rm d}=f_{\rm d}R_{\rm GI}$.

Finally, the field in most part of the disc should be approximately uniform due to strong diffusion. The field strength can be estimated with the field strength at the inner edge of the pseudodisc $B_z^{\rm pd}$ (Section \ref{sec:decoupling}), giving
\begin{equation}\label{eqn:Bzestimate}
B_z \sim B_z^{\rm pd}(R_{\rm d}).
\end{equation}
%

\subsubsection{Summary and discussion}

In summary, we can estimate the temperature $T_{\rm mean}(R)$, column density $\Sigma(R)$, and disc size $R_{\rm d}$ if the total mass and angular momentum of the protostar-disc system, $M_{\rm tot}$ and $L_{\rm tot}$, and the accretion rate, $\dot{M}_{\rm tot}$, are known. We can further estimate the (approximately constant) field strength $B_z$ inside the disc if the pseudodisc magnetic-field profile $B_z^{\rm pd}(R)$ is also known. This model contains only two free parameters, $Q_0$ and $f_{\rm d}$, both of which should be order unity constants; we choose $Q_0=1.5$ and $f_{\rm d}=1.5$ to match the typical $Q$ and $R_{\rm d} / R_{\rm GI}$ realised in the 3D simulation.

For completeness, we also highlight the applicability of the above 1D semi-analytic model. The main physical assumptions of the model are that (i) magnetic field is radially decoupled from the gas and remains dynamically unimportant in the disc (which also implies that accretion needs to be regulated by GI), and (ii) accretion rate is approximately constant throughout the whole disc.
These assumptions are likely valid for typical Class 0/I systems, as our 3D simulation suggests.
However, after the dispersal of the envelope and pseudodisc (Class II or later), these assumptions are probably no longer valid due to the lack of fast accretion driven by the infalling pseudodisc.
Additionally, these assumptions would become invalid at very small radii where the temperature is so high (${\gtrsim}2000~{\rm K}$) that thermal ionization becomes important and the field becomes well-coupled to the gas again. (Neither are thermal ionization and gas opacity included in our model at high temperatures.) Our 1D model should therefore not be used to predict disc properties in this region. 
We note that according to our 1D+pseudodisc model (which we present in the next subsection), the radius where temperature reaches ${\sim}2000~{\rm K}$ is always ${\lesssim}2~{\rm au}$. Due to the small size of this region, we do not expect the angular momentum contained within this radius or the angular-momentum transport and removal (by magnetorotational instability, magnetic braking, and/or outflow launching) that may occur within this radius to significantly impact the prediction of disc profile farther out. 

We comment that semi-analytic models for (partially or fully) gravitationally self-regulated discs have been commonly adopted in studies of protostellar or protoplanetary disc evolution \citep[e.g.,][]{LinPringle1987,Krumholz2010,Rice2010,Zhu2010,Kimura2021}. The main contribution of our work is not developing this particular model, but justifying its assumptions in the context of magnetized Class 0/I discs that form self-consistently in 3D simulations of protostellar core collapse (Sections \ref{sec:AMbudget}--\ref{sec:Bevolution}) and showing how such a model can be coupled to a pseudodisc simulation to give reliable predictions of disc evolution (Sections \ref{sec:1dpseudo} and \ref{sec:modelcomparison}).

\subsection{1D+pseudodisc model: evolving the 1D model with a pseudodisc simulation}\label{sec:1dpseudo}

In the previous subsection, we discussed how to estimate the disc profile using a 1D model when $M_{\rm tot}$, $L_{\rm tot}$, $\dot{M}_{\rm tot}$ and the magnetic field profile in the pseudodisc are known. Obtaining these inputs for the 1D model requires knowledge of pseudodisc evolution. Here we show that one should be able to accurately predict pseudodisc evolution using computationally inexpensive, large-inner-boundary simulations that do not directly resolve the disc (Section \ref{sec:hierarchical}) and discuss how to predict disc evolution by coupling the 1D model to such a pseudodisc simulation (Section \ref{sec:modelcoupling}).

\subsubsection{Hierarchical evolution and simulating the pseudodisc}\label{sec:hierarchical}

One interesting observation from \citetalias{XK21} is that the evolution is largely hierarchical. In section 6 of \citetalias{XK21}, we performed the same simulation in 3D and (axisymmetric) 2D. In axisymmetry, GI can no longer cause angular-momentum transport and disc spreading; disc evolution is therefore very different between 2D and 3D runs. However, the pseudodisc evolution in these two simulations appear to be nearly identical, suggesting that whatever is happening in the protostellar disc has little impact on the evolution at larger scales (e.g., in the pseudodisc).

One can argue that such hierarchical evolution is generally expected for young protostellar systems as follows. First, the inner pseudodisc is falling in at near free-fall, with a velocity that generally exceeds both the Alfv\'en speed and the sound speed in the pseudodisc, making it impossible for perturbations to travel outward from the disc into the pseudodisc. Therefore, if the disc were to affect the pseudodisc, it would have to do so indirectly by affecting the magnetic field above the pseudodisc (the gas there is dynamically unimportant). Now consider the magnetic field above the pseudodisc; the field there is nearly force-free, and the disc affects this field mainly by setting the inner boundary condition of this force-free region through the magnetic field in the flux tube threading the disc.
The importance of this inner boundary condition can be characterized roughly by the ratio $\Phi_{B}(R_{\rm d})/\Phi_{B}(R)$, where $\Phi_B$ is the magnetic flux through a circle with radius $R$ at the midplane; if $\Phi_{B}(R_{\rm d})/\Phi_{B}(R)\ll 1$, this inner boundary condition should not be important at $R$. In the pseudodisc, the slope of the magnetic field $B_z$ is shallower than ${\propto}R^{-2}$ (cf.~Fig.~\ref{fig:disc_1d}); therefore $\Phi_B(R)$ is dominated by the contribution around $R$, giving a small $\Phi_{B}(R_{\rm d})/\Phi_{B}(R)$ except very close to the disc. Altogether then, there is no obvious way for the disc to affect pseudodisc evolution significantly.

For such a hierarchical evolution, it is possible to obtain correct pseudodisc evolution at significantly reduced computational cost, since resolving disc evolution at small scales is no longer necessary. Consider a simulation setup identical to our 3D simulation except that the inner boundary size is larger ($\gtrsim$ the circularization radius of infalling material) and we no longer forbid angular momentum from being advected through the inner boundary. We can also use a 2D (axisymmetric) domain, as the initial condition and pseudodisc evolution are both axisymmetric.
In this case, a rotationally supported disc will be unable to form because the inner boundary is so large that all angular momentum just leaves the domain instead of being accumulated at small radii to allow disc formation and growth.
While this type of large-inner-boundary simulation apparently does not directly produce the correct disc evolution, we do expect it to predict the pseudodisc properties relatively accurately if the physical picture of hierarchical evolution holds.

\begin{figure*}
    \centering
    \includegraphics[scale=0.66]{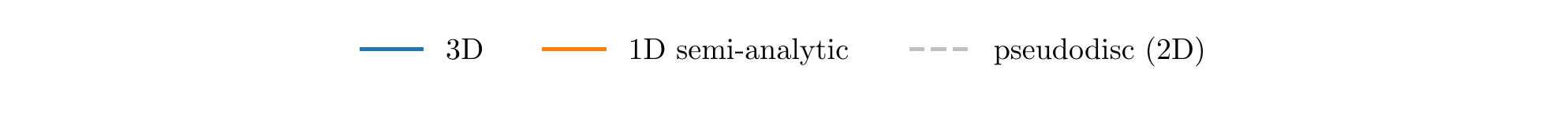}\\
    \vspace{-1em}
    \includegraphics[scale=0.66]{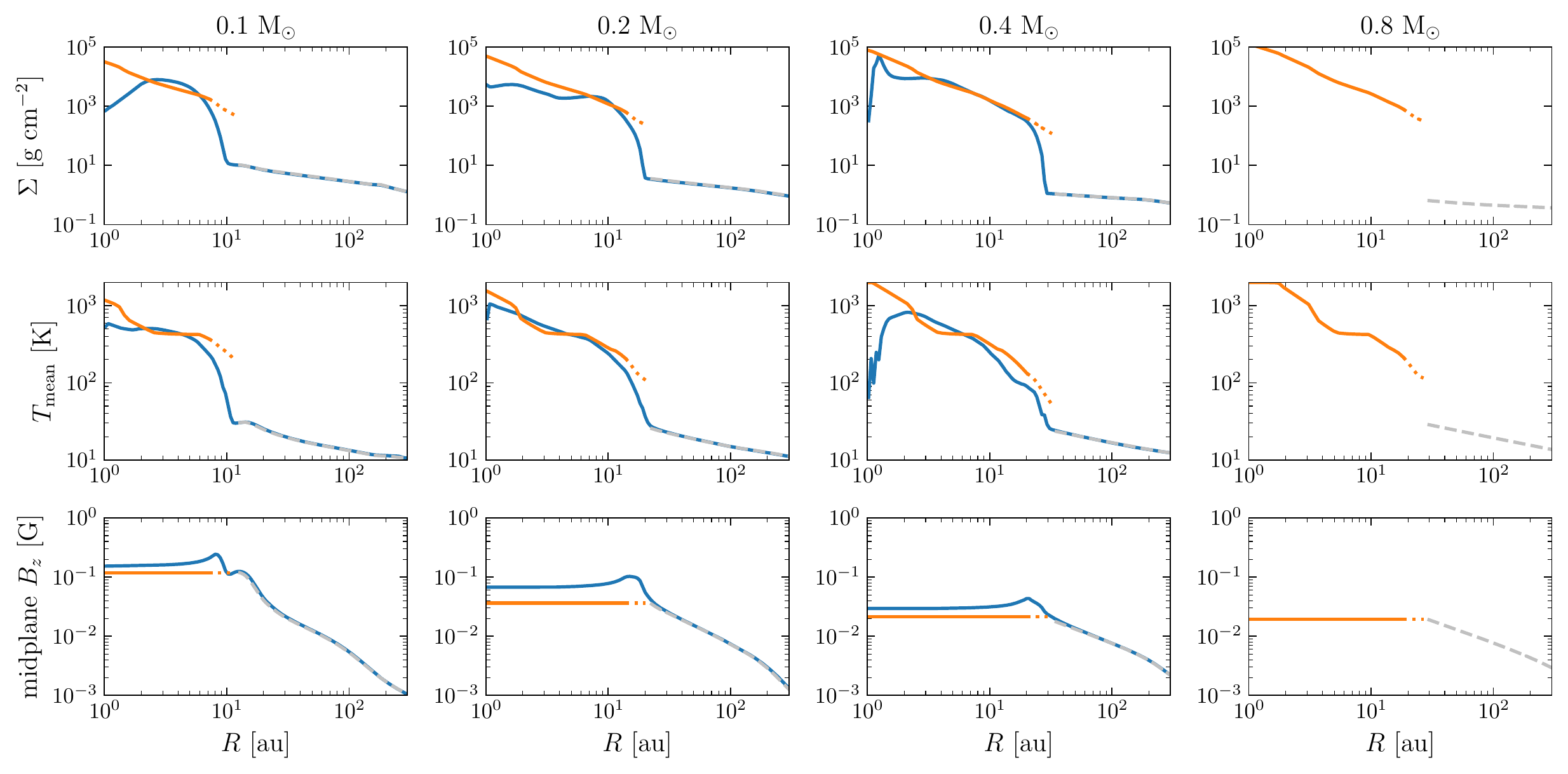}
    \caption{Comparison of disc profile between the 1D+pseudodisc model (orange) and our 3D simulation (blue) at a few different $M_{\rm tot}$. Solid lines correspond to $R\leq R_{\rm GI}$ and dotted lines mark $R_{\rm GI}\leq R \leq R_{\rm d}$ (where the assumption of $Q_{\rm K}=Q_0$ should no longer hold; we show this region only to mark disc size $R_{\rm d}$). The pseudodisc profile from our 2D, large-inner-boundary pseudodisc simulation (grey dashed) is also plotted at $R\geq R_{\rm d}$ for reference. The pseudodisc simulation agrees very well with the 3D simulation in this region.}
    \label{fig:1D_compare}
\end{figure*}

\begin{figure}
    \centering
    \includegraphics[scale=0.66]{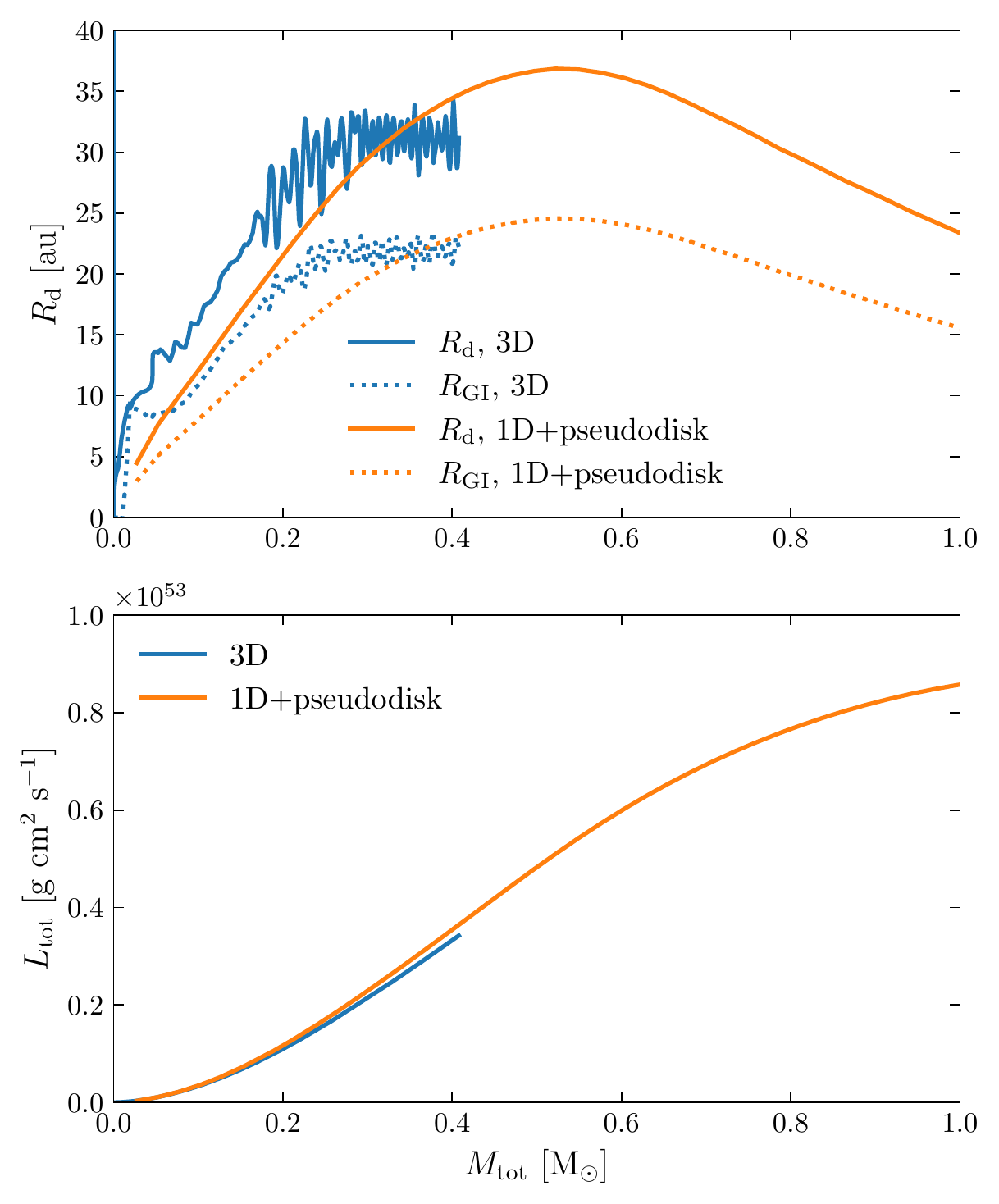}
    \caption{Comparison of the evolution of disc size (top panel) and angular-momentum budget of the protostar-disc system (bottom panel) between the 1D+pseudodisc model (orange) and our 3D simulation (blue). In the top panel, the outer boundary of the gravitationally unstable region $R_{\rm GI}$ (defined by the location where $Q_{\rm K}=2$ for our 3D simulation) is also plotted for reference. The 1D+pseudodisc model reproduces the evolution relatively well, and predicts disc shrinkage at late times.}
    \label{fig:1D_compare_Rd}
\end{figure}

\subsubsection{Coupling the 1D model to pseudodisc simulation}\label{sec:modelcoupling}

Now we outline how the pseudodisc evolution obtained with the aforementioned 2D large-inner-boundary simulation can be coupled to the 1D disc model to predict disc evolution.

Suppose we know $M_{\rm tot}$, $L_{\rm tot}$, and $R_{\rm d}$ at some time $t$; to evolve the system we want to get the value of these variables at $t+\Delta t$.
For given disc size $R_{\rm d}$, the total mass of the protostar-disc system is
\begin{equation}
M_{\rm tot} \approx M(r{<}R_{\rm d}) \approx M^{\rm pd}(r{<}R_{\rm d}).\label{eq:pd_Mtot}
\end{equation}
Here the superscript `pd' denotes the result obtained from the 2D pseudodisc simulation, which we expect to be a good approximation at $r\gtrsim R_{\rm d}$. [We use the mass within a spherical region $M(r{<}R_{\rm d})$ rather than a cylindrical region in order to exclude contributions from the low-density envelope at $z \gg R_{\rm d}$.]
Using this relation, we can estimate $M_{\rm tot}(t+\Delta t)$ using $M^{\rm pd}(r{<}R)$ at $t+\Delta t$ and $R_{\rm d}$ at $t$.
Then, for the angular-momentum evolution, we know that
\begin{equation}
\frac{{\rm d}L_{\rm tot}}{{\rm d}M_{\rm tot}} \approx \frac{F_L(R_{\rm d})}{F_M(R_{\rm d})} \approx \frac{F_L^{\rm pd}(R_{\rm d})}{F_M^{\rm pd}(R_{\rm d})},
\end{equation}
where $F_L$ and $F_M$ are mass and angular-momentum fluxes (evaluated on the sphere with $r=R_{\rm d}$), respectively, due to advection. Note that we do not include the contribution from the magnetic stress to $F_L(R_{\rm d})$, which corresponds to the assumption that magnetic braking in the disc is negligibly weak. This is true for our simulation; more generally, one can verify this assumption by estimating the strength of braking with the model described in Section \ref{sec:discbraking}.
Using this relation, we can update $L_{\rm tot}$ to $t+\Delta t$ as well. Next we estimate $\dot M_{\rm tot}(t+\Delta t)$ using $F_M^{\rm pd}(t+\Delta t)$. Finally, using the 1D model, we obtain $R_{\rm d}(t+\Delta t)$ from $M_{\rm tot}$, $L_{\rm tot}$, and $\dot M_{\rm tot}$ at $t+\Delta t$.

Using the procedure above (with initial condition $M_{\rm tot}$, $L_{\rm tot}$, $R_{\rm d}\to 0$), we can calculate the time evolution of the disc properties from psuedodisc simulation data.

\subsection{Comparison between the 1D+pseudodisc model and the 3D model}\label{sec:modelcomparison}

Figs \ref{fig:1D_compare} and \ref{fig:1D_compare_Rd} compare the prediction of the 1D+pseudodisc model outlined in the two previous subsections (with the pseudodisc simulation performed using a 5~au inner-boundary size) with our 3D simulation.
The comparison of disc profiles in Fig.~\ref{fig:1D_compare} shows that the evolution is indeed largely hierarchical, with the 2D pseudodisc simulation accurately predicting all properties of the pseudodisc. The 1D model for the disc also shows reasonable agreement with the 3D simulation. We note that the increased difference in the innermost region (${\lesssim}2~{\rm au}$) is likely because the disc properties are being affected by the inner boundary in the 3D simulation.
Also, the 1D model often shows some radial substructure in temperature, which is associated with a local minimum in the temperature-dependent opacity (cf.~Fig.~\ref{fig:opacity}); such a feature is not visible in 3D mainly because in that case the disc is not isothermal in the vertical direction. The comparison of disc-size evolution and angular-momentum budget in Fig.~\ref{fig:1D_compare_Rd} also shows relatively good agreement. At early times, the 1D+pseudodisc model returns a disc size slightly smaller than observed in the 3D simulation, mainly because the assumption of a constant $Q_{\rm K}=1.5$ would overestimate the column density at small radii (which in 3D is gravitationally stable for $M_{\rm tot}$ around $0.2~{\rm M}_\odot$; see Fig.~\ref{fig:Q_and_L_transport}) and reduce the disc size when the total angular momentum is fixed. But later on, both the 3D simulation and the 1D+pseudodisc model show the disc size saturating at ${\sim}30~{\rm au}$.

The 1D+pseudodisc model, which is evolved for a much longer time than is our 3D simulation, predicts the disc size to shrink eventually due to increasing magnetic braking in the pseudodisc (see Section \ref{sec:discbudget}). 
This trend is consistent with the recent survey of Class 0/I discs by \citet{Tobin2020} (see also \citealt{Segura-Cox2018} and \citealt{Andersen2019}), which suggests a slight decrease in mean disc radius over time (from 45~au in Class 0 to 37~au in Class I).
Our prediction of a gravitationally unstable disc of size ${\sim}30~{\rm au}$ is also consistent with a recent comparison between high-resolution multi-wavelength observations of a Class 0 disc and simulation-based mock observations, which suggests that the disc should be hot (at a few 100K) and gravitationally unstable \citep{Zamponi2021}. 
Our 1D disc model could facilitate similar comparisons in the future (especially for large populations of discs) and help determine whether hot, gravitationally unstable discs are in fact common amongst Class 0/I systems.

Overall, we find the 1D+pseudodisc model capable of predicting disc properties relatively accurately. The disc profile predicted from the 1D model serves as a good order-of-magnitude estimate in most cases, and the error in the prediction for disc size is no more than ${\sim}50\%$.
Given the computational difficulty of performing long-term 3D simulations at sufficiently high resolution (comparable to our 3D simulation, or at least high enough to ensure numerical convergence), the 1D+pseudodisc model might serve as a good alternative to study the long-term disc evolution during the Class 0/I phase.

\section{Comparison with previous studies}\label{sec:compare}

In our simulation, we observe the formation of a massive disc from a core with a realistic amount of initial rotation. Such behaviour is uncommon amongst similar disc-formation simulations in the literature (those with an isolated pre-stellar core as the initial condition), which often form less massive and smaller discs or fail to form any disc at all. In this section we discuss possible origins of this difference. As was mentioned in Section \ref{sec:discevolution}, the adoption of a realistic thermal model increases the disc temperature and, therefore, the column density required by gravitational self-regulation, thereby increasing the disc mass. (In \citetalias{XK21}, which adopted a barotropic equation of state, the disc mass is lower and is generally below the protostar mass.) 
In the following subsections we highlight two additional factors that could affect the outcome of disc formation (and might explain the difference between our work and previous studies) but have yet to receive much attention in the literature.

\subsection{The choice of the initial density profile}\label{sec:density_profile}

The initial density profile in our simulation corresponds to an approximately uniform sphere (with size $r_0$), plus an $n_{\rm n}\propto r^{-2}$ profile farther out that extends until $n_{\rm n}=500~{\rm cm}^{-3}$. The size and density of our `core' ($r\leq r_0$) region is chosen to match those of typical pre-stellar cores observed in NH$_3$ \citep[e.g.,][]{Jijina1999}, and thus should be representative of cores that have just begun their supercritical collapse. The `envelope' farther out mainly acts to provide the ambient pressure required for our relatively large and diffuse core to gravitationally collapse. Although this envelope is very massive, we do not count it as part of our pre-stellar core because most of the envelope is magnetically subcritical and will not gravitationally collapse.

Many disc formation simulations in the literature, however, adopt isolated initial density profiles that do not have such an envelope outside the core. Two common choices are isolated uniform spheres \citep[e.g.,][]{Li2011, Zhao2018, Hennebelle2020} and Bonnor--Ebert spheres with some density enhancement \citep[e.g.,][]{Machida2018, Machida2019, WursterBate2019, MachidaBasu2020}.\footnote{Both our density profile and the Bonner--Ebert profile consist of a nearly uniform inner region and an approximately ${\propto}r^{-2}$ outer region; their main difference is whether this outer region is considered as part of the core.}
These isolated profiles generally require significantly higher central density to gravitationally collapse (for given core mass and temperature), due to the lack of ambient pressure outside the core. In fact, simulations with these isolated initial density profiles often use initial central densities at least an order of magnitude higher than ours. 
When used together with uniform magnetic field and core rotation (which are adopted by most simulations, including ours), our more diffuse initial density profile better represents realistic pre-stellar cores than these isolated profiles, because the latter's high central densities would correspond to cores that have undergone a significant level of supercritical collapse, during which the density, magnetic field, and rotation should all become highly non-uniform.

We also comment that it is less clear how much impact the choice of initial density profiles can have on the outcome of disc formation, as the initial density profile is rarely, if ever, the only difference between any two simulations in the literature. This calls for a more systematic comparison of the evolution from different initial density profiles in future studies.

\subsection{The numerical treatment of angular momentum}

The outcome of disc formation can also be significantly affected by whether the numerical loss and diffusion of angular momentum is significant. In our simulation, we adopt a spherical-polar grid, which favorably reduces angular-momentum diffusion and eliminates angular-momentum dissipation at the grid scale; we also choose an inner boundary condition that prevents loss of angular momentum through the boundary \citepalias[see Section \ref{sec:bc} and discussion in][]{XK21}.

However, many simulations in the literature adopt Cartesian grids (with mesh refinement), which are more prone to numerical angular-momentum dissipation and diffusion, especially when the gird has low angular resolution.
While the typical angular resolutions at relatively large radii in these simulations are marginally sufficient to conserve angular momentum within an acceptable error during the pre-stellar collapse \citep{Commercon2008}, the resolution requirement for correctly capturing long-term disc evolution is likely more stringent (especially for small discs) due to the large separation between the disc's dynamical and evolutionary timescales.
More importantly, many simulations set a maximum level of mesh refinement to reduce computational cost, and this often significantly reduces the angular resolution at small radii.
Therefore, numerical angular-momentum dissipation and diffusion are likely non-negligible in most of these Cartesian-grid simulations.\footnote{One possible exception is \citet{Machida2019}, which used a large number of refinement levels to maintain relatively good angular resolution down to very small radii, and they obtained a massive, gravitationally unstable disc similar to those found in \citetalias{XK21} and in this paper. On the other hand, the high resolution allowed them to cover only the first ${\sim}2$~kyr of disc evolution after using ${\gtrsim}2$~yr of wall-clock-time.}

Excessive numerical angular-momentum dissipation and diffusion could significantly impact disc evolution and tend to inhibit the formation of massive discs with relatively high angular momentum. The numerical dissipation of angular momentum directly reduces the disc angular-momentum budget, and the numerical diffusion of angular momentum spreads the disc to reduce column density, which could increase the importance of disc magnetic braking and outflow and further suppress disc formation. (Meanwhile, in our simulation, gravitational self-regulation ensures that the disc column density is high enough to be gravitationally unstable, and the strong magnetic diffusion at high density and temperature suppresses magnetic braking and outflow.)

Overall, given the relatively low resolution of disc-formation simulations, the role of numerical angular-momentum dissipation and diffusion should receive more attention when performing and interpreting simulations.

\section{Summary and future work}\label{sec:summary}

In this paper we use a 3D non-ideal MHD simulation with ambipolar diffusion and Ohmic dissipation (Section \ref{sec:adod}) and a realistic model of radiative cooling (Section \ref{sec:radmodel}) to follow the evolution of a pre-stellar core for ${\sim}10~{\rm kyr}$ after protostar formation. We observe the formation of a massive, gravitationally unstable protostellar disc that grows to ${\sim}30~{\rm au}$ in radius (Section \ref{sec:overview}).

With the help of our simulation results, we obtain the following insights regarding the physical picture of Class 0/I disc evolution:
\begin{itemize}
\item As in \citetalias{XK21}, we find that the mass and angular-momentum budget of the disc is determined mainly by accretion through the pseudodisc (with angular momentum budget affected by magnetic braking in the pseudodisc), and that mass and angular-momentum removal from the protostar-disc system by outflow and disc magnetic braking are negligible (Section \ref{sec:discbudget}). The redistribution of angular momentum inside the disc is facilitated mainly by GI and the spiral waves it excites (Section \ref{sec:GItransport}); this causes the disc size to exceed greatly the circularization radius of pseudodisc material (Sections \ref{sec:discevolution}). We also argue that the gravitational self-regulation of accretion and the redistribution of disc angular-momentum should eventually leave most of the disc marginally unstable, with a Toomre $Q$ of order unity (Sections \ref{sec:GItransport} and \ref{sec:GIsize}).
\item The temperature profile of the disc is determined primarily by an equilibrium between heating and radiative cooling. While heating can come from a number of different sources, the total heating per radius can be well approximated (to within a factor of a few) by the accretion energy release $-g_R\dot M_{\rm tot}$ (Section \ref{sec:heatcool}). The relation between the cooling rate and disc temperature is well approximated by assuming a non-turbulent disc in thermal equilibrium (Section \ref{sec:coolT}).
The convection driven by the gravitationally driven spirals, although present, is not strong enough to significantly increase the efficiency of cooling. We provide an analytic estimate of the strength of spiral-driven convection and show that it should remain unimportant for all Class 0/I discs (Section \ref{sec:spiralconv}).
\item Inside the disc, strong non-ideal magnetic diffusion (mainly ambipolar) and a slow accretion timescale cause the magnetic field to decouple fully in the radial direction, producing a nearly uniform magnetic field inside the disc whose strength approximately matches the field strength in the pseudodisc near disc edge (Section \ref{sec:decoupling}). While the magnetic field remains relatively well coupled to the gas in the azimuthal direction, disc magnetic braking remains weak and barely affects disc evolution. We propose an analytic estimate of the disc magnetic braking rate, which is consistent with the simulation results (Section \ref{sec:discbraking}). Overall, the magnetic field is dynamically unimportant inside the disc.
\end{itemize}

Taken together, these insights lead us to a surprisingly simple, quantitative, physical picture of Class 0/I disc evolution. A 1D semi-analytic model (Section \ref{sec:1dmodel}) based on self-regulated GI, thermal balance, and magnetic decoupling predicts the time evolution of the disc size and the radial profiles of disc column density, mean disc temperature, and disc magnetic-field strength, provided inputs such as the total mass and angular momentum of the protostar-disc system.

These inputs can be obtained from a physics-rich yet inexpensive simulation of pseudodisc evolution, which uses a larger inner boundary and does not resolve the disc (Section \ref{sec:1dpseudo}). We argue that such a 1D+pseudodisc model can produce reasonable predictions because the evolution of the system is largely hierarchical: the small-scale (disc) dynamics do not greatly affect the large-scale (pseudodisc) evolution (Section \ref{sec:hierarchical}; also see section 6 of \citetalias{XK21}). We verify this idea of hierarchical evolution and show that our 1D+pseudodisc model agrees reasonably well with the 3D simulation, and the long-term evolution of disc size predicted using the 1D+pseudodisc model (which saturates at ${\sim}30~{\rm au}$ and eventually shrinks) is consistent with recent observations (Section \ref{sec:modelcomparison}).
Such a hierarchical modelling of disc formation and evolution circumvents the numerical difficulty of tracing the long-term evolution (e.g., throughout Class 0/I) with well-resolved (or at least numerically convergent) direct simulation.

In future work(s) of this series we plan to use the 1D+pseudodisc model to perform a large parameter survey to understand more systematically how the initial conditions of the pre-stellar core (including its cloud environment) and the disc chemistry and non-ideal MHD (including the Hall effect) impact disc formation and evolution. We also plan to use more 3D simulations to cover a small subset of this survey, in order to test whether the physical picture advanced in this paper and our 1D+pseudodisc model are widely applicable.

\section*{Acknowledgments}

The authors thank the expert referee for a constructive report. The simulations presented in this article were performed on computational resources managed and supported by Princeton Research Computing, a consortium of groups including the Princeton Institute for Computational Science and Engineering (PICSciE) and the Office of Information Technology’s High Performance Computing Center and Visualization Laboratory at Princeton University.

\section*{Data availability}

The data underlying this article will be shared on reasonable request to the authors.

\bibliographystyle{mnras}
\bibliography{XK21}

\newpage
\appendix

\section{Details of the radiation model}\label{A:radiation}

\subsection{Opacity tables}

\begin{figure}
    \centering
    \includegraphics[scale=0.66]{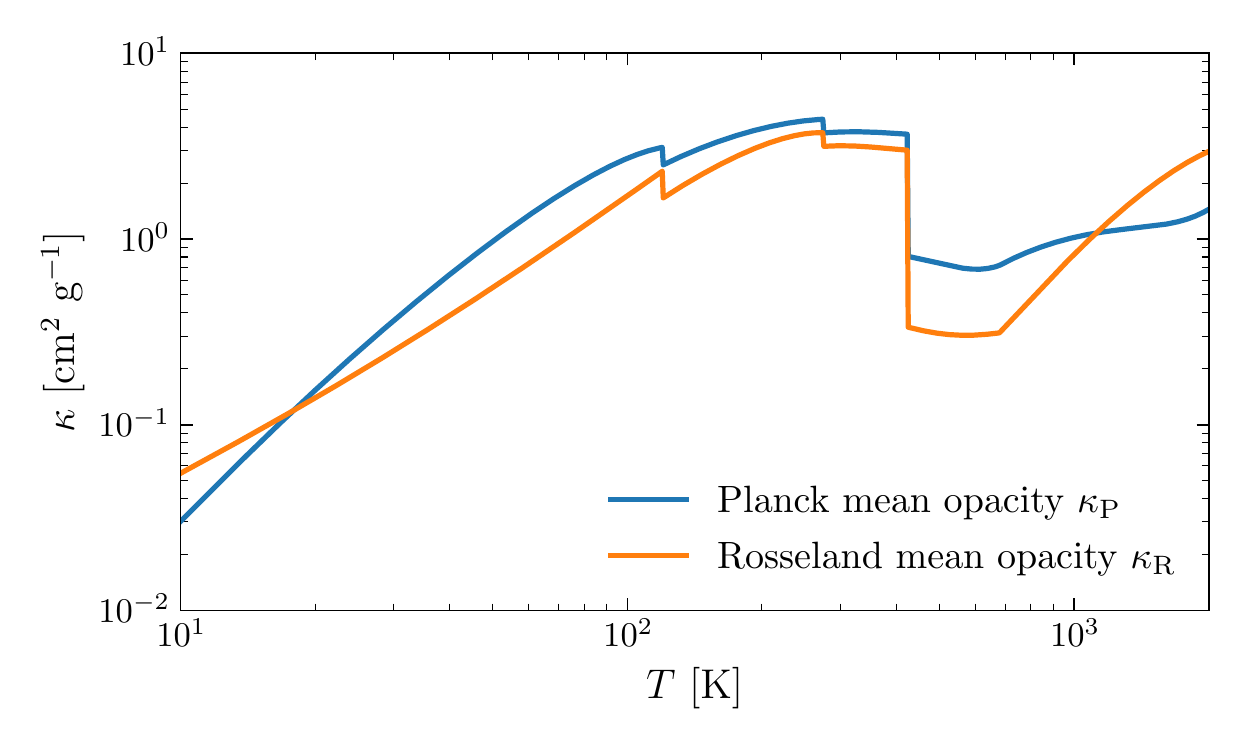}
    \caption{Opacity profiles used in our simulation.}
    \label{fig:opacity}
\end{figure}

For the range of temperatures and densities of interest here, the opacities are dominated by dust and depend only on temperature.
The $\kappa(T)$ profiles we use are given in Fig.~\ref{fig:opacity}.
These opacities are similar to those in the iron-poor model of \citet{Semenov2003} except that species abundances are chosen to better reflect conditions in molecular clouds (D.~Semenov, private communication).

\subsection{Optical depth estimate}

\begin{figure}
    \centering
    \includegraphics[scale=0.66]{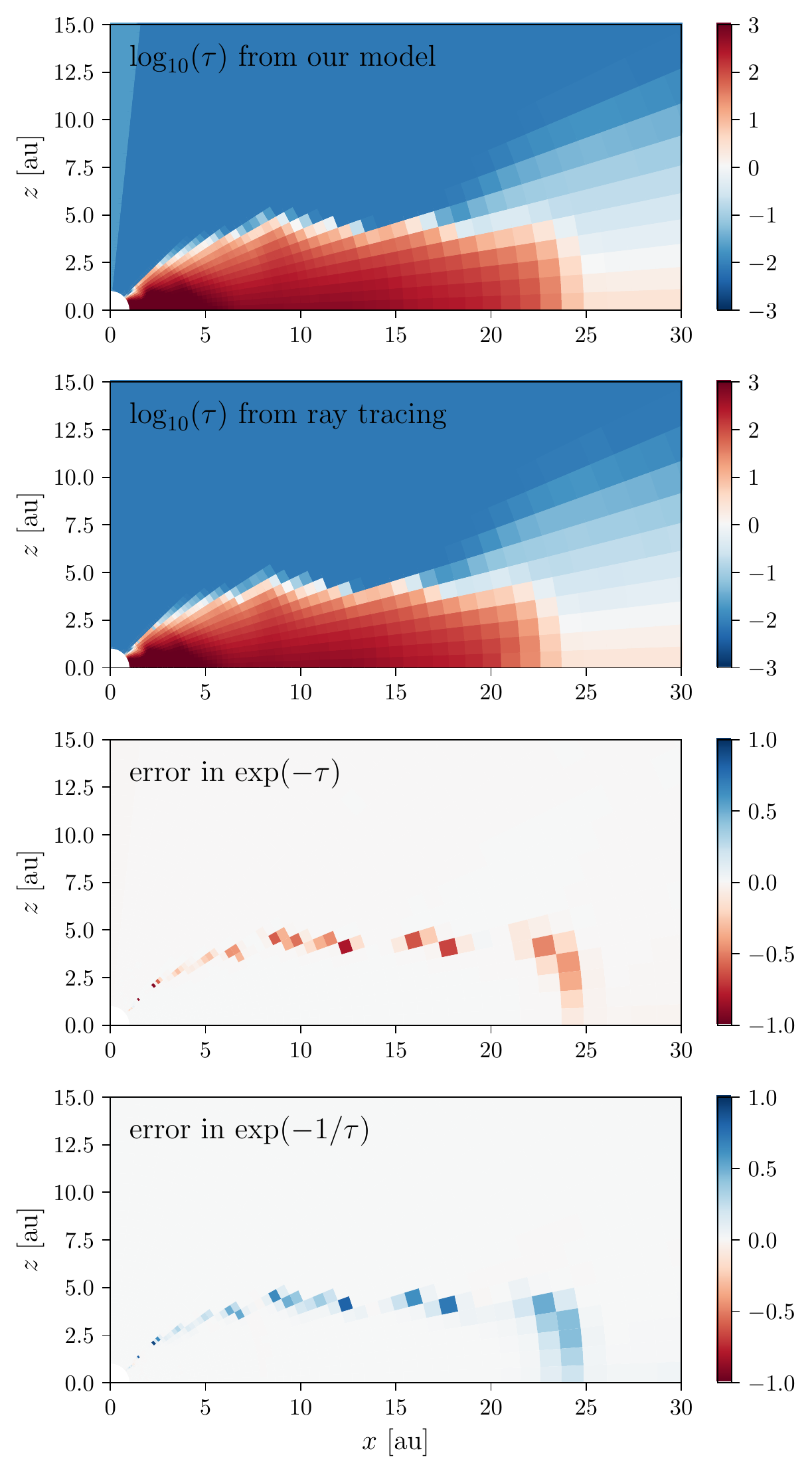}
    \caption{Comparison between optical depth $\tau$ estimated using our method and calculated via ray tracing. Here we use a simulation snapshot at $M_{\rm tot}=0.2~{\rm M}_\odot$ at $\phi=0$. The ray-tracing result shows the minimum optical depth among 160 randomly oriented rays for each cell.
    The absolute error in $\exp(-\tau)$ and $\exp(-1/\tau)$, which are used in our cooling model to scale the optically thin cooling term and the diffusion term, are also shown in the lower panels. The error remains ${\ll}1$ for most of the domain.}
    \label{fig:rt_comparison}
\end{figure}

We estimate the optical depth at a given point in the simulation domain as follows.
For a given cell, we define a collection of L-shaped rays that first go along the polar ($\theta$) direction towards the pole by $\geq 0$ cells, and then go out in the radial ($r$) direction to the outer radial boundary. We use the minimum optical depth (for Planck mean opacity) among these L-shaped rays as the optical depth estimate of the cell.

In practice, in order to find the L-shaped ray with the minimum optical depth $\tau_L$, we first calculate the optical depth at a given cell on the radial ray going from it to infinity $\tau_r$, and then perform the following iteration:
\begin{equation}
\begin{split}
    &\tau_L^{i,0} = \tau_r^{i,0},\\
    &\tau_L^{i,j} = \min\left\{\tau_r^{i,j}, \tau_L^{i,j-1}
    +\tfrac 12 r_i\Delta\theta_{j-1}\kappa_P^{i,j-1}
    +\tfrac 12 r_i\Delta\theta_j\kappa_P^{i,j}\right\}.
\end{split}
\end{equation}
The $i,j$ are indices for the $r$ and $\theta$ grid, where $j=0$ corresponds to the cell at the pole; $\Delta\theta$ denotes the grid spacing in $\theta$. The algorithm above can be efficiently parallelized, since in the first step (calculate $\tau_r$) each column of radial cells can be processed independently and in the second step (calculate $\tau_L$) each column of polar cells can be processed independently.

In Fig.~\ref{fig:rt_comparison} we compare the $\tau$ estimate using $\tau_L$ with the actual $\tau$ calculated using ray tracing with 160 randomly oriented rays from each cell. The error in $\exp(-\tau)$ and $\exp(-1/\tau)$, which are used in our cooling model to scale the optically thin cooling term and the diffusion term, are both small in most of the domain, suggesting that our estimate of $\tau$ is sufficiently accurate for the radiation model we adopt.

\subsection{Test problem: vertical temperature profile in thin disc}

\begin{figure}
    \centering
    \includegraphics[scale=0.66]{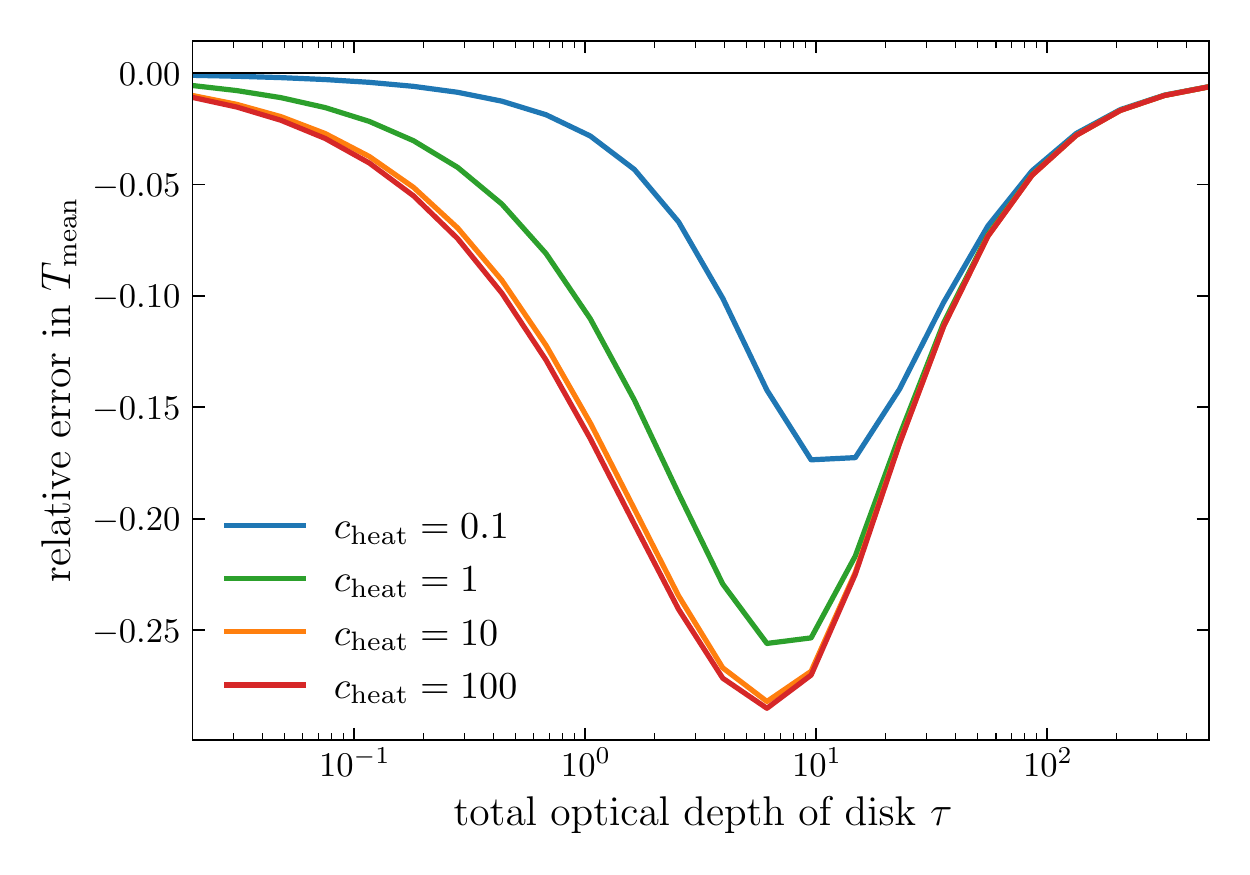}
    \caption{The error in mean temperature (when compared to full radiative transfer) for our test problem, across a wide range of  heating rates $c_{\rm heat}$ and disc optical depths $\tau$. The error remains small (${\lesssim}25\%$), and converges to zero at $\tau\ll 1$ and $\tau\gg 1$.}
    \label{fig:thin_disc_test_Tmean}
\end{figure}

\begin{figure}
    \centering
    \includegraphics[scale=0.66]{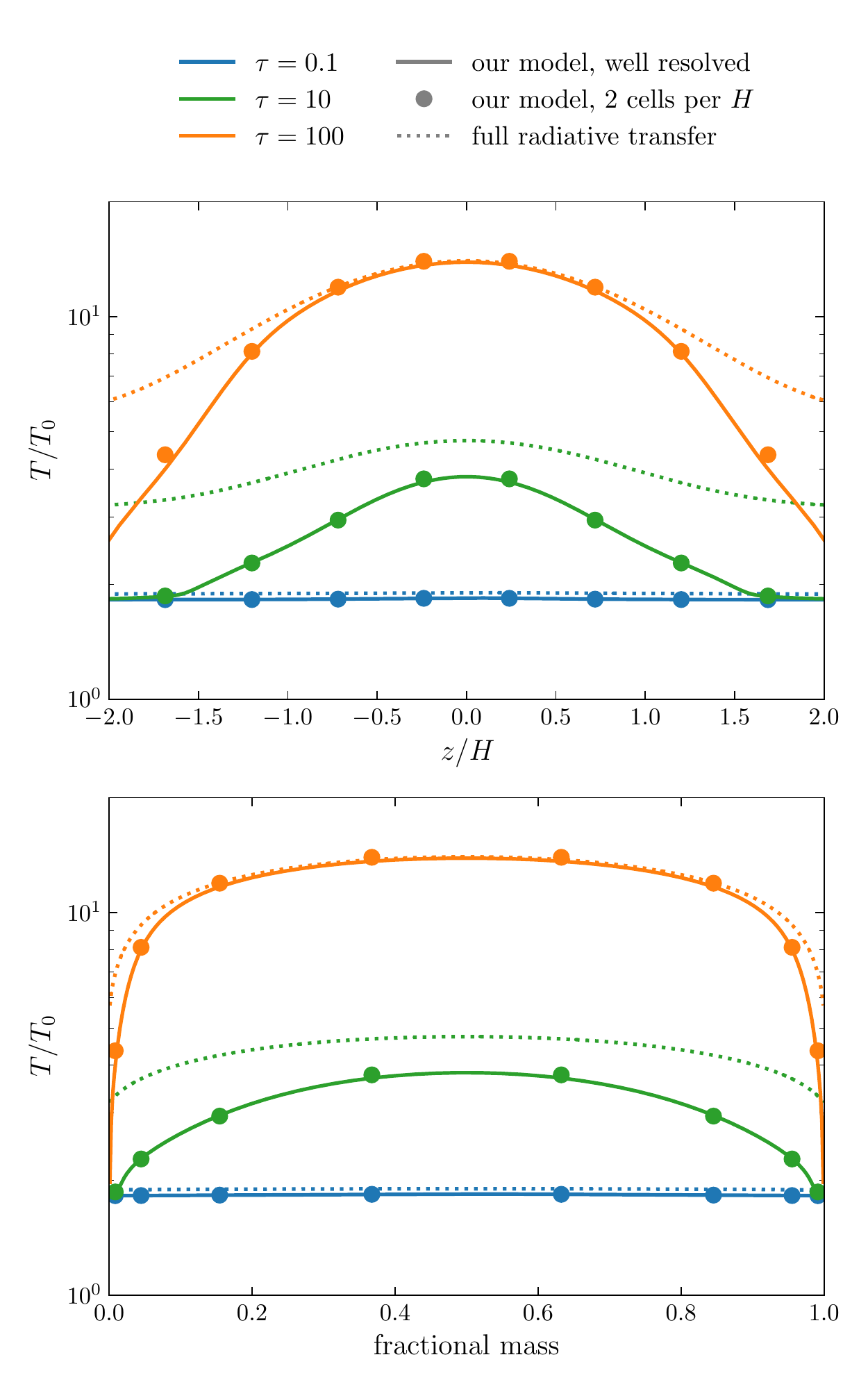}
    \caption{Top panel: Comparison of vertical disc temperature profile in thermal equilibrium for our test problem. All calculations have $c_{\rm heat}=10$, and different colours mark different disc optical depths $\tau$. Different line styles correspond to our model with high resolution (solid), our model with low resolution (2 cells per scale height $H$, circles), and the exact result from full radiative transfer (blue). Bottom panel: Same as top panel, except the horizontal axis now shows the fractional mass below a given height. For most of disc mass, our model gives small error.}
    \label{fig:thin_disc_test}
\end{figure}

To test the accuracy of our radiation model, we use the following test problem. Consider the equilibrium temperature profile of a heated,  geometrically thin disc, in which the typical length scale in the disc plane is effectively infinity when compared to the disc thickness, making this becomes a 1D problem in the vertical direction.
Let the disc have a density profile $\rho = \rho_0\exp(-z^2/H^2)$ and a constant, frequency-independent $\kappa$. We solve for the temperature profile and the mean temperature when the whole disc is in thermal equilibrium with a uniform heating per mass $Q/\rho$. We characterize the heating rate by a dimensionless parameter
\begin{equation}
    c_{\rm heat} \equiv Q/(4 \rho \kappa \sigma T_0^4),
\end{equation}
where $T_0$ is the ambient temperature at infinity.

Given that this problem is 1D, it can be solved semi-analytically with full radiative transfer without difficulty. In Figs \ref{fig:thin_disc_test_Tmean} and \ref{fig:thin_disc_test}, we compare the result of full radiative transfer with our radiation model. Fig.~\ref{fig:thin_disc_test_Tmean} shows the error in the mean disc temperature predicted by our model for a wide range of different heating rates and disc optical depths. We see that the error converges to zero at $\tau\ll 1$ and $\tau\gg 1$, and is maximal at $\tau\sim 10$. The error increases for increasing $c_{\rm heat}$, but saturates for $c_{\rm heat}\gtrsim 10$ (which can be considered as the limit of ambient temperature $T_0\to 0$). The error in mean temperature is always ${\lesssim}25\%$, suggesting that our model can give quite accurate predictions for the mean disc temperature.

Fig.~\ref{fig:thin_disc_test} provides a more detailed comparison of the vertical temperature profile at a given $c_{\rm heat}=10$ and a few different $\tau$. Note that the $\tau=10$ (green) curves are in the regime where the error of our model is maximal (see Fig.~\ref{fig:thin_disc_test_Tmean}). The temperature profile our model produces matches well with the full radiative transfer result near the midplane (which contains most of the mass), especially when $\tau$ is ${\gg}1$ or ${\ll}1$. The error in the atmosphere is often finite, and our model tends to underestimate the temperature there. This is mainly because our model ignores the re-absorption of the cooling radiation from the midplane by the optically thin atmosphere.
However, we do not expect this finite error to have a large impact on our simulation result. 
The temperature in the atmosphere affects the dynamics through pressure and chemistry (which eventually affects the non-ideal diffusivity). The low density in the atmosphere makes the pressure dynamically unimportant, even if the temperature is as high as the disc. Meanwhile, the non-ideal diffusivity there is always high (in fact usually above its numerical cap; see Appendix \ref{A:numerical}), and some order-of-unity modification to this already high diffusivity should barely affect the evolution.

An additional advantage of our radiation model is that it performs very well at low resolution. As shown in Fig.~\ref{fig:thin_disc_test}, at resolution as low as 2 cells per scale height (which is already lower than the typical resolution seen in our simulation), the prediction of the model still matches the well-resolved result very well.

\section{Other Details of numerical setup}\label{A:numerical}

\subsection{Polar averaging}

One major disadvantage of using spherical-polar coordinates for 3D simulations is that the cells become narrow wedges towards the pole, which may limit the timestep severely. To circumvent this problem, we `de-refine' cells near the pole in the $\phi$ direction through a polar-averaging technique that was discussed in \citetalias{XK21}.
In short, our polar-averaging technique smooths out fluctuations with large spatial frequencies (in $\phi$) near the pole to allow the adoption of a larger timestep.

The polar-averaging scheme we used in \citetalias{XK21} involves performing an FFT in the $\phi$ direction and truncating the high-frequency components, which is equivalent to convolving with a filter that takes the shape of a wavelet. However, this method is problematic when applied to non-barotropic simulations.
When using this scheme on a non-barotropic system, we are averaging total energy, density, momenta, and magnetic field, while internal energy is implicitly inferred from them. Because the filter for the convolution is not always non-negative, it is possible to obtain negative internal energy after the average. Since the code will force the pressure to be non-negative, this results in strong unphysical heating and can break the simulation.

The above problem can be resolved if we use a filter that is always non-negative. (One can mathematically prove that this guarantees the resulting internal energy to be positive.)
Therefore, for simulations in this paper we adopt a triangular filter instead. For example, if we are averaging a variable $y$ over $k_0$ cells, we have
\begin{equation}
    y_k{\rm~after~averaging} = \sum_{s=-k0}^{k0} y_{k+s} \frac{k_0-|s|}{k_0^2}.
\end{equation}
One caveat is that this filter damps all modes with nonzero (spatial) frequency. Since the filter is applied at every timestep, any non-axisymmetric perturbation will be strongly damped in the polar-averaging region, which covers the first two cells around the pole.
This is not too problematic for our simulation since the polar region is expected to be near axisymmetry. However, if in the future we want to simulate a system with a non-axisymmetric polar region (e.g., with misalignment between the magnetic and rotation axes), we will need to improve this treatment.

\subsection{Diffusivity caps}

In \citetalias{XK21} we capped the non-ideal magnetic diffusivities to avoid excessively small timesteps. Since the cell size increases at larger $r$, this cap is ${\propto}r^2$ and affects only the innermost region of our domain. In this paper we introduce another diffusivity cap (in addition to the cap used in \citetalias{XK21}) to tackle a different numerical problem.

Consider a region where the total energy is dominated by the magnetic energy and the magnetic diffusivity is very large. The low kinetic energy and pressure requires the magnetic field to be nearly force-free, i.e., $\bb{B}\btimes(\grad\btimes\bb{B}) \approx 0$. Meanwhile, large magnetic diffusivity also requires $\bb{B}\btimes(\grad\btimes\bb{B}) \approx 0$,
since otherwise that would create a large drift velocity (and therefore large $\partial_t \bb{B}$; see Section \ref{sec:decoupling}).
However, in \verb'Athena++', the interpolations in magnetic field used for calculating the Lorentz force and the non-ideal EMF are not exactly self-consistent; i.e., the small numerical interpolation errors in $\bb{B}\btimes(\grad\btimes\bb{B})$ are generally going to be different between the Lorentz force and non-ideal drift velocity. When $\bb{B}\btimes(\grad\btimes\bb{B})$ physically needs to be very small (below the typical interpolation error), this could lead to significant unphysical perturbations in the flow.

To avoid this problem, we cap the non-ideal diffusivities to $10 \sqrt{-g_r r^3}$, which corresponds to the diffusion timescale being 10$\times$ the local dynamical timescale.
This choice also ensures that the diffusivity is still strong enough to force the field to be nearly force-free in strongly diffusive regions.
In practice, this cap only affects part of the outflow cone, and increasing the prefactor of 10 barely affects the result.

\subsection{Cosmic-ray shielding}

When the column density is sufficiently high (${\gtrsim}10^2~{\rm g}~{\rm cm}^{-2}$), cosmic rays will be shielded appreciably and ionization will be reduced \citep{UmebayashiNakano1980}. In our simulation, a significant fraction of the inner disc has column densities above this threshold.
However, the chemistry model we adopt does not account for cosmic-ray shielding, and assumes constant $\zeta_{\rm CR}$ for the whole domain.
Therefore, our simulation underestimates (perhaps quite significantly) the level of magnetic diffusion inside the disc.

However, we argue that further increasing magnetic diffusion inside the disc should barely affect the evolution. In the radial direction, the magnetic field is already fully decoupled from the gas in the disc, and so further increasing magnetic diffusivity will not affect the evolution. In the azimuthal direction, increasing the magnetic diffusivity can decrease the amount of magnetic braking. Yet the impact of such a decrease on the disc angular-momentum budget should be no more than $10\%$ (cf.~Fig.~\ref{fig:ML_flux}), since disc  braking is already weak in our simulation.
We note in passing that the change in $B_\phi$ may also affect a magnetically launched outflow, which again is already negligible in our simulation (in terms of its contribution to disc mass and angular-momentum budgets) and is not expected to be amplified by increased diffusion.

\section{Numerical convergence}\label{A:convergence}

\begin{figure}
    \centering
    \includegraphics[scale=0.66]{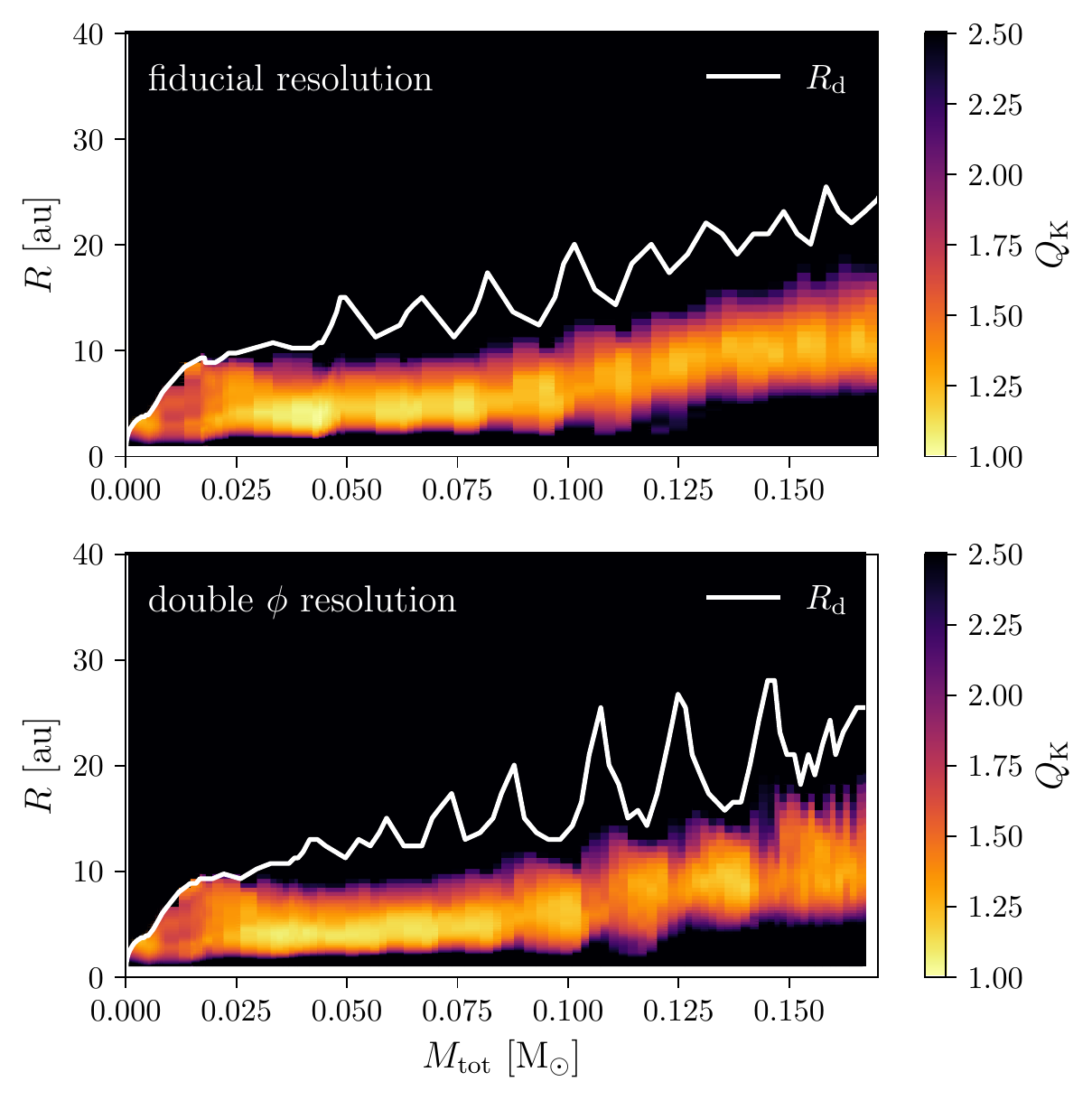}
    \caption{$Q_{\rm K}$ and disc-size evolution for the fiducial run and the run with doubled azimuthal resolution, similar to the bottom of Fig.~\ref{fig:Q_and_L_transport}. The two runs show largely similar evolution, with the double-azimuthal-resolution run exhibiting more oscillations in disc size.}
    \label{fig:phi_res_Q}
\end{figure}

\begin{figure}
    \centering
    \includegraphics[scale=0.66]{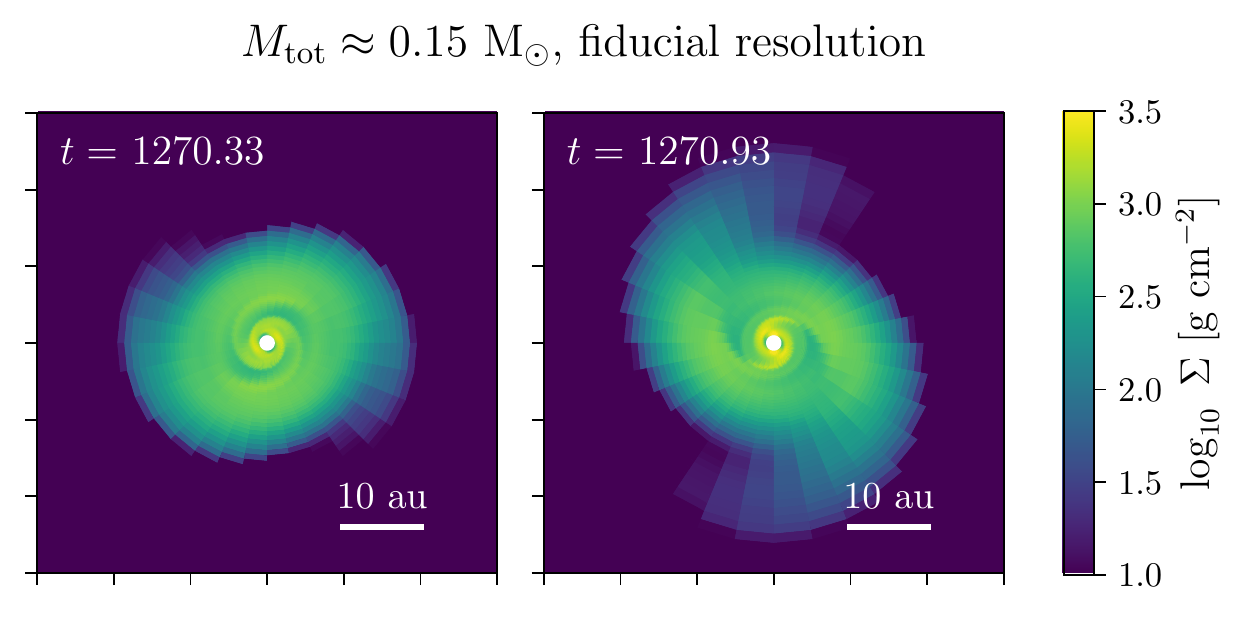}\\
    \includegraphics[scale=0.66]{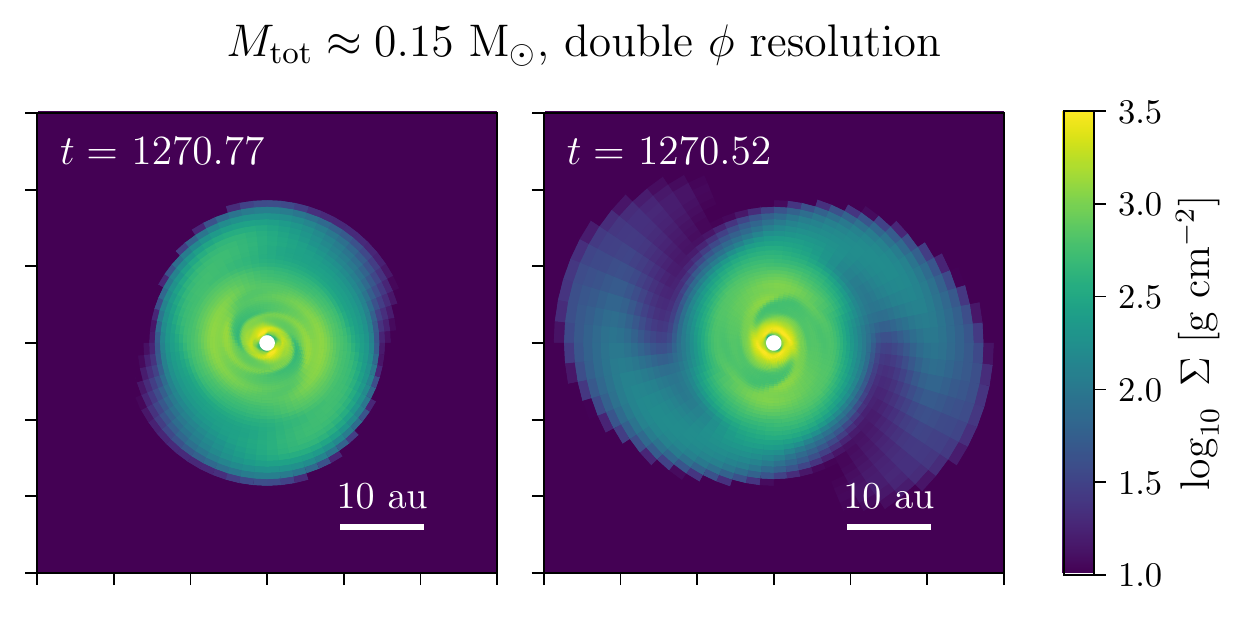}
    \caption{Midplane density snapshots similar to Fig.~\ref{fig:midplane} for fiducial (top) and double-azimuthal-resolution (bottom) runs around the same $M_{\rm tot}$. The discs are overall similar.}
    \label{fig:midplane_res}
\end{figure}

\begin{figure}
    \centering
    \includegraphics[scale=0.66]{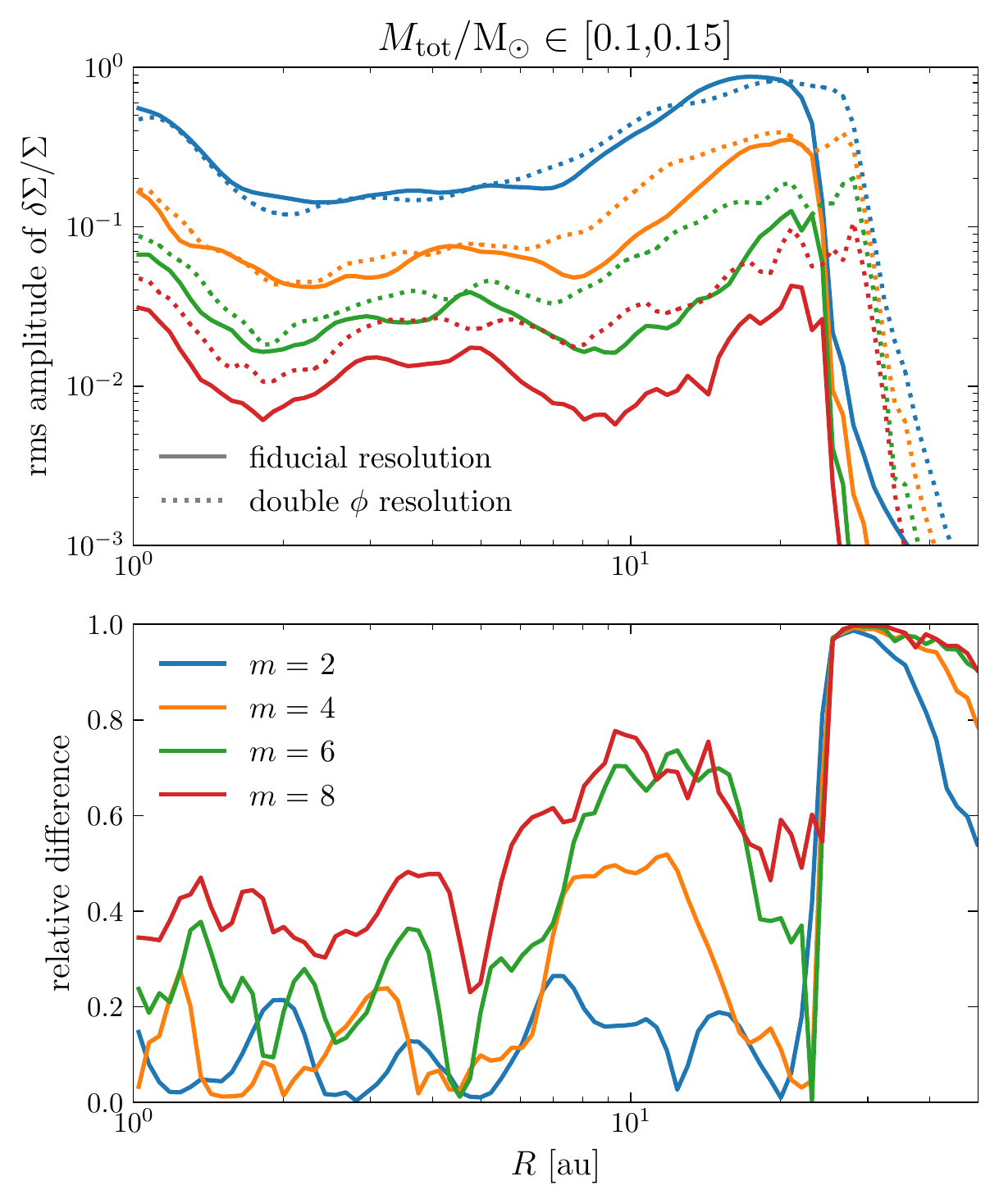}
    \caption{RMS relative amplitude of column-density perturbations $\delta\Sigma/\Sigma$ with different azimuthal wavenumbers $m$ (top panel) for the fiducial (solid) and double-azimuthal-resolution (dotted) runs. The relative difference (absolute value) is shown in the bottom panel for reference. The fiducial run accurately predicts the amplitude of the dominant $m=2$ mode, but the error increases for higher $m$.}
    \label{fig:phi_res_spiral}
\end{figure}

In \citetalias{XK21} we demonstrated that, at a resolution similar to that used in this paper, the results (mainly the angular-momentum budget and disc morphology) remain similar when we increase the radial ($r$) or polar ($\theta$) resolution or decrease the inner-boundary size $r_{\rm in}$.
Here, we test whether our azimuthal ($\phi$) resolution is also sufficient by re-running our simulation (for a shorter duration) with double the azimuthal resolution (32 cells in $\phi\in[0,\pi]$) and comparing the results with our fiducial simulation.

The double-azimuthal-resolution run covers the evolution up to $M_{\rm tot}\approx 0.17~{\rm M}_\odot$. We find the two runs to be very similar in the azimuthally averaged diagnostics shown in the main text, such as the 1D profiles in Fig.~\ref{fig:disc_1d}.
To demonstrate the overall similarity and highlight the minor differences, we show a comparison of the $Q_{\rm K}$ profile and disc size evolution in Fig.~\ref{fig:phi_res_Q}.
The evolution of both $Q_{\rm K}$ and disc size are overall similar, but the double-azimuthal-resolution run shows larger oscillations in disc size.
This is likely because the resulting decrease in numerical dissipation makes the spiral waves decay slower at the outer edge of the disc.
We also compare column density snapshots from the two runs (Fig.~\ref{fig:midplane_res}), which are again largely similar, except that the double-azimuthal-resolution run shows more small-scale fluctuations and spirals reaching slightly larger radii due to decreased numerical dissipation.

A more detailed comparison regarding the properties of the gravitationally-excited spiral waves is given in Fig.~\ref{fig:phi_res_spiral}, in which we compare the amplitudes of column-density perturbations with different azimuthal wavenumbers $m$. (Here we only have even $m$ because we only simulate $\phi\in[0,\pi]$.)
The two runs show relatively good agreement for the dominant $m=2$ perturbation, with a relative error ${\lesssim}20\%$ in the disc.
At higher $m$, the error increases, as the corresponding length scale gets closer to grid scale. Still, the fiducial resolution run agrees qualitatively with the double-azimuthal-resolution run for all modes shown in the figure.
We also note that the double-azimuthal-resolution run shows much higher rms amplitude at the edge of the disc and farther out, likely because the (decaying) spirals at disc edge can reach larger radii at higher resolution (which is visible in the disc-size evolution in Fig.~\ref{fig:phi_res_Q}).

In summary, the evolution remains overall similar when we increase the azimuthal resolution, with the only differences being the oscillation of the disc edge (controlled by the dispersal of spirals) and the amplitude of small-scale (high-$m$) perturbations. These differences should not impact the main results of this paper.

\section{Detailed estimate of disc thermal budget}\label{A:thermal_budget}

In this appendix we analyze the thermal budget of an accreting disc (subject to no external irradiation) to show that the cooling rate required for thermal equilibrium is generally of order $-g_R\dot M$. 
This result does not require very specific assumptions on how energy and angular momentum are transported in the disc, and is still applicable when the disc thickness and disc-to-star mass ratio are finite.

\subsection{Definitions}

For our problem, we can identify three independent dimensionless small parameters:
\begin{align}
    \epsilon_1 &\equiv \frac{\dot M}{\pi R^2 \Sigma \Omega},\\
    \epsilon_2 &\equiv \left(H/R\right)^2,\\
    \epsilon_3 &\equiv \frac{\pi R^2\Sigma}{M_{\rm tot}}.
\end{align}
Here $\epsilon_1$ corresponds to the rate of column density evolution (in units of $\Omega$), $\epsilon_2$ is the square of disc aspect ratio (and is comparable to the ratio between thermal and kinetic energy), and $\epsilon_3$ is comparable to the disc-to-star mass ratio. Thin-disc models typically simplify the problem by assuming $\epsilon_2,\epsilon_3\to 0$; such an assumption is not necessarily valid in young protostellar discs.
To ensure the generality of our results, we do not assume any specific ordering between these parameters, except that $\epsilon_1,\epsilon_2\ll 1$, $\epsilon_3\lesssim 1$, and $\epsilon_3\lesssim \epsilon_2^{1/2}$. The last relation comes from the assumption that gravitational self-regulation should be able to maintain $Q\gtrsim 1$.
For brevity we use the normalization $\Sigma, R, \Omega\sim 1$ when labeling the size of each term.

We expect each variable to be the sum of a slowly varying axisymmetric component, corresponding to the mean disc profile whose characteristic rate of evolution is ${\lesssim}\epsilon_1\Omega$, and a rapidly varying (and generally non-axisymmetric) component, corresponding to turbulent fluctuations whose characteristic rate of variation is ${\sim}\Omega$. For any variable $a$, we separate these two components by dividing it into a mean component $\bar a$ and a fluctuating component $a'$ as follows:
\begin{equation}
    \bar a(t,R,z) \equiv \mc{F}(\langle a\rangle_\phi),\quad a' \equiv a-\bar a,
\end{equation}
where $\mc{F}$ is a low-pass filter (in time) with a cutoff frequency between $\Omega$ and $\epsilon_1\Omega$, and $\langle\,\dots\,\rangle_\phi$ denotes an average in $\phi$.
We assume that the mean and fluctuating components are well separated in frequency space. More specifically, we assume that there exists a filter $\mc{F}$ such that (i) $\bar a$ and its time derivative are dominated by contribution from frequencies ${\lesssim}\epsilon_1\Omega$, and $a'$ and its time derivative are dominated by contributions from frequencies ${\gtrsim}\Omega$; and (ii) the fluctuating component of one variable $a'$ is uncorrelated with the mean component of any other variable $\bar b$, or some function of the mean component of multiple variables (e.g., $\bar b\bar c,\bar b/\bar c$).

For variables that are approximately constant inside the disc in vertical direction at a given radius, such as velocity and gravitational potential, we can further split the mean component $\bar a$ into a $z$-independent piece and a $z$-dependent piece,
\begin{equation}
    a_0(t,R) \equiv \langle\overline{\rho a}\rangle_z/\langle\bar\rho\rangle_z, \quad a_1\equiv \bar a-a_0.
\end{equation}
Here $\langle\,\dots\,\rangle_z \equiv \int_{-\infty}^\infty {\rm d}z\, (\,\dots\,)$. Our earlier assumption that the mean and fluctuating components are well-separated in frequency implies that $a_0,a_1$ are both uncorrelated with any fluctuating variable $b'$.

\subsection{Basic equations}

The evolution of mass, momentum, and energy in the disc can be described by the following equations:
\begin{align}
    &\partial_t\rho + \grad\bcdot(\rho\bb{v}) = 0,\label{eq:hydro_before_avg_first}\\
    &\partial_t(\rho\bb{v}) + \grad\bcdot(\rho\bb{v}\bb{v}) = \rho \bb{g},\\
    &\partial_t(\rho u + \tfrac 12\rho v^2) + \Phi\partial_t\rho + \grad\bcdot [(\rho u + \tfrac 12 \rho v^2+\rho\Phi + p)\bb{v}] = Q_{\rm rad}.\label{eq:hydro_before_avg_last}
\end{align}
where $\Phi$ includes contributions from both the protostar and the disc. To get the last equation we have used that $\rho \bb{v}\bcdot \bb{g} = -\grad\bcdot(\rho \bb{v}\Phi) + \Phi\grad\bcdot(\rho\bb{v}) = -\grad\bcdot(\rho \bb{v}\Phi) - \Phi\partial_t\rho$. These equations ignore the effect of the magnetic field for simplicity; this is a reasonable approximation for our simulation, since the magnetic field in the disc is very weak and mostly straight due to magnetic diffusion. Meanwhile, for discs in which the magnetic field affects the transport of angular momentum and energy, one can show that the heating associated with these processes is at most ${\sim}-g_R\dot M$ per radius \citep[e.g.,][]{bgh94}.

Next, we use the following operator to average the equations:
\begin{equation}\label{eqn:avgoperator}
    \langle \,\dots\, \rangle \equiv \frac{1}{\Delta t}\int_{t-\Delta t/2}^{t+\Delta t/2}{\rm d}t'\int_{-\infty}^{\infty}{\rm d}z\int_0^{2\pi}R{\rm d}\phi \, (\,\dots\,) .
\end{equation}
This averaging corresponds to integrating over the cylindrical surface at $R$ and averaging over a time window $\Delta t$, which we choose to be ${\lesssim}(\epsilon_1\Omega)^{-1}$ but ${\gg}\Omega^{-1}$. This averages out turbulent perturbations while retaining the long term evolution of the mean disc. The operator \eqref{eqn:avgoperator} has the following properties:
\begin{align*}
    &\langle \partial_t \,\dots\,\rangle = \partial_t\langle\,\dots\,\rangle,\\
    &\langle \grad\bcdot \bb{A}\rangle = \partial_R\langle A_R\rangle,\\
    &\langle \grad\bcdot \msb{T}\rangle \bcdot \hat{\bb{\phi}} = R^{-1}\partial_R\langle R \msf{T}_{R\phi}\rangle.
\end{align*}
Here $\bb{A}$ is any vector and $\msb{T}$ is any tensor; $\hat{\bb{\phi}}$ is the unit vector in $\phi$.
Given our earlier assumption that the mean and turbulent components are uncorrelated, we also have that
\begin{equation}
    \langle a\rangle = \langle\bar a\rangle,~~~\langle ab\rangle = \langle\bar a \bar b + a'b'\rangle,~~~\langle\rho a\rangle = \langle\overline{\rho a}\rangle = \langle\bar\rho a_0\rangle.
\end{equation}

After applying $\langle\,\dots\,\rangle$ to equations \eqref{eq:hydro_before_avg_first}--\eqref{eq:hydro_before_avg_last} (where we only take the $\phi$ component of the momentum equation), we obtain
\begin{align}
    &\partial_t\langle\bar\rho\rangle + \partial_R\langle \bar\rho v_{R,0}\rangle = 0, \label{eqn:avgmass}\\
    \begin{split}
    &\partial_t\langle R\bar\rho v_{\phi,0}\rangle + \partial_R\langle R\bar\rho v_{R,0}v_{\phi,0} + R\overline{\rho v_R}v_{\phi,1} \\
    &\quad \mbox{} + R(\rho v_R)'v_\phi' + (4\pi{\rm G})^{-1}Rg_R'g_\phi'\rangle=0,
    \end{split}\label{eqn:avgAM}\\
    \begin{split}
    &\partial_t\langle\bar\rho (u+\tfrac 12 v^2)_0\rangle + \langle\Phi_0\partial_t\bar\rho + \Phi_1\partial_t\bar\rho + \Phi'\partial_t\rho'\rangle\\
    &\quad\mbox{}+ \partial_R\langle \bar\rho v_{R,0}e_{\rm tot,0}+\overline{\rho v_R}e_{\rm tot,1}+(\rho v_R)'e_{\rm tot}'\rangle = -2\pi R\Lambda_{\rm cool}.\label{eqn:avgE}
    \end{split}
\end{align}
Here $\Lambda_{\rm cool} \equiv \langle -Q_{\rm rad} \rangle/(2\pi R)$ and can be interpreted (approximately) as the radiative cooling rate per unit area, since the transport by radiative diffusion in the radial direction is much weaker when the disc is geometrically thin.
We can use the mass equation \eqref{eqn:avgmass} to eliminate a few terms from the angular-momentum and energy equations, \eqref{eqn:avgAM} and \eqref{eqn:avgE}, and rewrite them as
\begin{align}
\begin{split}
    &\langle R\bar\rho\partial_t v_{\phi,0}\rangle + \langle\bar\rho v_{R,0}\partial_R(Rv_{\phi,0})\rangle \\
    &\quad\mbox{}+ \partial_R\langle R\overline{\rho v_R}v_{\phi,1} + R(\rho v_R)'v_\phi' + (4\pi {\rm G})^{-1}Rg_R'g_\phi'\rangle = 0,
    \label{eq:angular_momentum_terms}
\end{split}\\
\begin{split}
    &\langle\bar\rho v_{R,0}\partial_R e_{\rm tot,0}\rangle + \langle\bar\rho\partial_t(u+\tfrac 12 v^2)_0\rangle + \langle \Phi_1\partial_t\bar\rho + \Phi'\partial_t\rho'\rangle\\
    &\quad\mbox{}+ \partial_R\langle \bar\rho v_{R,0}(p/\rho)_0+\overline{\rho v_R}e_{\rm tot,1}+(\rho v_R)'e_{\rm tot}'\rangle = -2\pi R\Lambda_{\rm cool}.
    \label{eq:energy_terms}
\end{split}
\end{align}
Because $\dot M = -\langle\rho v_R\rangle = -\langle\bar\rho v_{R,0}\rangle$, the first term in the energy equation \eqref{eq:energy_terms} is $\mc{O}(\epsilon_1)$ and corresponds to the accretion heating we discussed in Section \ref{sec:heatcool}. In the discussion below, we will show that the remaining terms on the left-hand side of equation \eqref{eq:energy_terms} are all ${\lesssim}\mathcal O(\epsilon_1)$ (or, in dimensional form, ${\lesssim}-g_R\dot M$).

\subsection{Perturbation amplitude estimates}

In order to know the amplitude of each term in the energy equation, we need to first estimate the amplitudes of the vertical variation in the mean disc profile $a_1$ and the turbulent fluctuations $a'$.

We start by estimating the vertical variation in the mean disc profile, $\Phi_1, u_1, (p/\rho)_1$ and $v_{\phi,1}$. $\Phi_1$ can be divided into contributions from the central protostar and from disc self-gravity; the former is of order $(H/R)^2\Phi_0 \sim \epsilon_2$, while the latter is of order $H{\rm G}\Sigma \sim \epsilon_2^{1/2}\epsilon_3$ (here ${\rm G}\sim\epsilon_3$ under our normalization because ${\rm G}M_{\rm tot}\sim 1$ and $M_{\rm tot}\sim \epsilon_3^{-1}$), which is ${\lesssim} \epsilon_2$ since $\epsilon_3\lesssim \epsilon_2^{1/2}$.
The relative vertical variation of $u$ and $p/\rho$ can both be order unity, so $u_1,(p/\rho)_1 \sim \bar u,\overline{(p/\rho)} \sim \epsilon_2$.
Finally, for $v_{\phi,1}$, because it leads to a vertical variation of centrifugal acceleration of ${\approx}2v_{\phi,0}v_{\phi,1}/R$ that must be canceled by the vertical variation of gravity and pressure gradient (both are ${\sim}\epsilon_2$) , we expect $v_{\phi,1}\sim \epsilon_2$. In summary,
\begin{equation}
    \Phi_1, u_1, (p/\rho)_1, v_{\phi,1} \sim \epsilon_2.
\end{equation}

Now we estimate $\overline{\rho v_R}$ and $\boldsymbol v'$ using the angular momentum equation \eqref{eq:angular_momentum_terms}.
Consider its first term. Since $v_{\phi,0}$ evolves at a timescale of order $M_{\rm tot}/\dot M$, we expect
\begin{equation}
    \langle R\bar\rho\rangle\partial_t v_{\phi,0} \sim \dot M/M_{\rm tot} \sim \epsilon_1\epsilon_3.
\end{equation}
Then, for the second term, since $\dot M = -\langle\bar\rho v_{R,0}\rangle$ suggests that $v_{R,0}\sim \epsilon_1$, we have
\begin{equation}
    \langle\bar\rho v_{R,0}\partial_R (Rv_{\phi,0})\rangle \sim \epsilon_1.
\end{equation}
Given the amplitude of the first two terms, each of the three terms in $\partial_R\langle\,\dots\,\rangle$ has to be ${\lesssim}\epsilon_1$ (assuming that they do not exactly cancel out). Therefore,
\begin{equation}
    (\overline{\rho v_R}/\bar\rho) v_{\phi,1}\lesssim\epsilon_1,\quad [(\rho v_R)'/\bar\rho] v_{\phi}' \lesssim \epsilon_1.
    \label{eq:vv}
\end{equation}
Because $v_{\phi,1}\sim \epsilon_2$ (as shown earlier), the first part of equation \eqref{eq:vv} gives
\begin{equation}
    \overline{\rho v_R}/\bar\rho \lesssim \min(1,\epsilon_1/\epsilon_2).
\end{equation}
Here we cap it to $\mc{O}(1)$ because the radial velocity should remain smaller than the rotational velocity. Now consider $(\rho v_R)'$ (which is ${{\gtrsim}\bar\rho v_R'}$, assuming that the fluctuations in $\rho$ and $v_R$ do not exactly cancel) and $v_{\phi}'$, which corresponds to turbulent fluctuations. Assuming that these turbulent fluctuations are relatively well-correlated (so that $\langle(\rho v_{R})' v_{\phi}'\rangle$ and $\langle|(\rho v_{R})'| |v_{\phi}'|\rangle$ are of the same order of magnitude) and $v_{R}'\sim v_{\phi}' \gtrsim v_{z}'$, the constraint in equation  \eqref{eq:vv} gives
\begin{equation}
    |\boldsymbol v'| \lesssim \epsilon_1^{1/2}.
\end{equation}
This also allows us to estimate $e_{\rm tot, 1}$:
\begin{equation}
    e_{\rm tot, 1}\approx u_1+(p/\rho)_1+\Phi_1+v_{\phi,0}v_{\phi,1}+\tfrac 12 (\boldsymbol v'\cdot\boldsymbol v')_1 \lesssim \max(\epsilon_1,\epsilon_2).
\end{equation}

Next we consider the amplitudes of $\rho'$ and $\Phi'$. We can assume that the perturbations have characteristic length scale ${\sim}H$ and characteristic frequency ${\sim}\Omega$. (This is true for both spiral waves and local turbulence; in the former case the characteristic scale is ${\sim}H$ in $R,z$ but can be ${\sim}R$ in $\phi$.) Considering the fluctuating component of the continuity equation gives
\begin{equation}
    \rho'/\bar\rho \sim \Omega^{-1}H^{-1}v' \lesssim \epsilon_1^{1/2}\epsilon_2^{-1/2}.
\end{equation}
Under the normalization we use ($\Sigma,r,\Omega \sim 1$), $\bar\rho$ is of order $\Sigma/H \sim \epsilon_2^{-1/2}$ at midplane, and the gravitational constant ${\rm G}$ is $\mathcal O(\epsilon_3)$ because $M_{\rm tot}$ is $\mathcal O(\epsilon_3^{-1})$ and ${\rm G}M_{\rm tot}$ is $\mathcal O(1)$. The Poisson equation then gives
\begin{equation}
    H^{-2}\Phi' \sim {\rm G}\rho',~~~\Phi' \lesssim \epsilon_1^{1/2}\epsilon_3.
\end{equation}

Then we consider the amplitude of $(p/\rho)'$ and $u'$; these two perturbations should have comparable amplitudes. They can come from two sources: compression, which gives $u'/\bar u,(p/\rho)'/\overline{(p/\rho)} \sim \rho'/\bar\rho$, and non-uniform heating, which at most gives $u', (p/\rho)' \sim \delta \Lambda_{\rm heat} \Omega^{-1}\Sigma^{-1}$ (since the heating rate varies at a characteristic rate of at least $\Omega$). Here $\delta \Lambda_{\rm heat}$ is the amplitude of variation in heating rate, and should satisfy $\delta \Lambda_{\rm heat}\lesssim \Lambda_{\rm heat} \approx \Lambda_{\rm cool} \sim \epsilon_1$.\footnote{Strictly speaking this is a circular argument, since we cannot assume that heating is $\mathcal O(\epsilon_1)$ yet. More rigorously, one could argue that even when $u', (p/\rho)' \sim \Lambda_{\rm heat} \Omega^{-1}\Sigma^{-1}$, terms with $u', (p/\rho)'$ should remain small compared to the cooling term.} Therefore,
\begin{equation}
    u', (p/\rho)' \lesssim \max(\epsilon_1, \epsilon_1^{1/2}\epsilon_2^{1/2}).
\end{equation}

Finally, for $e_{\rm tot}'$ we have
\begin{equation}
    e_{\rm tot}'\approx u' + (p/\rho)' + \Phi' + \bar {\boldsymbol v} \cdot \boldsymbol v' \lesssim \epsilon_1^{1/2}.
\end{equation}

In summary, the amplitudes of the terms in the energy equation \eqref{eq:energy_terms} are:
\begin{equation}
\begin{split}
    &\underbrace{\langle\bar\rho v_{R,0}\partial_R e_{\rm tot,0}\rangle}_{\sim \epsilon_1}
    + \underbrace{\langle\bar\rho\partial_t(u+\tfrac 12 v^2)_0\rangle}_{\sim\epsilon_1\epsilon_2 + \epsilon_1\epsilon_3}
    + \underbrace{\langle \Phi_1\partial_t\bar\rho\rangle}_{\sim \epsilon_1\epsilon_2}
    + \underbrace{\langle \Phi'\partial_t\rho'\rangle}_{\lesssim\epsilon_1\epsilon_2^{-1/2}\epsilon_3\lesssim\epsilon_1}\\
    &\mbox{} + \partial_R\langle \underbrace{\bar\rho v_{R,0}(p/\rho)_0 }_{\sim\epsilon_1\epsilon_2}
    +\underbrace{\overline{\rho v_R}e_{\rm tot,1}}_{\lesssim \epsilon_1}
    +\underbrace{(\rho v_R)'e_{\rm tot}'}_{\lesssim \epsilon_1}\rangle
    = -2\pi R\Lambda_{\rm cool}.
\end{split}
\end{equation}
For the terms inside the operator $\partial_R\langle\,\dots\,\rangle$, the amplitudes labelled below them include the contribution from $\partial_R\langle\,\dots\,\rangle$.
We see that all terms on the left-hand side are indeed $\mathcal O(\epsilon_1)$ or less, making $\epsilon_1$ (or $-g_R\dot M$) a good order-of-magnitude estimate for the cooling rate required at thermal equilibrium.

We also note that when disc is infinitely thin ($\epsilon_2,\epsilon_3\to 0$) and all transport is local (i.e., self gravity is negligible), the energy equation consists only of the mean accretion term, $\langle \bar\rho v_{R,0}\partial e_{\rm tot,0}\rangle$, and the divergence of a turbulent energy flux, which is simply $\Omega$ times the turbulent angular-momentum flux. In that case one can obtain a more precise estimate of the heating rate. (This result still holds when other local angular-momentum transport mechanisms, such as turbulent magnetic transport through Maxwell stress, are present; cf.~\citealt{bgh94}.) In our case, the problem is complicated greatly by the presence of terms that transport only energy but no angular momentum (through advection of thermal energy and pressure work) or transport energy and angular momentum non-locally (through self-gravity). In this case, a more precise estimate of the heating rate is hardly possible without additional assumptions on the details of energy and angular-momentum transport mechanisms.

\section{Testing the analytic estimate of energy transport by spiral convection}\label{A:spiral_convection}

\begin{figure}
    \centering
    \includegraphics[scale=0.66]{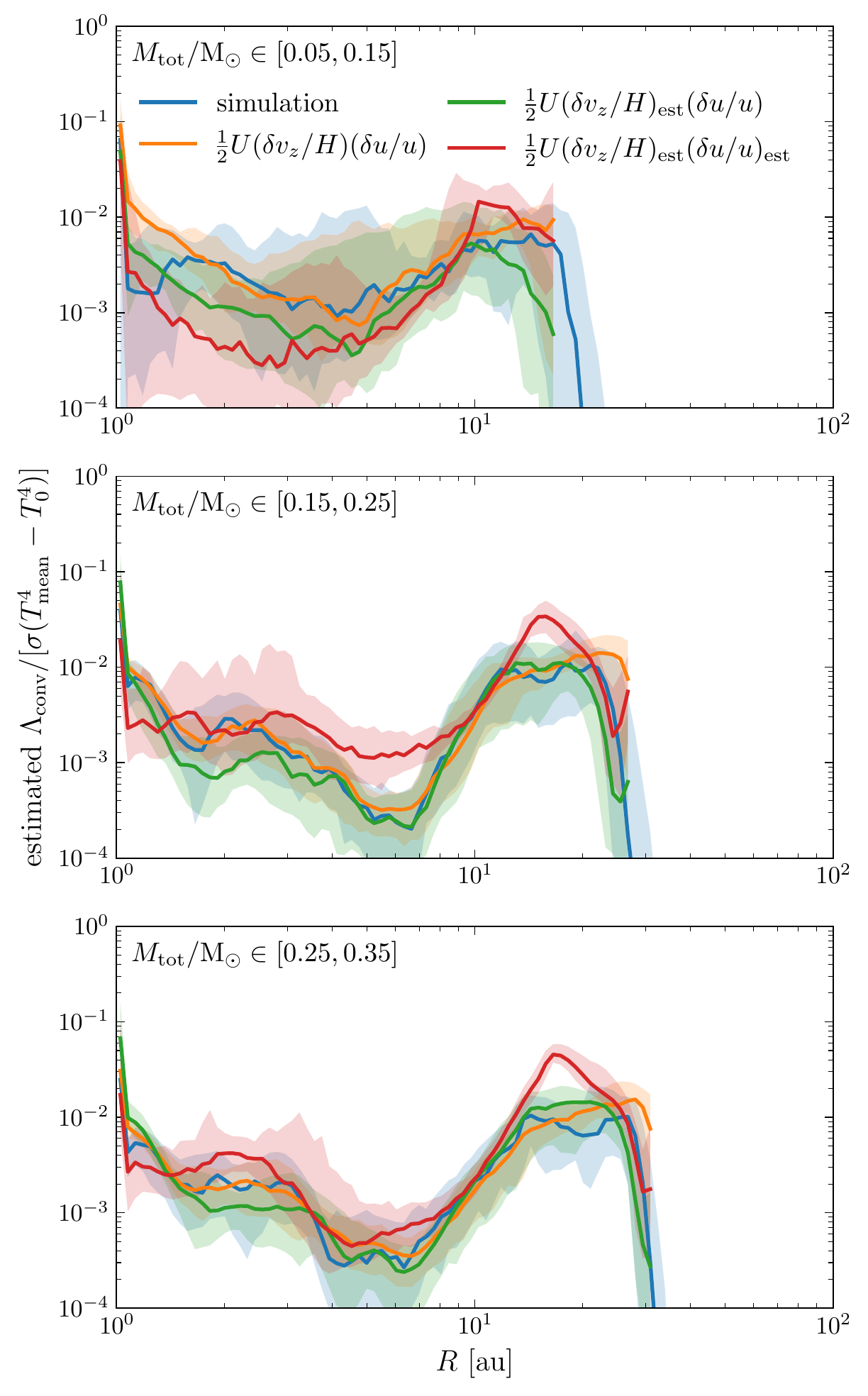}
    \caption{Vertical energy transport by convection, as measured from the simulation (blue) and estimated under different levels of approximation. The analytic estimates are only plotted for $R\leq R_{\rm d}$. All estimates are fairly good inside the disc.}
    \label{fig:cooling_rate_convection}
\end{figure}

In this appendix we test the accuracy of the analytic estimate of the rate of convective energy transport by spiral convection, equation \eqref{eq:convection_est}. Fig.~\ref{fig:cooling_rate_convection} compares the vertical energy transport by convection with estimates under different level of approximations. In this figure, $(\delta v_z/H)_{\rm est}$ uses equation \eqref{eq:dvz_est} and $(\delta u/u)_{\rm est}$ uses equation \eqref{eq:du_est}, where $\Lambda_{\rm heat}$ is estimated with $\Lambda_{\rm cool}$ measured from the simulation.\footnote{
These estimates depend on $\Omega_{\rm s} \equiv |\Omega - \Omega_{\rm p}|$. For the curves in Fig.~\ref{fig:cooling_rate_convection} the spiral-wave pattern speed $\Omega_{\rm p}$ is measured from the largest spirals in the outer part of the disc.
Once the GI region becomes extended (covering a large range in $\log R$), spiral waves in the inner disc are excited more locally ($\Omega_{\rm p}\sim\Omega$), and our choice of $\Omega_{\rm p}$ is no longer correct there.
However, this should not significantly affect the value of $\Omega_{\rm s}$, as $\Omega_{\rm s} \sim \Omega$ for the inner disc either way.
}
The comparison shows that the analytic estimates in Section \ref{sec:spiralconv} are all fairly accurate inside the disc.

\bsp	
\label{lastpage}
\end{document}